# Melt homogenization and self-organization of chalcogenides glasses: evidence of sharp rigidity, stress and nanoscale phase separation transitions in the $Ge_xSe_{100-x}$ binary

*S. Bhosle, K. Gunasekera, P. Boolchand*

School of Electronics and Computing Systems, College of Engineering and Applied Science

University of Cincinnati, Cincinnati, OH 45221-0030

*M. Micoulaut*

Laboratoire de Physique Theorique de la Matiere Condensee, Universite Pierre et Marie Curie

Boite 121, 4 Place Jussieu, 75252 Paris Cedex 05, France

A Raman profiling method is used to monitor growth of $Ge_xSe_{100-x}$ melts and reveals a two step process of homogenization. Resulting homogeneous glasses show the non-reversing enthalpy at $T_g$, $\Delta H_{nr}(x)$, to show a square-well like variation with x, with a rigidity transition near $x_c(1) = 19.5(5)\%$ and stress transition near $x_c(2) = 26.0(5)\%$ ) representing the boundaries of the rigid but stress-free Intermediate Phase (IP). The square-well like variation of $\Delta H_{nr}(x)$ develops sloping walls, a triangular shape and eventually disappears in glasses having an increasing heterogeneity. The $\Delta H_{nr}$ term ages over weeks outside the IP but not inside the IP. An optical analogue of the reversibility window is observed with Raman spectra of *as-quenched melts* and *$T_g$ cycled glasses* being the same for glass compositions in the IP but different for compositions outside the IP. Variations of Molar volumes, display three regimes of behavior with a global minimum in the IP and a pronounced increase outside that phase. The intrinsic physical behavior of dry and homogeneous chalcogenides glasses can vary sharply with composition near *elastic* and *chemical* phase transitions, showing that the physics of network glasses requires homogeneous samples, and may be far more interesting than hitherto recognized.

## 1. INTRODUCTION

Bulk glass formation occurs in insulating, semiconducting, metallic and H-bonded materials systems, but in a select range of chemical compositions. What is so special about these select melt compositions that can be cooled slowly to bypass crystallization and yield large (gram sized) bulk glasses? Important clues to understanding this unusual behavior evolved from Rigidity theory[1], which showed that bulk glass formation usually occurs when networks become isostatic[1] at an optimal connectivity. The theory starts from fundamental interactions including bond-stretching and bond-bending forces between atoms, and identifies metastability of glassy networks in terms of *non-local internal network stress*, and not intensive free energies. Thorpe independently identified[2] a new class of cyclical or floppy modes in simulations of realistically compacted yet fully disordered 3D mean-field models. By establishing the count of floppy modes as a function of the number of central and non-central valence bond forces – Phillips and Thorpe discovered the *Stiffness Transition*- the connectivity related flexible to stressed-rigid elastic phase transition, which has become the focus of modern theory of network glasses. To test these elegant ideas much experimental work[3-5] has been done in the field starting from the mid 1980s. And as data on several families of covalent (chalcogenides) and ionically modified covalent (modified oxides) systems evolved, it emerged starting in the late 90s that there are actually <u>two</u> distinct elastic phase transitions[6-8] and not just one as predicted[9]. These two transitions now widely recognized[3], are the *rigidity* transition followed by a *stress* transition observed at a slightly higher network connectivity. In *random* networks these two transitions coincide, which is to say that rigidity and stress both nucleate once the network connectivity exceeds the stiffness transition value of $r$ = 2.40 for 3D systems. Here $r$ represents the mean coordination number of a network. In real systems, networks apparently reconnect to minimize stress when $r$ is near 2.40, with the opening of an intervening region between the onset of *rigidity* and that of *stress now recognized as the Intermediate phase. The phase represents a* rigid but stress free state of disordered matter , also called self-organized[10, 11]. Experiments have also shown that the Intermediate Phase (IP) glass compositions possess unexpected physical properties[12] they are characterized by thermally reversing glass transitions[13, 14], barely age, possess characteristic elastic power-law, and form space filing networks. In this context, it has recently been shown[15] that the thermally reversing character of the glass transition in the IP may well be the consequence of the isostatic nature of networks. And in spite of substantial progress[16, 17] in understanding IPs in several families of chalcogenides and modified oxide glasses, challenges remain.

Experimentally, one of the more formidable challenges rests in synthesizing pure and homogeneous glass compositions at closely spaced compositional intervals to reproducibly probe the nature of the rigidity and stress elastic phase transitions. The width of these phase transitions in laboratory experiments appears to be limited largely by the heterogeneity of melts used in synthesis of non-stoichiometric glass compositions. The fact is that in glass science, since its inception, we have not had a diagnostic structural probe to experimentally track in real time heterogeneity of melts during synthesis. The result has been that melts have been reacted and equilibrated above the liquidus, typically for times ranging from a few hours to a several tens of hours [18-28]. Recently we introduced a Raman profiling method[29] to address the issue, and found that 2 gram sized melts of the well studied[7, 20, 22, 23, 30-33] $Ge_xSe_{100-x}$ binary take at least 168 h (7 days) to homogenize on a scale of 10 μm when reacted at 950°C. Why is the process of melt homogenization slow? In this paper we address the issue. We will also show that traces of water doping enhance homogenization by a factor of 3 or more; however, the resulting homogeneous glasses possesses physical properties that are measurably different from those of their dry counterparts. Differences in $T_g$, molar volumes, and enthalpy of relaxation of $T_g$ between dry and wet glasses is traced to some of bridging Se sites in the former replaced by dangling [OH] and [H] ends in the latter. To establish the intrinsic physical behavior of chalcogenides glasses it is necessary to synthesize dry samples. The present experiments show that once glasses have been homogenized they self-organize and display rather abrupt rigidity and stress transitions with widths less than 0.5% in Ge content(x). This is a remarkable finding because it illustrates that the intrinsic physical properties of covalent glasses can change abruptly particularly near elastic phase transitions. The finding of sharp elastic phase transitions, we anticipate will stimulate discussions amongst theorists and experimentalists alike, and will assist in unraveling the fundamental nature of these critical points including the elusive nature of glass transition[34].

In this work we provide details of synthesis of homogeneous binary $Ge_xSe_{100-x}$ glasses, and rather complete optical, thermal and mechanical characterization. The binary chalcogenide is perhaps one of the most well studied glass forming systems [7, 20, 22, 23, 30-33, 35]. The diagnostic role of the non-reversing enthalpy at $T_g$ as a probe of homogeneity and purity of batch compositions has come to the fore as we illustrate here. These findings permit identifying physical properties of glasses that are *intrinsic* to these materials, to be distinguished from those that are extrinsic caused either due to incomplete homogenization of a batch composition, or presence of water traces, or lack of complete relaxation of a sample. Variation in the glass transition temperature, $T_g(x)$, in dry and homogeneous $Ge_xSe_{100-x}$ samples is now accurately described in terms of a polynomial that can be used to predict a glass composition x, if $T_g$ is measured. In these homogeneous glasses, the jump in $C_p$ near $T_g$ from the glass to the liquid ( $\Delta C_p(x) = C_p(liquid) - C_p(glass)$ ), deduced from the reversing heat flow in modulated DSC experiments, is found to be *independent of x* over a wide range of composition, $10\% < x < 33.33\%$. These $\Delta C_p(x)$ results do not support the suggested[36] correlation between melt fragilities m(x) at $T > T_g$, with $\Delta C_p(x)$ of glasses at $T < T_g$. One the other hand, melt-fragilities correlate well with enthalpy of relaxation $\Delta H_{nr}(x)$ at $T_g$ deduced from the non-reversing heat-flow in m-DSC experiments[37]. The correlation between m(x) and $\Delta H_{nr}(x)$ appears to be a promising avenue to understanding the fundamental nature of the glass transition[13, 14].

After a discussion of the equilibrium phase diagram (section 2), we describe experimental results in section 3. In section 4 we discuss these results, and conclude with the principal findings in section 5.

## 2. EQUILIBRIUM PHASE DIAGRAM OF Ge-Se BINARY

The equilibrium phase diagram of the $Ge_xSe_{100-x}$ binary taken from ref [38] appears in Fig. 1. In the Se-rich domain, there is a eutectic near x = 5.5(5) % of Ge at a T = 212ºC. At the eutectic temperature a liquid of $Ge_{5.5}Se_{94.5}$, c-Se and c-GeSe$_2$ coexist as suggested by the solidus (horizontal line) at 212°C. The phase diagram shows that the liquidus ($T_l$) steadily increases (broken line) from 212°C at x = 5.5% to 742°C at x = 33.3%, and serves as a guide in synthesizing glasses as we discuss in section 3. Congruently melting stoichiometric crystalline compounds exist at x = 0 ( c-Se)[39], x = 33.3% ( c-GeSe$_2$)[32], x = 50% (c-GeSe)[40]. In addition, there is a metastable crystalline composition c-Ge$_4$Se$_9$[41] formed at x = 30.7%. The structure of this metastable crystalline phase has a close bearing to the 2D form of GeSe$_2$. The metastable form present in the phase diagram plays a role in the aging experiments performed on the bulk glasses in the present binary, and we discuss the issue in section 4.

The phase diagram shows that when a melt of $Ge_{15}Se_{85}$ composition is cooled past the liquidus to T = 300°C, it will decompose into a liquid of $Ge_{10}Se_{90}$ composition and c-GeSe$_2$. Cooling it further to the Eutectic temperature of 212°C, will result in segregation of the sample into two phases, which can be described by the following stoichiometric relation,

$$Ge_{15}Se_{85} = \alpha\,(Se) + (100 - \alpha)\,GeSe_2 \qquad (1)$$

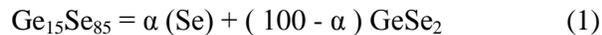

with α = 55%. Thus, it is quite reasonable to expect $Ge_xSe_{100-x}$ melts in the 0 < x < 33.33% range when cooled across $T_l$ to either (i) completely segregate to form c-Se and c-GeSe$_2$ upon slow cooling, or (ii) upon a fast quench to form a completely homogeneous glass of the melt stoichiometry and avoid decomposition. Alternatively, an intermediate circumstance can occur; the bulk glass formed may possess Se-rich and GeSe$_2$-rich regions that would result in microscopic heterogeneities (MH) determined by the considerations above (Fig.1). In the next section we will describe synthesis of bulk glasses and show that, in general, at short reaction times (< 6 hours), melts are indeed, quite heterogeneous, and do indeed possess Se-rich and Ge-rich glassy regions, and even crystalline GeSe$_2$ –rich regions. But as they are reacted for extended time, melts homogenize rather slowly if they are dry, and quickly if they are wet, a process that we have monitored by Raman profiling the quenched melts.

## 3. EXPERIMENTAL

In this section we provide details on synthesis of bulk chalcogenide glasses paying particular attention to homogenization of melts (3.1.). These data are then followed by characterization of bulk glasses by Raman scattering (3.2.), modulated differential scanning calorimetry (3.3.) and molar volume experiments (3.4.) in indicated sections.

### 3.1. *Synthesis and homogenization of $Ge_xSe_{100-x}$ melts*

$Ge_xSe_{100-x}$ melts of 2 grams in size were synthesized using 99.999% Ge and 99.999% Se pieces (3-4 mm diam) from Cerac Inc. The starting materials were mixed in the desired ratio by weight, sealed in evacuated ($10^{-7}$ Torr) quartz tubing (5mm ID) using a hydrogen/oxygen torch. The vacuum line consisted of a LN2 trapped High Vacuum pumping system. Prior to sealing, quartz tubes were held in a vacuum oven at 80°C for 24 hours. A total of 21 sample compositions spread in the 10% < x < 33.33% range of Ge were prepared. The quartz ampoules were held vertically in a T-programmable box furnace for varying time periods, $t_R$, ranging from 6 hours to 168 hours at 950 °C. Upon heating melts to 950°C, the liquid column was noted to reflux vigorously. Periodically *melts* were quenched from 50°C above $T_l$ (liquidus, Fig. 1) and examined in Raman profiling experiments (see below). Once homogenized, *as-quenched melts* in quartz tubes were taken to $T_g+ 20°C$ in the box furnace, held there for 10 minutes, and then slow cooled to room temperature at 3°C/min to realize *homogeneous bulk glasses*. We describe the homogenization of *melts* in 3.1.1. In section 3.1.2 we elucidate the role of laser spot size in monitoring homogenization of melts. In section 3.1.3 we address the issue of melt size and homogenization time, finally in section 3.1.4, the role of rocking melts on the homogenization process.

### *3.1.1. Raman profiling of melts*

All Raman profiling measurements on melts made use of a Thermo Nicolet FT-Raman 960 bench, using 1.064 micron radiation from an Nd-YAG laser with a laser spot size of 50 microns. Spectra were acquired at 9 locations, spaced about 2.5 mm apart covering 25 mm length of the melt column. At each location, an acquisition took 7 minutes, and used 100 mW of laser power, with 200 scans yielding 2 cm$^{-1}$ resolution. A typical profiling scan, involving spectra taken at 9 locations, took about an hour of accumulation time. Profiling experiments were performed at all the 21 compositions after reacting them for about 192 h, and in each case melts were found to be homogeneous on a scale of 10µm (see below). At three compositions, x = 15%, 19% and 33.33%, data were recorded as a function of $t_R$ over 192h to monitor growth of melt homogeneity. These data provide a unified view of how melts homogenize.

**Ge$_{19}$Se$_{81}$ melt :** Such melts in the initial stages ($t_R$= 6 hours, Fig. 2), were found to be quite heterogeneous. We observe modes at the tube bottom (location 1), which are identified[32] with α-GeSe$_2$. The scattering strength of the modes decreased as one moves up from (location) 1 to 3, and vanish at 4.The scattering strength of the Se-chain mode (near 250 cm$^{-1}$) increases from 4 to 9, while that of the symmetric stretch of GeSe$_4$ tetrahedra (200 cm$^{-1}$) decreases in the same position range, showing that melts become steadily Ge deficient in going from 1 to 9. The Raman spectra are superimposed in Fig 3a, and provide a pictorial view of melt heterogeneity. Continued reaction ($t_R$= 24 h) of melts ( Fig. 3a), leads the α-GeSe$_2$ (Fig. 3b) phase to dissolve into the melt, but it is only after $t_R$= 96 h (Fig. 3c) that the phase has completely dissolved. At that point melt stoichiometry narrows to vary from 17% at 1 to 21% at 9. Note that the sequence of position colors in the inset of Fig 3c is replicated in the scattering strength increase of the Se- chain mode indicating that the Se content of glasses increases as one goes from position 1 to 9. It is useful to mention that the FT-Raman software normalizes spectra to the highest peak (Fig. 3c), and for that reason no scattering is noted on the Ge mode at 200 cm$^{-1}$. Our data show that a fully homogenized melt is realized only after $t_R$= 168 hours (Fig. 3d) when all the 9 line-Raman spectra coalesce into a single spectrum. Noteworthy is the fact that scatter in the low frequency range (100-180 cm$^{-1}$) (Fig. 3c) , near the Se-chain mode, and near the 300 cm$^{-1}$ (ES mode) is absent in Fig. 3d as melts homogenize. These data show that the small spread in Ge-stoichiometry of 4% prevailing across the melt at $t_R$= 96h (Fig 3c) took an additional 72 hours of reaction (Fig. 3d) to completely disappear, and the melt to homogenize. The feature of slow homogenization of melts appears also at x = 15% and 33.33%.

**$Ge_{15}Se_{85}$ :** The homogenization behavior of such melts was similar to the one discussed above for $Ge_{19}Se_{81}$. At a lower Ge content, less α-$GeSe_2$ formed at the tube bottom, and after 24 h of reaction nearly most of it dissolved in the melt (fig 4b). Continued reaction leads ($t_R$= 96 hours) of melts, lead to increased homogeneity (Fig. 4c) with the variation in Ge stoichiometry across the melt narrowing to about 2% ( 14% at 1 and 16% at 9), as suggested by the scattering strength variation of the 200 $cm^{-1}$ relative to the 250 $cm^{-1}$ mode. But a fully homogenized melt was realized only after $t_R$=168 hours (Fig. 4d) when all the 9 Raman spectra coalesce into a single one, and with the scattering strength variation in the low frequency (100-180 $cm^{-1}$) and high frequency (320 cm-1) domain vanishing. These data along with the one at x = 19%, illustrate that the process of homogenization has broadly two steps, one going from Fig. 4a to 4c leading to the appropriate local structures being formed, and the second step from Fig. 4c to 4d, resulting in a global homogenization of the melt.

**$Ge_{33.3}Se_{66.66}$ :** At high Ge concentration new features appear in melts not seen earlier, as illustrated in Fig. 5-7 that reproduce the profiling scans after 6h, 24h and 96 h. Now we observe modes near 170 $cm^{-1}$ and 230 $cm^{-1}$ not seen earlier at lower x. Their origin comes from Ge-rich amorphous phases such as ethanelike $Ge_2Se_6$ and distorted rocksalt GeSe phase (section IV). Few sharp modes are observed at 5 (Fig. 5) towards the center of the melt column, and these are readily identified with α-$GeSe_2$. After 24 hours of melt reaction (Fig. 6), modes of the Ge-rich amorphous phases decrease, and now those of α-$GeSe_2$ appear at tube bottom (1, Fig. 6) and persist until the middle of the melt column (6, Fig. 6). The upper half of the melt column shows modes of CS and ES units, but with the Ge-content of melts steadily decreasing from positions 5 to 9. Continued reaction ($t_R$= 96h) of melts, promotes homogeneity (Fig. 8) as α-$GeSe_2$ phase, and the two Ge-rich amorphous phases dissolve into melts, but with a surprising result. We observe large scattering at low frequency near 120 $cm^{-1}$. The scattering strength increases as we move towards the melt column center (Fig. 7). The increased scattering near 120 cm$^{-1}$ ( 1-6 in Fig. 7) is not a specific mode but really the quasi-elastic scattering increasing at low frequency and being cut off by the spectral response of the FT system near 100 cm$^{-1}$. The buildup of quasi-elastic scattering after 96h of reaction is the signature of a long scale heterogeneity of melts particularly towards the melt center, features that are strikingly observed in Fig. 8c. The data of Fig. 8d, unambiguously show that GeSe$_2$ melt homogenizes globally only after reaction time $t_R$ = 192 h. These data at x = 33.33% (Fig. 8) along with those at x = 19% (Fig. 3) and 15% (Fig. 4),reinforce the view that once melts acquire the characteristic local structures of underlying glasses in the first step of homogenization, it takes an additional 80-100 hours of reaction at 950°C for a batch composition in the second step to globally homogenize.

An interesting observation was made during homogenization of GeSe$_2$ melts. After reacting melts for $t_R$ = 180 hours, Raman profiling data showed that the batch had homogenized (Fig. 9), except for just one position, 7, where α-GeSe$_2$ had formed. Further inspection of the sample showed location 7 to coincide with the meniscus cavity tip, a singularity that apparently nucleated crystallization. By rocking the melt for an additional few hours we could completely homogenize the sample. We discuss the observation in section 4.

### 3.1.2. *Raman profiling, laser spot size and spatial homogeneity of melts*

The Thermo Nicolet FT-Raman system, has in its micro-setting provisions for two laser spot sizes, 250 μm and 50 μm. In an earlier study[42, 43] we had probed Ge$_x$Se$_{100-x}$ melts using a laser spot size of 250 μm. In those experiments, we found melts to homogenize in $t_R$ = 96h (Fig. 11), i.e., in a shorter time than in the present work. However, measurements of the reversibility window in the earlier study[13, 42] revealed walls that were quite wide in relation to the ones in the present work as we discuss later. Clearly, laser spot size in these Raman profiling experiments intrinsically sets the spatial resolution at which melts are probed for homogenization. And it appears dry chalcogenide melts homogenize slowly, and must be homogenized on at least a 50μm scale to observe their intrinsic physical behavior in compositional studies.

Once homogenized, melts were separately examined using a Dispersive Raman system model T64000 from Horiba Inc.. In this system the scattering was excited using 647 nm radiation (Kr- laser), and the laser beam brought to a 10 μm laser spot size using a confocal microscope attachment with a 10X objective. In Fig. 10, we compare the Dispersive Raman profiled data on a $GeSe_2$ glass sample taken at 10 μm resolution with FT-Raman profiled data taken at 50 μm resolution. These data unequivocally show that reaction of melts at 950°C at 168 h leaves them homogenized on a scale of 10 μm.

### 3.1.3. *Melt size and reaction time to homogenize*

Do melt sizes play a role in the kinetics of homogenization? To address the issue we synthesized a ¼ gram sized melt of $Ge_{19}Se_{81}$ and monitored its structural evolution as a function of $t_R$ in FT-Raman profiling experiments. We found the melt homogenized in only 6h (fig. 12c) while the 2 gram melt took 168h (Fig. 12b) to homogenize. For comparison, in Fig 12, we compare Raman profiling results on a 2 gram melt with the ¼ gram melt both reacted for 6h. It is abundantly clear from these data that melt sizes play a crucial role in the kinetics of homogenization, and we discuss the issue in section IV.

### 3.1.4. *Rocking of melts and homogenization*

At two compositions (x = 19% and 25.5%) we investigated the effect of rocking on melt homogenization. Quartz tubes were positioned horizontally in a muffle furnace, and the furnace rotated continuously in a vertical plane by 45 degrees at a rate of 1/10 cycles per second. After reaction of the elements at 950°C for 48 hours, Raman profiling data on these rocked melts showed that the appropriate local structures had formed, i.e., no crystalline phases were observed suggesting that the 1$^{st}$ step of homogenization had been speeded up (Fig. 13a and d). However, a significant variation in Ge stoichiometry across the batch composition still persisted. At that point samples were transferred to a box furnace and kept vertical and reacted further. At $t_R$ = 120 h, melts were profiled and found still not to be completely homogeneous (Fig. 13b and e). However, at $t_R$ = 168h, melts did homogenize completely as shown in Fig. 13c and f. Note that the lineshape spread near the Se- chain mode (250 cm$^{-1}$), in the region of the low frequency band (100-180 cm$^{-1}$) and near the ES mode (320 cm$^{-1}$) disappeared after 168h. It is this slow step 2 of homogenization of chalcogenides melts that is a recurring theme in the equalization of melt stoichiometry across batch compositions globally. We discuss it in section IV.

*3.2. Dispersive Raman scattering in glasses*

All Raman scattering measurements on glasses made use of a dispersive system (Model T 64000 , Horiba, Jobin Yvon Inc). A 5mW quantity of 647 nm radiation from a Kr-ion laser with a 50 μm spot size was brought to a line focus on a glass samples contained in evacuated quartz tubes used in their synthesis. The typical laser power density on samples was at 10W/cm$^2$. The back scattered radiation was analyzed in the triple subtractive mode using a CCD Detector. An accumulation typically lasted 2 mins. The rationale for using Dispersive Raman measurements rather than FT-Raman measurements to probe the physics of glasses is discussed in Appendix 1 and elsewhere[44]. The observed Raman lineshapes in glasses were analyzed as a superposition of Gaussian profiles using Peak Fit software. Examples are provided later. In section 3.2.1., we elucidate the role of water as an impurity in glasses, in section 3.2.2., we compare Raman lineshapes of as-quenched melts with $T_g$-cycled glasses, and finally in section 3.2.3, we provide the observed lineshapes in the homogenized glasses.

*3.2.1. Water as a dopant in glasses*

We synthesized a pair of samples (x = 19% and 33.33%) of 2 gram in size, this time using finely powdered (< 5μm) elemental Ge and Se, which were left in the laboratory ambient environment (45% Relative Humidity) for 24 hours. In our case we found a 2 gram batch size of finely ground $Ge_{20}Se_{80}$ glass picked up 4.5 mg in weight after 24 hours exposure to laboratory environment. These data suggest an uptake of 1 water molecule for 85 atoms of the glass. These starting materials were encapsulated in evacuated (10$^{-6}$ Torr) quartz tubing and reacted at 950C the usual way. Melts were homogenized and Raman profiled. Surprisingly, after reacting the melt at x = 19% for $t_R$= 42h, it completely homogenized as illustrated in Fig 14a. The behavior is in sharp contrast dry melts that took 168h to completely homogenize (Fig 3c). Melt at x = 33.33%, took a 72 h to completely homogenize as illustrated in Fig.

14d. These data unambiguously show that traces of water speed up the kinetics of melt homogenization rather remarkably, an issue we will discuss in section IV.

Dispersive Raman scattering on these wet melts when compared to their dry counterparts show the presence of *residual scattering* as shown in Fig. 15. For example, at x = 19% (Fig. 15 a) we observe that modes of the CS and ES tetrahedra and CM sit on a baseline that is measurably higher than the corresponding modes in the lineshape of dry samples. A parallel behavior is noticed at x = 33.33% (Fig. 15b). Furthermore, in Fig. 15a, we note that the Se chain mode scattering strength in the wet sample is lower than in the dry sample (Fig. 14a). Presence of water impurities in melts leads to residual scattering of the laser light. We return to discuss these Raman data along with those on calorimetric and molar volumes to elucidate the role of water impurities in chalcogenides in Section 4.

### 3.2.2. *Raman scattering of "as quenched" melts compared with " $T_g$ cycled" glasses*

Once melts were found to be homogenized in Raman profiling experiments, we undertook to compare Raman scattering of the *as-quenched melts* with those of the $T_g$-*cycled glasses* slow cooled to room temperature making use of the Dispersive system. The $T_g$-*cycled glasses* were obtained by taking the *as-quenched melts* in quartz tubes, and heating to $T_g$ for 10 min., and then slow-cooling to room temperature at 3°C/min. in a box furnace. It is useful to mention that samples were always retained in the same quartz tubes used in their synthesis until the glasses were thermally relaxed and brought to 23°C, i.e., they were not exposed to the laboratory environment. The purpose in undertaking such Raman scattering investigations was to establish in what way, if any, does glass molecular structure of the *as-quenched* samples differ from their $T_g$ *cycled* counterparts, hopefully shedding some light on changes of structure accompanying a cool down across $T_g$. The results are summarized in Figures 16-18. These dispersive macro-Raman scattering measurements used 647 nm radiation, with samples contained in original quartz tube used for synthesis, to avoid photo-oxidizing or degrading samples due to exposure to humid laboratory environment. Each panel in these figures compares two lineshapes, the *as-quenched* melt with the $T_g$-*cycled* glass, at a total of 9 compositions.

Starting at x = 15% (Fig. 16), we find that the scattering strength of the Ge centered CS and ES tetrahedra decrease in going from the melt to the glass, a feature also observed in T-dependent Raman measurements by Murase[45]. But as x increases to 19%, close to the onset of the IP, lineshapes become quite similar, and in fact vanish as one goes across the IP glass compositions (x = 22%, 23% and 24%, Fig. 17), only to grow again as x increases to 29%, 31% and 33%. Remarkably, the absence of a change in the scattering strength ratio of the CS mode to the Se chain mode across $T_g$ at x = 25% was also noted by Murase in T-dependent Raman experiments[45]. These data show that molecular structure changes across $T_g$ are minuscule for IP glass compositions, demonstrating that these represent the optical analogue of the reversibility window (see below).

Raman profiling scans at 18 of the 21 melt compositions synthesized in the present work appear in Fig. 19. At each composition, 9 Raman spectra were recorded, and they completely overlap into a single lineshape, which is truly representative of these melt compositions, and we discuss these in Section 4.

### 3.2.3. *Raman scattering results on dry homogeneous glasses*

Dispersive Raman data on $T_g$-*cycled* homogeneous glasses appear in Fig. 20 at select compositions. As mentioned earlier, these data were acquired using a macro-Raman configuration with glass samples in the original evacuated quartz tubes used for synthesis to avoid hydrolysis. Furthermore use of the cylindrical surface of a quartz tube brings laser light to a line focus and thereby reduces the laser power density on samples, and suppresses photo-structural effects. The $T_g$ cycled samples, as mentioned earlier, were obtained by heating homogenized melts in a box furnace to Tg and then slow cooling to RT at 3°C/min. Lineshape evolution with x shows a characteristic pattern, which is qualitatively similar to earlier reports[7, 23, 46, 47]. There is broad agreement on mode assignments; Se-chain mode near 250 cm$^{-1}$, the Corner-sharing Ge tetrahedral mode (200 cm$^{-1}$), the ES GeSe$_2$ mode (217 cm-1, 320 cm-1), the ethanelike[48] mode (180 cm-1), outrigger[46] Se-Se mode (245 cm-1). The pair of modes (arrow locations), one at 245cm$^{-1}$ and another at 180 cm$^{-1}$ both appear first once x > 31.5%, and both grow simultaneously as x increases to 33.33%. We will discuss these data in section 4.

Lineshapes were deconvoluted to extract mode-centroids, -widths and -scattering strengths. An example of a typical lineshape fit appears in Fig. 21, corresponding to a glass sample at x = 20%. A plot of the CS mode frequency, $\nu_{CS}(x)$, with glass composition x ( Fig 22) displays three distinct regimes; at low x (15% < x < 20%) $\nu_{CS}(x)$ variation is rapid, at intermediate x ( 20% < x < 26%) the $\nu_{CS}(x)$ variation is slower than at low x and almost quasi-linear, and finally at high x (26% < x < 33.3%), we observe a power-law variation in x.

The frequency and scattering strength of the Se-Chain Mode (CM) appear in (Fig. 23). On these plots we have included corresponding data obtained by Jin et al.[49] on samples that were reacted at 950°C for 4 days. The samples used by Jin et al. [49] were homogenized using a 250 micron spot size in FT - Raman profiling experiments, and we now know were not as homogeneous as the present ones. The CM frequency, $\nu_{CM}(x)$, blue shifts steadily with x, but at x > 31.5% the behavior is reversed- the mode starts to red-shift. The scattering strength of CM normalized to the CS mode, $I_{CM}(x)/I_{CS}(x)$ (Fig. 23b) decreases steadily with increasing x and , as expected, approaches zero when x increases to 33.33%.

The frequency and scattering strength variation of the ES mode with x (Fig. 23) show a rather rich behavior. The mode frequency variation, $\nu_{ES}(x)$, displays three distinct regimes of variation- at low x (15% < x < 20%), the mode frequency increases rapidly, at intermediate x (20% < x < 26%), we observe a quasi-linear behavior, and at high x ( 26% < x < 33.33%) the mode frequency increases as a power-law in x. The variation $\nu_{ES}(x)$ is reminiscent of the variation in $\nu_{CS}(x)$ (Fig. 22) discussed above.

A plot of the ES mode integrated scattering strength normalized to the CS one, $I_{ES}(x)/I_{CS}(x)$, appears in Fig. 23d. In the low x range, i.e., 10% < x < 20%, the ratio starts out near 0.222) and decreases to 0.18 (2) near 20%. At higher x, 20% < x < 26% range, the ratio increases as a power-law in x. To deduce the ES/CS fraction, $N_{ES}/N_{CS}$, from the Raman scattering strength ratio, $I_{ES}(x)/I_{CS}(x)$, one must fold in matrix element effects[50] as we discuss in section 4.

It is useful to mention here that the $N_{ES}/N_{CS}$ fraction has been deduced from neutron structure factor measurements[51]. On the plot of Fig. 23, we project these data and find that they are in reasonable agreement with the Raman $I_{ES}(x)/I_{CS}(x)$ mode scattering strength ratio. The neutron results were obtained by a first principles modeling of the glass structure at two stoichiometric composition $GeSe_4$ and $GeSe_2$, and fitting the observed neutron structure factors to deduce the concentration of these species[21, 51, 52]. The $N_{ES}/N_{CS}$ fraction has also been reported recently by $^{73}Ge$ NMR[23] and those data are also quite similar to the present Raman scattering strength ratio. On the other hand, analysis of the electronic density of states from XPS measurements[18] reveal the $N_{ES}/N_{CS}$ fraction to be at 3/7 and to be independent of x in the 20% < x < 30% range.

*3.3. Calorimetric measurements on homogenized glasses*

A model 2920 MDSC from TA Instruments Inc., operated at 3° C/min scan rate, 1°C T- modulation, and 100s modulation period was used to study the enthalpy of relaxation at $T_g$. In these experiments one deconvolutes the total heat flow into reversing- and non-reversing heat flow components[53, 54] ( Fig. 25). The reversing heat flow captures quasi-equilibrium thermodynamic properties of the metastable glass state, specifically its heat-capacity jump ($\Delta C_p$), and from the inflexion point of the heat flow the $T_g$. Compositional trends in $T_g(x)$ and $\Delta C_p(x)$ appear in Fig 26(a) and (b) respectively. We find $T_g$ increase monotonically as x increases, while $\Delta C_p(x)$ terms remains fixed near 0.35 cal/gm in the present set of samples. Data on $\Delta C_p(x)$ on the $Ge_xAs_xSe_{100-2x}$ ternary[55] also reveal a similar behavior and are included in the plot.

We have fit the $T_g(x)$ variation in our homogeneous glasses to a polynomial and the results are given by equation (2) below.

$$T_g(x) = 39.781 + 8.702x - 0.271x^2 + 0.011x^3 \qquad (2)$$

The smooth line in the plot of fig. 26a is a plot of equation 2, and it reproduces the observed $T_g$s to an accuracy of typically 2°C in most cases except near the composition x = 31.5%. The slope $dT_g/dx$ reaches a maximum near x = 31.5% (Fig. 26a) and has a bearing on the misfit near that composition. We also observe a cusp in the $\Delta H_{nr}(x)$ term near the same composition x=31.5%. These data are a signature of nanoscale phase separation[56] of the present glasses at x > 31.5%, and we will comment on the issue in section 4.

The non-reversing heat flow captures non-equilibrium effects[54], including network configurational changes that occur upon softening of a glass. This component usually shows a peak as a precursor to $T_g$. By subtracting the integrated area under the peak observed upon cooling (exotherm) from the peak observed upon heating (endotherm), one obtains the frequency corrected $\Delta H_{nr}(x)$ term[53]. For glasses in the composition range, 20% < x < 26%, we find the $\Delta H_{nr}(x)$ term to show a global minimum ( ~0) (Fig. 26 c), and to abruptly increase at x < 20% and x > 26%, displaying a square well like behavior. The global minimum in $\Delta H_{nr}(x)$ term in the 20% < x < 26% range, is the reversibility window, and it is a feature of associated with isostatic networks[15].

We also examined the effect aging samples at room temperature and at 240°C. For these measurements samples hermetically sealed in Al pans were rerun after 2 weeks of aging at room temperature, and these data (A1) appear in Fig. 26c as the open circles (red). All compositions except those in the 20% < x < 26%, show a general increase in the $\Delta H_{nr}(x)$ term upon aging, with the step near x = 19.5% becoming abrupt, but not the one near 26%. Samples at higher x (>26%) possess a $T_g$ that exceeds 260°C. These glasses were aged at 240°C for two weeks, and the data ( A2) reveal the $\Delta H_{nr}(x)$ term to now show an abrupt increase near x = 26%. Some glass compositions, particularly at x = 29%,30%,31.5%, 32%, partially crystallized upon aging at 240°C. These sample compositions were examined in XRD investigations and we shall comment on these data in section 4.

The thermal properties of wet glasses differ remarkably from their dry counterparts. Thus, for example, $T_g$'s of wet samples are found to be lower than dry ones; Furthermore, the $\Delta H_{nr}(x)$ term for wet samples is measurably larger than for dry ones. These data are illustrated in Table 1. These variations in calorimetric and optical data between wet and try samples form part of a general behavior that can be traced to bonded water in the network producing dangling ends as we discuss in section 4.

*3.4. Molar volumes of homogenized glasses*

Molar volumes of the homogeneous bulk glasses (Fig. 27) were measured using a Mettler Toledo model B154 balance (with 0.1mg resolution) and a quartz fiber suspended from the pan. Bulk glasses were weighed in air and Ethyl alcohol (200 Proof, Anhydrous ACS/USP grade) and the density obtained using Archimedes principle. In these experiments use of samples greater than 200 mgms was sufficient to achieve an accuracy of ¼% in density. A silicon wafer was used to calibrate the Alcohol density and a single crystal of Ge used to ascertain the accuracy of the density measurements. Molar volumes were calculated from the measured density, and show a rather striking variation with glass composition: in the 20% < x < 26% range, the *reversibility window* , we observe a broad minimum, while outside this window molar volumes increase precipitously by about 4% in the present homogeneous samples. We have also projected in the plot of Fig. 27, the $V_m(x)$ data for the two wet samples synthesized at x = 19% and 33.33%, and find that, in general, $V_m(x)$ decrease in wet samples, a behavior noted earlier in oxides[57] as well. In the case of the composition, x = 33.33%, the $V_m(x)$ reduction is about 2.6%. At the composition x = 19%, close to the reversibility window, the reduction in $V_m(x)$ is much smaller, about 0.3%.

We are aware of three previous studies [20, 30, 31] where rather complete $V_m(x)$ trends on the present binary are reported. Broadly speaking our $V_m(x)$ trends bear similarity to previous reports with some caveats. In two cases[20, 30] a broad minimum in $V_m(x)$ is observed in the reversibility window, except the increase of $V_m(x)$ observed for glass compositions outside the window is nearly halved. The latter, most likely, is a manifestation of an *intrinsic* heterogeneity of glass samples used in the previous reports. The $V_m(x)$ data of Feltz et al.[30] is about 2% lower than the Mahadevan data[20] across the board, and the difference could either reflect samples of Feltz et al. being either partially wet or even a calibration offset in the density of the liquid standard used for density measurements

4. **DISCUSSION**

In this section we will discuss the following issues: variation of glass transition temperature (section 4.1.), melt-homogenization and nanoscale mixing (section 4.2.), Rigidity and Stress Transitions in $Ge_xSe_{100-x}$ glasses (section 4.3.), Ideal glasses, melt-fragilities and IP glasses (section 4.4.), and finally onset of Nanoscale Phase separation at x > 31.5% (section 4.5.)

*4.1. Variation of glass transition temperature*

Glass transition temperatures, $T_g$, of bulk materials are determined by several factors including melt quench rates[58-60], network connectivity[61], sample purity[57] and the scan rates[54] used to measure them. In the present study all 21 melt compositions, after a water quench, were cycled through Tg and then cooled at 3°C/min to room temperature. Clearly factors other than quench rates contribute to the observed increase of Tg with Ge content "x" of the glasses. As mentioned earlier, the $T_g$'s were established from the inflexion point of the reversing heat flow in mDSC experiments, and the values reported (Fig. 26a) are the average of $T_g$ measured scanning up and then scanning down in temperature at a rate of 3°C/min. At these low scan rates, kinetic shift in $T_g$ due to finite scan rates are minuscule (~1°C). More importantly, by scanning up and then down in T, we have obtained the mean $T_g$, which is independent of scan rate.

The compositional variation of $T_g$ (Fig. 26a) observed in our experiments directly reflects the increased connectivity of the network backbones. The idea first emerged from Stochastic Agglomeration theory[62] which has been used to quantitatively establish $T_g(x)$ trends in the stochastic or low Ge alloying regime in the present binary. Previous work[56] in the field has demonstrated that at x < 10%, the measured slope, $dT_g/dx$, is in excellent agreement with the parameter free SAT prediction of this slope. In this range of composition the crosslinking of $Se_n$ chains by Ge atoms proceeds in a *random* fashion. At higher x (> 10%), new building blocks contribute to the agglomeration process, with the result that the slope changes and the agglomeration process ceases to be stochastic. In the composition interval, 20% < x < 26%, $T_g(x)$ become thermally reversing as glasses self-organize (see below), and the variation $T_g(x)$ cannot be described adequately by SAT. And as x increases to 31.5%, the slope $dT_g/dx$ maximizes corresponding to a network that is fully polymerized. At x > 31.5%, excess Ge first begins to segregate as Ge-Ge signatures first appear, and the backbone becomes partially polymerized (see below).

The variation of $T_g(x)$ in the 10% < x < 33.33% range can be well described in terms of a third order polynomial (equation 2). The nature of $T_g$ changes with composition however, and this is an issue that comes to the fore as established by the enthalpy of relaxation[10, 14], which displays characteristic regimes, which have emerged from an interplay between rigidity theory[15] and mDSC experiments[13].

To facilitate a more direct comparison of the present results with earlier DSC work in the field, we have also performed DSC measurements on the present homogeneous samples using a 10°C/min scan rate. We find that the DSC derived $T_g(x)^{DSC}$ (Fig. 28) to be about 6°C lower than the mDSC derived $T_g(x)^{mDSC}$ (Fig. 26a), a difference that results from the method used to define $T_g$. In mDSC, $T_g$ is defined from the inflexion point in the reversing heat flow, while in DSC $T_g$ is defined from the inflexion point of the (total) heat flow. In Fig.28, data of Guin et al.[63] are projected, and their $T_g$ about 19°C lower than the present results, while those of Sreeram et al. [64] are much closer to the present work at x = 10% and 22% but much less so at other compositions where data is available. On the plot of Fig. 28 we have also projected $T_g(x)$ for a wet sample at x = 19% synthesized in the present work. Our $T_g(x)$ data for the wet sample nicely agrees with the trends reported by Guin et al. [63] These data suggest the possibility that samples of Guin et al. [63] may have water traces.

Within Rigidity Theory other descriptions of the variation of $T_g(x)$ have emerged. Naumis et al.[65] have shown the Lindemann melting criteria in conjunction with the mean-squared displacement of atoms can be used to account for the variation in $T_g(x)$ in the present binary. The result provides a neat origin of the usual Gibbs-Di Marzio equation which predicts the variation of the glass transition temperature in cross-linked polymeric glasses (such as Ge-Se). The parameters of the Gibbs Di Marzio equation are found to be related to the underlying topology of the glass network, as already demonstrated by other authors[61, 62]. Compositional trends in glass transition temperatures have also been used for quantitative design of glassy materials using temperature –dependent constraints within rigidity theory [66].

*4.2. Melt homogenization and nanoscale mixing*

The Raman profiling data presented earlier (Figures 2-11) provides new insights into physical processes that lead to homogenization of chalcogenides melts. The reaction of elemental Ge with Se in evacuated quartz tubing at 950°C is broadly consistent with *two* steps that underlie the homogenization process of melts. The presentation in this sub-section is in three parts as follows; First step- Formation of characteristic melt local structures (section 4.2.1.), Second step-Nanoscale mixing and global homogenization of a melt composition (section 4.2.2.),and finally, why wet melts homogenize quicker than dry ones? (Section 4.2.3.)

### *4.2.1. First step: Formation of characteristic melt local structures*.

In the initial stages, when Ge-Se mixtures are heated to 950°C, the sloshing liquid running up and down the reaction tube is predominantly molten Se, ($T_m$= 221°C), and with increasing reaction time it alloys ts with increasing amounts of Ge. Because the density of liquid Ge ( $\rho_{Ge}$ = 5.60 gms/cm$^3$ exceeds that of Se ($\rho_{Ge}$ = 3.99 gms/cm$^3$) , in the initial stages the tube bottom largely contains Ge-rich melts that may have crystalline inclusions. In melts containing low Ge content, such as $Ge_{15}Se_{85}$, we observe a small amount of α-$GeSe_2$ to form at the tube bottom (Fig. 4) in the initial stages ($t_R$ = 24-48h), an illustration of Microscopic heterogeneity (MH). With increasing Ge content of melts ( as in $Ge_{19}Se_{81}$), more of the α-$GeSe_2$ forms (Fig. 3) in the initial stages at the tube bottom. In melts at x = 33.33% (Fig. 5), new Ge-rich crystalline phases appear in the initial stages. Thus, for example, in the spectra of Fig. 5 we observe modes near 175 cm$^{-1}$ and near 220 cm$^{-1}$ , which are replicas of the $A_g$ (170 cm$^{-1}$) and $B_{2g}$ (230 cm$^{-1}$) phonons of the distorted rocksalt structure of c-GeSe.[40]. More exactly the feature near 175 cm$^{-1}$ is really a composite of two vibrational modes- one from the distorted rocksalt structure (170 cm$^{-1}$) and the other from ethanelike $Ge_2Se_6$ local structures (180 cm$^{-1}$).These Ge-rich phases form near the tube bottom (loc 1,2,3) as expected. Moving up along the tube, we then observe a mode near 180 cm$^{-1}$ of ethanelike units ($Ge_2Se_6$) at locations 5,6,7 and a sharp mode near 210 cm$^{-1}$ from α-$GeSe_2$. Towards the top of the tube, we observe a broad mode near 200 cm$^{-1}$, which is the symmetric breathing mode of CS mode of Ge(Se$_{1/2}$)$_4$ tetrahedra. Clearly, the equilibrium phase diagram provides guidance in understanding the phases formed in the initial stages of reaction in these Ge-Se melts(Fig. 1), and these contribute to MH in <u>partially</u> reacted melts as shown by the Raman profiling data (Fig. 3,4,8).

Upon progressive reaction, these Ge-rich crystalline and/or amorphous phases dissolve in melts, and appropriate local structures of glasses evolve. These local structures include CS-, ES- tetrahedra and Se$_n$ chain fragments[7, 46]. At the end of step 1, melts are not homogeneous however, as is illustrated in the data of Figures 3c,4c and 8c. Melt stoichiometry, measured in terms of Ge content x, typically varies anywhere from 3 to 6 % across the length of the liquid column. In Raman scattering (Fig. 3c, 4c and 8c) evidence for such heterogeneity is deduced from the scattering strength ratio of the Se$_n$ chain mode to the CS mode. In Fig. 3c, for example, we find melt stoichiometry varies from x = 21% at location 1 (tube bottom) to x = 17% at location 9 (tube top). The 4% spread in Ge content across the length of the tube for a melt of Ge$_{19}$Se$_{81}$ average stoichiometry is a typical result. Parallel results are observed for Ge$_{15}$Se$_{85}$ and GeSe$_2$ melts in Fig. 4c and 8c. In dry melts, the first step usually entails reaction times, t$_R$, of up to 96 hours.

### *4.2.2. Second step: Nanoscale mixing and global homogenization of a melt composition.*

We view the second step of homogenization to be *nanoscale mixing* as Ge atoms diffuse up and Se atoms down the reaction tube and melts globally homogenize. The process involves a sequence of bond-breaking and bond-forming steps as Ge diffuses, and the concentration gradients dissipate and batch composition homogenizes. The case of GeSe$_2$ melts is diagnostic (Fig 7, location 4,5,6) in this respect. We observe growth in scattering near 110 cm$^{-1}$, which is due to quasi-elastic scattering caused by melt heterogeneity at intermediate and extended range scales. Here we have to remember that in these FT-Raman data, the spectral response cuts off near 100 cm$^{-1}$ because of notch filters, and the peak observed at the arrow location near 110 cm$^{-1}$ (Fig. 7) is not a vibrational mode but it is evidence of increased quasi-elastic scattering as one approaches the laser line. Melts in the lower-half of the tube, and especially at locations 3,4 and 6 appear to display significant quasi-elastic scattering, which we suppose is due to heterogeneity of melts on an extended scale. In sharp contrast, as melts are reacted for $t_R$ = 192 h, not only does the quasi-elastic scattering vanish but also the spread in lineshape at higher frequencies also disappears as melts homogenize (compare Fig. 8c with d) as all 9 spectra completely coalesce to yield a solitary lineshape. The result is not peculiar to GeSe$_2$ melts, it is observed in Ge$_{15}$Se$_{85}$ melts ( compare Fig. 3c and d) and also in Ge$_{19}$Se$_{81}$ ( compare Fig. 4c and d). Our experiments show that in dry melts, nanoscale mixing of Ge with Se typically takes about 96h of reaction time (192h- 96h) , as the Ge/Se fraction across the 2 gram batch composition <u>globally</u> homogenizes. From these data we can estimate a diffusion constant ($D_{exp}$) of Ge and Se atoms in liquid GeSe$_2$ at 950°C by using the Einstein relation,

$$D = (1/6)(x^2/t) \qquad (3)$$

Taking a path length (x) of 2.5 cm for Ge and Se atoms to diffuse in an amount of time (t) of 96h in the reaction tube, one obtains a $D_{exp}$ = 3 x 10$^{-6}$ cm$^2$/s. From MD simulations, the viscosity data on liquid GeSe2,[41] Micoulaut and Massobrio have obtained[67, 68] a Diffusion constant of Ge and Se of 10$^{-5}$ cm$^2$/s in liquid GeSe$_2$ at 950°C. In bulk glasses realized by $T_g$ cycling such homogeneous melts, we have found that the calorimetric properties are quite uniform. In particular, the $\Delta H_{nr}$ term appears not to display variations across a batch composition, as noted in glass samples that were synthesized by reacting the starting materials for 48 hours and were clearly heterogeneous. Our experience reveals that the enthalpy of relaxation $\Delta H_{nr}$ deduced from mDSC experiments serves as a rather diagnostic probe of glass sample homogeneity.

### 4.2.3. Why wet melts homogenize quicker than dry ones?

An important finding of the present work is that *wet* melts synthesized at x = 19% and 33.33% homogenized much quicker than their dry counterparts. In synthesizing wet melts, the starting elements (Ge,Se) were finely powdered and left in laboratory environment for just 24 hours prior to sealing them in evacuated quartz tubes. It is widely known that in such powders the large surface to volume ratio of the micron sized particles leads to adsorption of water from ambient air. And it is difficult to remove water from such starting materials by merely pumping even with a high vacuum line at room temperature.

Viscosity of pure Se melts is found to reduce upon alloying chain terminators such as halogens and Tl [69] but it increases measurably upon alloying chain cross-linkers such as Ge as networks polymerize. Melts containing traces of water vapor will transform bridging Se sites , i.e., Ge-Se-Ge signatures to Ge-[OH] .... [H]-Se- Ge ones, creating [OH] and [H] dangling ends . Monovalent [OH] and [H] species will also serve as $Se_n$ chain terminators and assist Ge to react with $Se_n$. For this reason a wet 19% melt completely homogenizes in only 42h (Fig. 14a), while its dry counterpart (Fig. 3d) took nearly 168h to homogenize. A parallel circumstance occurs at x = 33.33%, where a wet melt completely homogenized in 72h (Fig. 14(d)) while its dry counterpart took 192h to completely homogenize (Fig. 8d). These data underscore the crucial role of water impurities in promoting melt nanoscale mixing.

While it is tempting to add traces of water vapor to accelerate homogenization of melts in the present chalcogenides, the fact is that the presence of traces of water impurities measurably <u>alters</u> the thermal, optical and mechanical properties of melts/glasses. Thus, we find $T_g$(x = 19%) of a dry sample of 171.6°C, is 13.6°C higher than the $T_g$ of its wet counterpart (158°C). A parallel circumstance occurs at x = 33.33% , where $T_g$ of a dry sample (425.7°C) exceeds that of its wet counterpart ($T_g$ = 420.6 °C). The lower $T_g$s of the wet samples compared to dry ones is due to a loss in connectivity of the Ge-Se backbone as dangling ends form. And their presence lowers $T_g$ as network connectivity decreases, a finding that is entirely consistent with SAT[62]. A perusal of the data of Table 1 also shows that the $\Delta H_{nr}$ term at $T_g$ in wet glasses is significantly larger than in dry countereparts. We understand the increased enthalpy of relaxation near $T_g$ as the rocking of dangling [OH] and [H] ends, an entirely non-ergodic process, as a glass softens near $T_g$, and is therefore manifested in the $\Delta H_{nr}$ term exclusively. Molar volumes of wet glasses (Table 1) are found to be lower than their dry counterparts because of a loss in network structure due to cutting of the network. The behavior is observed in oxide glasses as well [70]. Thus, the physical properties of wet glasses are distinctly different from their dry counterparts. These findings lead naturally to the notion that if Ge-Se melts of 2 gram in size homogenize in less than 68h of reaction time they are probably wet.

*4.3. Rigidity and stress transitions in $Ge_xSe_{100-x}$ binary*

We begin the discussion here by first commenting on the existence of an optical analog of the reversibility window (section 4.3.1.). This is followed by identification of the three elastic phases in the present glasses (section 4.3.2.). Comments on the Intermediate Phase are then presented in section 4.3.3. A model description of the variation of the non-reversing enthalpy with sample heterogeneity is presented in section 4.3.4. Finally, we discuss variation of molar volumes and Intermediate Phase in section 4.3.5.

*4.3.1. Optical analog of the reversibility window*

The first indications that batch compositions in the 19.5% < x < 26% range, behave quite differently from those outside the compositional window emerged in comparing the observed Raman lineshapes of the "as quenched" melts with their "$T_g$ cycled" glass counterparts (compare Fig. 16 with 15 and 17). In figure 16, we compare the Raman scattering of " as quenched" x = 22%,23% and 24% melts with their "$T_g$ cycled" glass counterparts, and find that the observed lineshapes of the two are quite similar. These data suggest that melt compositions in the window upon cooling across $T_g$, show little or no change in molecular structure. These are features that one identifies with <u>strong</u> melts. The behavior is in sharp contrast to the one encountered for melt compositions outside the IP as exemplified by melt compositions at x > 26% (Fig. 17), which show not only increased residual scattering and but also systematically higher concentration of ES units compared to their " $T_g$ cycled" counterparts. In these compositions, upon cooling to T < $T_g$, more ordered structures evolve. A similar pattern occurs for glass compositions x < 20%. In particular for the composition at x = 19%, close to 20%, differences in lineshapes between as quenched melts and $T_g$ cycled glasses become minuscule. But at x = 17% and at x = 15%, the fraction of the CS mode decreases substantially in the $T_g$ cycled glass. Melts of these batch compositions that reside outside the IP range clearly undergo substantial configurational changes upon structural arrest at $T_g$ and are viewed as fragile [41], leading to increased activation energy of viscosity as one goes away from the window.[15, 37].

*4.3.2. Three Elastic Phases*

The rigidity and stress transitions in amorphous networks has been demonstrated to be percolative in nature by the Pebble game algorithm[71-73]. These simulations show isostatically rigid clusters first percolate at the Rigidity transition. With a further increase of network connectivity, redundant bonds first onset near the stress transition in the network at the second transition. In between these two elastic phase transitions, the IP, one has rigid but stress-free networks populated with unusual properties. Numerical experiments have also shown that elasticity in the stressed-rigid phase increases as a power-law in *r* or network connectivity [74, 75]. The power-law prediction of optical elasticity has been observed in Raman scattering and IR reflectance experiments as discussed elsewhere [76, 77].

***Corner-Sharing mode.*** The variation in the CS mode frequency (Fig. 21) has been analyzed to extract optical power-laws in the stressed-rigid (Fig. 29a) and IP (Fig. 29b) using the relation,

$$\nu^2 - \nu_c^2 = A(x - x_c)^p \qquad (4)$$

and yields a power $p_2 = 1.50(3)$ in the stressed-rigid and $p_1 = 1.1(1)$ in the IP respectively. In equation (4) $\nu_c$ represents the frequency at the elastic threshold $x_c$. To reliably ascertain the power-law we have analyzed the CS mode frequency variation using two methods, a polynomial fit and separately a log-log plot to the $\nu_{CS}(x)$ data using equation 4. The polynomial fit approach is not sensitive to $x_c$ but the log-log plot approach most certainly is. The starting value of $x_c$ is varied so that the elastic power-law using both approaches converges to the same value. The dual approach to analyzing the $\nu_c(x)$ gives reliable elastic thresholds of $x_c(1) = 19.5\%$ for the rigidity- and of $x_c(2) = 26.0(3)\%$ for the stress- transitions respectively.

The power-law of $p_2 = 1.50(3)$ in the stressed-rigid phase may be compared to the predicted value of 1.50[74] and 1.40[75]. These results strongly support that glassy networks at $x > x_c(2)$ are stressed-rigid. There are currently no theoretical predictions available for the elastic power-law in the IP, although this is not because of a lack of an attempt to obtain them[78].

Variation in the scattering strength of the CS mode measured at a fixed laser-power density as a function of glass composition x is plotted in Fig. 24a. These data display an almost linear variation with the intensity extrapolating to zero at $x = 0$. This is not an unexpected result given that alloying Ge in the base Se glass increases the fraction of CS tetrahedral units in the backbone in proportion to the Ge cross-linking fraction 'x'.

*ES-sharing Mode.* The mode frequency and mode scattering strength of the ES mode in our homogeneous glasses have also been analyzed. The mode scattering strength at a <u>fixed</u> laser-power, is found to display three distinct regimes (Fig. 24(b)), it increases linearly in the $10\% < x < 26\%$ region, increases as a power-law at $x > 27\%$ with a power $n_1 = 2.30$ in the $26\% < x < 31.5\%$ range, and a power-law $n_2 = 1.40$ at $x > 31.5\%$. For reference purposes, we have also indicated the rigidity and stress transitions compositions, $x_c(1)$ and $x_c(2)$ in the plot of Fig. 24b.

The ES mode frequency variation, $\nu_{ES}(x)$ (Fig 22c), also displays three distinct regimes of variation (Fig 22c). The data at $x > 26\%$ in the stressed-rigid regime has also been analyzed in our homogeneous glasses, and shows an elastic power-law $p_2$ of 1.50(3) (Fig. 30a) from the polynomial plot and $p_2 = 1.47(3)$ from the log-log plot (Fig. 30b) with both approaches converging to yield the same value of the stress transition $x_c(2) = 25.5(5)\%$. It is thus comforting to see that the Raman mode frequency variation of both CS and ES tetrahedra yield the same elastic power-law ($p_2$) and stress transition ($x_c(2)$) phase boundary, thus confirming the details of the elastic phase transition at $x_c(2)$.

*Variation of the ES/CS fraction in $Ge_xSe_{100-x}$ glasses*. The Raman Scattering strength ratio of ES/CS (Fig.22d) shows a value of about 0.20 in the $10\% < x < 20\%$ range, it increases slowly to 0.22 in the $20\% < x < 26\%$ range, and then increases as a power-law in the $26\% < x < 33.3\%$ range to acquire a value of 0.37. The Raman scattering strength ratio must be normalized by matrix element effects to obtain the ES/CS fraction in the network. These were estimated[50] by K.Jackson et al. using cluster calculations, who found the Raman cross sections to excite the ES and CS mode to be respectively 40.5 $A^4$/amu and 47.9 $A^4$/amu. It is difficult to reliably ascertain the uncertainty in these cross sections, but if one assumes an error of 10%, then these cross-sections are about the same, suggesting that the Raman scattering strength ratio also represent the ES/CS fraction in the glassy networks. We have projected in Fig. 22d the neutron scattering results for the ES/CS fraction at x = 20%[33] and at x = 33.33%[21, 51] from recent ab-initio analysis of the neutron structure factors. The ES/CS fraction has also been inferred from preliminary $^{77}$Se NMR experiments[23], and the reported results are in reasonable accord with the Raman results.

### 4.3.3. Intermediate Phase

The power of mDSC in probing thermal transitions in general, and the glass transition in particular has come to the fore in recent years [53]. The variations in $T_g(x)$ and $\Delta H_{nr}(x)$ in the present glasses (Fig 28), independent of scan rate, provides invaluable information on network structure and dynamics respectively. The $\Delta H_{nr}(x)$ term provides a measure of the hysteretic nature of the glass transition. The vanishing of the term in the 19.5% < x < 26% composition range provides the thermal signature that liquid and glass structures are configurationally very close to each other as one goes across $T_g$, and that glass transitions become <u>thermally reversing</u> in character. Earlier, we had noted that Raman lineshapes of melts upon cooling across $T_g$ show little or no change in this special compositional window. The present calorimetric data confirm the finding that glass compositions in this window possess liquid-like entropy.

The square well like variation of $\Delta H_{nr}(x)$ term in the present samples of proven homogeneity, is accentuated by aging. In the present samples, the $\Delta H_{nr}(x)$ term already shows a step-like variation near x = 20% in the rejuvenated (F) samples, but a more gradual increase of the term at higher compositions, x > 26%. Upon aging the samples at room temperature for 2 weeks (A1 in Fig. 28c), we find that the variation in $\Delta H_{nr}(x)$ near x = 20% now becomes abrupt, while the term increases for each composition at x > 26%, but the increase is small. This is expected because for these stressed-rigid glass compositions, room temperature T= 23°C < $T_g$ ( > 260°C, fig. 26c), one expects the kinetics of aging to be very slow. By aging these glass compositions at 240°C for 2 weeks, one can speed up the kinetics, as is observed in the variation of $\Delta H_{nr}(x)$ term, which now becomes abrupt (A2 in Fig. 26c) near x = 26%. It is quite remarkable that glass composition in the window show little or no aging attesting to the rather special state of matter in the intermediate phase.

Why does the variation of $\Delta H_{nr}(x)$ become abrupt near the rigidity (x = 19.5%) and stress transitions (26.0%) ? In the present homogeneous samples, we fundamentally observe the percolative nature of the two elastic phase transitions [71-73]. The vanishing of the $\Delta H_{nr}(x)$ term in the IP reflects the intrinsically isostatic nature of that phase. Recently Micoulaut has modeled[15] the enthalpy of relaxation at $T_g$ and shown that the term can be expected to be large in flexible and stressed-rigid networks but to show a minimum when networks are optimally constrained. From a general viewpoint, such a result can be understood qualitatively from energy landscape approaches. At low connectivity (in the flexible Se-rich compositions), the energy of the system is dominated by the presence a few principal minima corresponding to the bond energy between Se atoms and between Ge and Se. The number of these minima is obviously proportional to <r>, and contains also a contribution arising from floppy modes, because for each deformation mode, there are a certain number of energy minima. Overall, the complexity (the number of local minima) of the energy landscape is proportional to <r> which decreases when the concentration of Ge increases. At the other end, one has a rough energy landscape with an increased number of principal minima proportional to <r>. In between, there is a compositional region where <r> is optimal and the fraction of floppy modes small, leading to a simple energy landscape where relaxation is optimized.

If one follows the models proposed for the IP[11, 79] one finds changes at the thresholds defining the IP to be controlled by specific structural parameters. In the SICA approach [79], the jump at the stress transition is controlled by the fraction of edge-sharing (ES) units, with a smaller jump when the ES fraction is increased. The present findings suggest that the sharpness of the reversibility window upon ageing may result from a conversion of ES tetrahedral to CS ones. Fig. 16 shows that the intensity of the 217 cm$^{-1}$ mode corresponding to ES tetrahedra to decrease when a glass is $T_g$ cycled, i.e. relaxed (Figures 16-18). These findings suggest that a decrease in the fraction of ES units upon ageing correlates with the sharpness of the IP boundaries.

### *4.3.4. Variation of ΔH$_{nr}$(x) and glass heterogeneity*

The reversibility window in Ge$_x$Se$_{100-x}$ glasses was examined in two previous reports. The data in Fig. 31a are taken directly from the first report on the subject by X.Feng et al[7] and those in Fig. 31b from the second report [68]. In ref.[7], melts were reacted at 950°C for 2 days, while in ref. [68] they were reacted at 950°C for 4 days (Fig. 11). Finally, Fig. 31c shows the observed reversibility window in the present work where melts were homogenized at 950°C for 7 days and found to be homogeneous when FT-Raman profiled using a 50 μm laser spot size (Fig. 3,4,8). It is useful to mention that the source and handling of starting materials, elemental Ge and Se, including purity and lump sizes, the batch size (2 grams), and vacuum sealing of the starting materials was kept the same in all the three set of investigations. The only variable was the reaction time, $t_R$ of the starting materials at 950°C. These data of Fig. 31 highlight the crucial role of batch homogeneity on the variation of the ΔH$_{nr}$(x) term in glasses.

The square well like variation of ΔH$_{nr}$(x) with a width in W = 6.5% in the present glasses of proven homogeneity ( Fig. 31c), suggests a simple model for the correlation of the window width W with heterogeneity, Δx, of Ge-content across a melt composition. Here Δx represent the spread in the mean value of x across a Ge$_x$Se$_{100-x}$ batch preparation that is accessed from the Raman profiling experiments (Fig. 3,4,8). In this simple model, we assume ΔH$_{nr}$(x) to have a bimodal value as sketched in Fig. 32d, i.e., ΔH$_{nr}$(x) = 0 in the reversibility window and 1 outside the window in arbitrary units. One can then predict the variation of the IP width W as a function of the heterogeneity parameter Δx. In present glasses for simplicity, we take $\Delta x = 0$, and $W = 6.0\%$, starting from $x = 20\%$ to $x = 26\%$. Clearly then, if the heterogeneity parameter, $\Delta x = 3\%$, one expects $\Delta H_{nr}(x)$ term to vanish <u>only</u> at the window center, i.e., $x_{cen} = 23\%$, and the term to increase linearly as one moves away from the center composition towards the IP edges as contributions to the heat flow term from the flexible phase on the low side ($x < 20\%$), and the stressed rigid phase on the high side ($x > 26\%$) are manifested. The model predicts the variation in $\Delta H_{nr}(x)$ to be triangular (Fig. 32a). The behavior is reminiscent of the findings (Fig. 31a) of X.Feng et al.[7] The model calculation also predicts that if $\Delta x = 2\%$, $W = 2\%$ (Fig. 32b). And as $\Delta x$ decreases, one expects W to increase linearly as shown in Fig. 33. The linear variation of the model, $W = 6 - 2(\Delta x)$, provides a useful means to quantitatively characterize the expected heterogeneity of Ge-content of melts/glasses, $\Delta x$, if the width W of the reversibility window is measured. For the case of the glasses studied[80] in 2009, the observed reversibility window (fig. 31b) width W was found to be 4%. The plot of Fig. 33, predicts that for these melts/glasses, $\Delta x_{pre} = 1\%$. This particular set of melts were reacted for $t_R = 4$ days, and the present Raman profiling data of Figures 3, 4 and 8, place the heterogeneity parameter at $t_R = 96h$, $\Delta x_{obv} = 1\%$ at $x = 15\%$ and $\Delta x_{obv} = 2\%$ at $x = 19\%$, yielding an average variation of melt stoichiometry $\Delta x_{obs} = 1.5\%$, which may be compared to the model predicted value of, $\Delta x_{pre} = 1\%$. The $\Delta x_{obv}$ was obtained by taking the spread in scattering strength ratio of either the CS/ES or the CS/CM vibrational modes from the observed lineshapes. Thus, the model introduced here provides a useful means to directly correlate the calorimetrically measured value of the reversibility window width W with the Raman profiling experiments deduced melt/glass heterogeneity $\Delta x$. These data underscore the need to synthesize homogeneous samples to access the IP width W in Chalcogenides. An interesting spinoff of the model is that even when samples are <u>not</u> homogeneous, i.e., $\Delta x$ is finite and small (<3%), one can reliably infer the width W of the reversibility window by taking the separation between the mid-points of the walls ($W_{app}$, see Fig. 32). We find that $W_{app} = 1.1\ W$ for a wide range of glass sample heterogeneity in the $1\% < \Delta x < 5\%$ (Fig. 32) range. Finally, the plot of Fig. 32a shows that as glass samples become more heterogeneous, i.e., $\Delta x > 5\%$, the shape of the reversibility changes qualitatively and may even not be observed.

Our $T_g(x)$ results also reveal that glass samples that possess a small heterogeneity, $\Delta x = 2.5\%$, the glass transition temperature $T_g$ of such samples do not vary measurably from those of very homogeneous samples. This can be seen by comparing the variation of Tg in the present homogeneous samples ( smooth red line in Fig. 26a) with the green open circles giving the $T_g(x)$ variation in the samples reported by Feng et al. [7] in 1997. Thus, presence of some heterogeneity of glass composition is <u>not</u> reflected in $T_g$, but it reflected in pronounced fashion in the enthalpy of relaxation at $T_g$.

On the other hand, presence of water impurities in glasses, such as at x = 19% and x = 33.33% (Fig. 26a, c) investigated here, leads to a rather large depression of $T_g$ (171(1)°C to 158(1)°C at x = 19%, and from 425 (1)°C to 420 (1) ° C at x = 33.33%, and to a measurable increase in $\Delta H_{nr}(x)$ term (0.36(5) cal/gm to 0.55 cal/gm at x = 19%, and 0.52(5) cal/gm to 0.75 cal/gm at x = 33.33%). These data are summarized in Table 1. The reduction in $T_g$ is the natural consequence of a loss in network connectivity as OH- groups bond to Ge and H to Se replacing bridging Se in the network as discussed earlier. On the other hand the same loss of connectivity that produces the Ge-OH and Se-H dangling ends in the network also contribute to the $\Delta H_{nr}(x)$ term as the glass softens upon heating as T approaches $T_g$. The heat intake upon glass softening due to these dangling ends is non-ergodic in nature, and it contributes to the non-reversing enthalpy as expected. These data underscore the crucial role of melt purity and homogeneity in synthesis of glasses if extrinsic effects are to be minimized. Once care is taken to address these issues in synthesis of glasses, one may access the intrinsic physical properties of glasses rather than the impurity modified behavior.

### 4.3.5. *Variation of molar volumes and Intermediate phase*

A recurring theme in glass structure is the space filling nature of disordered networks such as proteins[81] and chalcogenide glasses. Molar volumes provide a direct measure of space filling of a glass, and in Fig. 27 we compare data on the present homogeneous $Ge_xSe_{100-x}$ glasses with earlier reports.

Remarkably, we find the IP glass compositions to possess the lowest $V_m(x)$, and the term to increase both in the stressed-rigid and the flexible phases. In the $V_m$ plot we also include data on two wet glass compositions at x = 19% and 33.33%, and find $V_m$ to lower drastically, a feature we have noted earlier in oxides[57] as well. Presence of bonded water in glassy networks cuts the backbone and, in general, lowers the molar volume. Presence of bonded water in these network glasses also lowers $T_g$ and increases $\Delta H_{nr}(x)$. The optical, calorimetric and mechanical properties examined in the present covalent glasses are suggestive of networks that are delicately nano-structured. Solid electrolytes show similar features with calorimetric, vibrational and electrical properties being strongly influenced by the presence of water impurities[57].

In the present chalcogenides the short range covalent forces determine the nature of local structures such as long- and short- $Se_n$ chains, ES and CS tetrahedral units and ethanelike $Ge_2Se_6$ units that form at different Ge content x. In addition, there are long range forces (Coulombic and van der Waals) that come into play as evidenced by the rather systematic variation in $V_m(x)$. The global minimum in $V_m(x)$ in the IP compositional window is most likely the consequence of a minimal count of floppy modes and redundant bonds that permits the network as a whole to adapt and reconnect and expel stress globally. The stress-free character of glass compositions in the IP was elucidated earlier in Pressure Raman experiments[82], in which a critical externally applied pressure $P_c(x)$ could be identified when Raman modes first blue shift. The pressure $P_c(x)$ provides a measure of network stress, and displays a trend that closely mimics that of $\Delta H_{nr}(x)$. In particular, one finds $P_c = 0$ in the IP range and to increase as one goes away from the IP both in the stressed-rigid regime and the flexible regime, mimicking closely the behavior of $\Delta H_{nr}$ term. These data all strongly point to a new functionality of adaptation acquired by networks in the narrow IP window.

The Molar volumes reported by Mahadevan et al. [20] are generally in good accord (Fig. 27) with those reported in the present work. But both set of $V_m$ results are measurably higher than those reported Feltz et al.[30] and Senapati et al.[31]. This could possibly be just an issue of recalibrating the standards used by Feltz et al. and Senapati et al. or simply due to samples that may not have been as dry as the ones used in the present work. Nevertheless, the range of variation in $V_m$ outside the IP are measurably higher in the present glasses than in the earlier reports, a feature that most likely derives from the intrinsic homogeneity of the present glasses.

*4.4. Ideal glasses, Melt Fragilities and IP glasses*

The correlation between the global minimum in $\Delta H_{nr}(x)$ with the Raman lineshapes that show little or no change upon cooling across $T_g$ for the window compositions (Fig. 16-18) is a profound result and raises fundamental issues. Both the calorimetric and optical data point to the fact that the configurational entropy change across $T_g$ for these privileged IP compositions is minuscule, i.e., glassy networks in the IP possess liquid -like entropy. Not surprisingly, melt fragilities also reveal a minimum for IP compositions. Here the melt fragility data at x > 25% is difficult to obtain from viscosity measurements because of the tendency of such melts to crystallize. Melts in the reversibility window range not only display high glass forming tendency [1, 35], but also form rigid but stress-free networks that age minimally. We associate these properties with self-organized glasses that are also *ideal* glasses which do not need a high cooling rate to avoid crystallization. This feature was qualitatively reported a long time ago [35] with slow cooling allowing only glass formation at network connectivity somewhat lower than the critical mean coordination number $r = r_c = 2.40$ of the rigidity transition. An increase of the cooling rate (from air quench to water quench) increased the glass-forming region up to $r = 2.67$. Similarly, it has been stated[83, 84] that ideal glasses are those which are able to increase their melt viscosity down to lower temperatures. For this reason, glass forms more easily at eutectics because freezing-point depressions bring the system to lower temperatures and higher viscosities with a weaker driving force to crystallization. However, when comparing Fig. 1 and the IP shown in Fig. 26, there is clearly no correlation between the location of an eutectic and the one of the IP. These findings show that it is, indeed, the rigidity of the system which controls the ease of glass formation.

Our calorimetric results on the present binary show that $\Delta C_p(x)$ is independent of x in the 10% < x < 33.33% range (Fig. 26b). In Fig. 26b, we compare the $\Delta C_p(x)$ data on the present samples with those reported by Feng et al. [7], and find that for both sets of data, within the errors of measurement, $\Delta C_p(x)$ = 0.035(5) cal/gm/C or 1.17R at $Ge_{20}Se_{80}$, where R represents the Gas constant of 8.3 Joule/mole/K. The $C_p$ glass (x = 20%) = 0.06 cal/gm/°C below $T_g$, yields a value of $C_p^{glass}$ (x = 20%) = 2.35R, and a $C_p^{liquid}$ (x = 20%) = 3.51R at T > $T_g$, a value somewhat greater than the Dulong Petit value for $C_v$ = 3R in monotomic solids. This is as it should be given that $C_p = C_v + \alpha^2 TV/K_T$, where α and $K_T$ represent respectively the thermal expansion and isothermal compressibility of the melt. A value of $\Delta C_p(x)$ = 0.035(5) cal/gm/°C was noted [55] earlier in the $Ge_xAs_xSe_{100-2x}$ ternary several years ago over a wide range of composition x (Fig. 26b). It is useful to emphasize that these $\Delta C_p(x)$ data were obtaining by analyzing the step in the reversing heat flow in mDSC experiments. If, on the other hand we were to deduce the jump in the specific heat, $\Delta C_p(x)$, from the total heat flow, as one normally does in a DSC experiment, our data shows a shallow minimum ( a change from 0.04(2) cal/gm/c to 0.02(2) cal/gm/C) in the reversibility window (Fig. 34 ). We believe that extraction of $\Delta C_p(x)$ from the total heat flow is polluted by the glass transition endotherm overshoot that is manifested in the total heat flow outside the IP. The large increase in the overshoot of the $T_g$ endotherm[85] in a DSC experiment, is actually a manifestation of an increasing $\Delta H_{nr}(x)$ term. The presence of this overshoot apparently influences a measurement of the $\Delta C_p(x)$ term in DSC experiments as demonstrated above. These difficulties are circumvented in using mDSC as a method to examine glass transitions, since contributions to the reversing and non-reversing heat flow are completely separated.

The mDSC results are compelling in suggesting that there appears little or no correlation between melt fragilities and $\Delta C_p(x)$ in the present chalcogenides. The $\Delta C_p$ term remains independent of x over a wide composition range ( 10% < x < 33.3%), a finding that is at odds with the prevailing view [36]. On the other hand, the present finding suggests that melt fragilities (Fig. 35) correlate well with the $\Delta H_{nr}(x)$ term. The correlation is physically appealing since both $T_g$ and $\Delta H_{nr}$ are of non-ergodic origin, underscoring the non-equilibrium nature of the glass transition. The $\Delta C_p$ term is of ergodic origin and most likely of vibrational character, which should be distinguished from the $\Delta H_{nr}$ term that is largely configurational in nature.

The present work on chalcogenides glasses shows that *ideal* glasses rarely occur in monolithic stoichiometric systems like $SiO_2$, $As_2S_3$, $B_2O_3$ or $GeSe_2$. They form at non-stoichiometric compositions and particularly in multi-component systems [55, 86, 87] where numerous isostatic local structures can open a wide compositional window of self-organization as networks adapt to expel stress from the backbone. These new ideas are in contrast to the prevailing view of an ideal glass[36] realized by slow cooling stoichiometric melts to approach the configurational entropy that is close to the corresponding crystal at a low temperature, usually identified as the Kauzmann temperature.

*4.5. Onset of Nanoscale phase separation in $Ge_xSe_{1-x}$ glasses at $x > 31.5\%$*

A chemically ordered continuous random network (COCRN) model structural description[47] of the present glasses appeared in the early 80s and gained popularity. Such a model was recently also favored based on FT-Raman scattering results [88]. A limitation of FT-Raman scattering as a probe of $Ge_xSe_{100-x}$ glasses is that the near IR excitations uses 1.06 μm radiation, and it largely excites the mid-gap states in these glasses where defects are manifested. The observed lineshapes in FT-Raman scattering are significantly broader than Dispersive Raman studies (see Appendix I). This has the consequence that compositional variation of CS and ES mode frequency in the FT-Raman experiments are generally washed out compared to those observed in Dispersive Raman experiments. For this reason, in the present work optical properties of glasses were deduced exclusively from Dispersive Raman studies. And as physical properties of these glasses were investigated more intensively, one found a non-monotonic compositional behavior of Molar volumes, Mossbauer site intensity ratios[56, 89], Non-reversing enthalpy of relaxation at $T_g$. These findings provided first indications that a COCRN model may be too simplistic a description <u>at all x</u>.

A chemically ordered continuous random network (COCRN) model requires that Ge-Ge bonds first appear once x > 33.33%, the chemical threshold. The observation of broken chemical order of $GeSe_2$ glass[89, 90], which initiates at x > 31.5% as Ge-Ge bonds first appear[56] in the network, again represent features of experimental data that are difficult to reconcile with a COCRN model. The maximum in the slope $dT_g/dx$ near $x_c(3)$ in the present glasses is the signature of segregation of Ge-Ge bonds in the network once they first nucleate near $x = x_c(3) = 31.5\%$. The structural evidence first emerged from $^{119}$Sn Mossbauer spectroscopy and Raman scattering experiments, which have revealed signature of homopolar Ge-Ge homopolar bonds to first appear in these glasses once $x > x_c(3)$. These Ge-Ge bonds form part of ethanelike units that apparently decouple or nanoscale phase separate from the backbone. The decoupling is suggested by the sudden decrease of the slope $dT_g/dx$ and the non-reversing enthalpy $\Delta H_{nr}(x)$ once $x > x_c(3)$ (Fig. 26c). Both $T_g$ and $\Delta H_{nr}$ are network topology determined properties of glasses, and the lowering of the slope $dT_g/dx$ and the $\Delta H_{nr}$ term at $x > x_c(3)$ reflects loss of network character due to demixing of some of the excess Ge ( at $x > x_c(3)$ ) from the backbone. In a COCRN model of these glasses, one expects Ge-Ge bonds to first appear at x > 33.33%, the chemical threshold, and the stoichiometric glass to be chemically ordered.

Why is the chemical threshold shifted to x < 33.33%? As the Ge content of the binary glasses increases to x = 31%, a small but finite fraction of Se-Se homopolar bonds form at the edges of a characteristic outrigger raft cluster, a moiety first introduced to account for chemical phase separation of the stoichiometric $GeSe_2$ glass[91]. The evidence for this moiety has come from Raman scattering[46] and $^{129}$I Mossbauer emission spectroscopy. The Raman vibrational signature of these edge Se-Se dimers was first correctly established by Murase et al. [46], who showed the Se-Se stretch mode occurs near 246 cm$^{-1}$ in a $GeSe_2$ glass. A perusal of the Raman lineshapes of the present glasses (Fig. 19), shows that this particular mode, and a corresponding mode associated with Ge-Ge bonds [48] simultaneously grow in the 31% < x < 33.33% range as the network steadily demixes. These optical data confirm the nanoscale phase segregation of these glasses noted earlier in $^{129}$I Mossbauer spectroscopy measurements [89] that showed evidence of a finite concentration of I-Se bonds persisting all the way to x = 33.33%. In these experiments, $^{129m}$Te tracer was alloyed in the binary glass, and one probes the Se environments and finds evidence of a bimodal (A,B) site distribution. The most unexpected finding of these experiments was that the $^{129}$Te dopant selects the chemically disordered outrigger Se (B) sites over the chemically ordered bridging Se sites (A) in the cluster interior by a factor of 70 or more. The observed integrated site intensity ratio $I_B/I_A$ = 1.7 in the stoichiometric glass, is due to the oversized Te segregating to cluster edges to minimize strain energy. In summary, the present thermal measurements of both $T_g$ and $\Delta H_{nr}(x)$ along with earlier Raman and Mossbauer spectroscopy[56, 89, 92] results provides persuasive evidence for onset of nanoscale phase separation of the present binary once $x > x_c(3) = 31.5\%$.

## 5. CONCLUSIONS

An FT-Raman profiling method is used to track structural heterogeneity of $Ge_xSe_{100-x}$ melts along the length of a reaction tube used in synthesis, and we find that a 2 gram sized melt takes about 196 hours of reaction time at 950°C to homogenize on a scale of 10 μm. The process of homogenization is viewed to consist of two broad steps; in the first step aspects of local structures characteristic of melts/glasses emerge after 72 hours of reaction of the starting materials. At the end of the first step, a variation in Ge stoichiometry of about 5% persists along the length of the tube, with the low end being Ge rich and the upper Ge deficient. An additional 96 hours of reaction time renders melts completely homogeneous across a batch composition as revealed by Raman lineshapes taken at 9 locations along the tube length completely overlaping. Such homogenized bulk glasses at 21 compositions spanning the 10% < x < 33.33% range were synthesized, and their physical properties further examined in Dispersive Macro-Raman, modulated DSC and Molar volume measurements. These data, on glasses of proven homogeneity, reveal sharply defined rigidity transition near $x_c(1)$ = 19.5(5)% and stress transition near ($x_c(2)$ = 26.0(5)%, with optical elastic power-laws in the <u>I</u>ntermediate <u>P</u>hase (IP: 19.5% < x < 26.0%) of $p_1$ = 1.10(5), and in the Stressed rigid phase ( x > 31.5%) of $p_2$ = 1.50(3). These experiments supported by theory show present glasses to be intrinsically nanostructured displaying several distinct regimes of variation; at low x (< 20%), Ge randomly cross-links $Se_n$ chains in the elastically flexible phase. At intermediate compositions (20% < x < 26%), networks acquire new functionalities including dynamical reversibility and non-aging, physical properties associated with self-organization. At high x (26% < x < 31%) networks continue to form fully polymerized networks that are now elastically stressed-rigid. At still higher x (> 31.5%) they first segregate into Ge-rich and Se-rich nanophases. Melts containing traces of water homogenize much quicker but their physical properties including $T_g$, $\Delta H_{nr}$, $V_m$ are found to be measurably different from their dry counterparts. Rigidity theory has proved to be a valuable tool to understanding the complex structural behavior displayed by this prototypical binary chalcogenide glass system.

**Acknowledgements**

It is a pleasure to acknowledge conversations with Dr. B.Goodman, Dr. D.McDaniel during the course of this work. The present work represents in part the MS Thesis work of Siddhesh Bhosle submitted to University of Cincinnati, and was supported in part by NSF grant DMR- 08-53957.

**Table 1**

|  | $T_g$ (°C) | $\Delta H_{nr}$ (cal/g) | $V_{mol}$ (cm$^3$mol$^{-1}$) | Comments |
|---|---|---|---|---|
| **19% Dry** | 171.6 | 0.360 | 18.34 | $t_R$= 168 h |
| **19% Wet** | 158.0 | 0.55 | 18.03 | $t_R$= 42 h |
| **33.33% Dry** | 425.7 | 0.52 | 18.87 | $t_R$= 192 h |
| **33.33% Wet** | 420.6 | 0.74 | 18.14 | $t_R$=72 h |

**Appendix I.**

**Comparing and contrasting Raman scattering on $Ge_xSe_{100-x}$ bulk glasses excited in the NIR (1.064 μm) with scattering excited in the red (0.647 μm).**

In the present work we have used FT-Raman to monitor heterogeneity of *quenched melts*, and Dispersive Raman to examine the vibrational density of states in homogenized *bulk glasses*. We have also performed FT-Raman experiments on the homogenized *bulk glasses* and in this Appendix compare these results with corresponding Dispersive Raman ones. It is useful to mention outright that frequency calibration of Raman modes in Dispersive measurements made use of the atomic transitions in Ne as discussed elsewhere[44]. On the other hand, the FT-Raman system uses the frequency of the Nd-YAG laser to calibrate the mode frequencies.

Fig. 36 gives a summary of the variation in the CS- mode frequency and CS-mode width as a function of Ge-content of the bulk glasses for both the FT- and Dispersive measurements. At x = 10%, we find the FWHM of the CS mode in the FT measurements (15.2 $cm^{-1}$) is nearly 30% larger than in the Dispersive one ( 11.7 $cm^{-1}$) (Fig. 36b). At x = 33.33%, in the stressed–rigid glasses, the CS-mode line width is 17.6 $cm^{-1}$ in FT-Raman, and 15.0 $cm^{-1}$ in Dispersive Raman experiments, i.e., about 17% larger. In the Dispersive Raman measurements, the compositional variation in the CS- and ES-mode frequency ($\nu_{CS}$) displays three distinct regimes of variation (Fig. 36), but that variation is generally smeared in the FT Raman measurements, and one, in general, one observes a smoother monotonic variation of both $\nu_{CS}(x)$ and $\nu_{ES}(x)$ (Fig 36). Recently Sen et al. [88] have reported the CS mode frequency variation in the present binary using a Bruker FT Raman system. Their data on the variation of $\nu_{CS}(x)$ is projected in Fig 36, and their results are quite similar to those obtained in the present work using the present FT- Raman set up.

The optical band gap ($E_g = E_{53}$, corresponding to an optical absorption coefficient $\alpha = 5 \times 10^3$ cm$^{-1}$) in amorphous $Ge_xSe_{100-x}$ thin-films is found to increase monotonically from 2.15 eV at x = 15% to $E_g$ = 2.38 eV[49] at x = 33.33% . These data place the midgap energy in these binary selenides at about 1.27 eV. In the Dispersive Raman measurements we made use of the 647 nm excitation ( 1.96 eV) radiation from a Kr-ion laser. Here the radiation excites electronic states in the conduction band tail states. On the other hand, in the FT-Raman measurements the exciting radiation was 1.064 μm ( 1.14 eV), and excited electronic states reside near the midgap region. The mid-gap region is also the region where 'defect' states are principally populated[93, 94] in amorphous materials. The defects include Valence alternation pairs, homopolar bonds. Thus, in present $Ge_xSe_{100-x}$ binary , because of the larger optical gap, the FT experiments serve to largely probe the "defected regions" of the network.

We can thus understand the broader linewidth of the CS mode in the FT Raman experiments compared to the Dispersive ones. The broader linewidth of the CS mode also leads to some washing out of the $\nu_{CS}(x)$ variation that appears to increase smoothly with x, in sharp contrast to the three regimes of variations with kinks near the elastic thresholds in the Dispersive Raman measurements.

It is also useful to mention that the integrated intensity of the CS mode ( $I_{CS}(x)$) at a fixed laser power density in the Dispersive measurements shows a linear variation (Fig.24) in the examined range of Ge content , 10% < x < 33.33%. As the Ge content of the binary glasses increases ( Fig 24a) , one expects the concentration of the CS units to increase linearly. These data do not allow for a "resonant" behavior of the Raman scattering as x increases to 33.33%, an observation that is largely consistent with the photon energy $E_{ph}$ = 1.96 eV being sufficiently below $E_g$ = 2.38 eV. On the other hand, the integrated intensity of the ES mode, ($I_{ES}(x)$) (Fig 24b) shows a more interesting behavior: $I_{ES}(x)$ is linear in the 10%, x < 26% range, but becomes super-linear in the stressed-rigid regime (27% < x < 31.5%) with a power-law (~ $x^{n1}$) behavior of n1 = 2.30(5). The super-linear behavior continues in the NSPS regime (31.5% < x < 33.33%) with a somewhat reduced power-law $n_2$ = 1.40(5). These data reflect that ES units growth contributes to the evolution of the stressed-rigid phase. These data again do not support a 'resonant Raman behavior'.

Finally, in Fig. 37, we directly compare the Raman lineshapes observed in FT-measurement with the one in the Dispersive measurement at two compositions, x = 20% (Fig 35a) and x = 33.33% (Fig 35b). Several features of the data become transparent, (i) the widh of the CS mode in FT experiments is larger than in the Dispersive measurements, and (ii) frequency of the CS and ES modes in the FT experiments is somewhat greater than in the Dispersive measurement. We have already commented the role of midgap defect states that may contribute to line broadening of the CS mode in the FT-measurements for the case of the present binary.

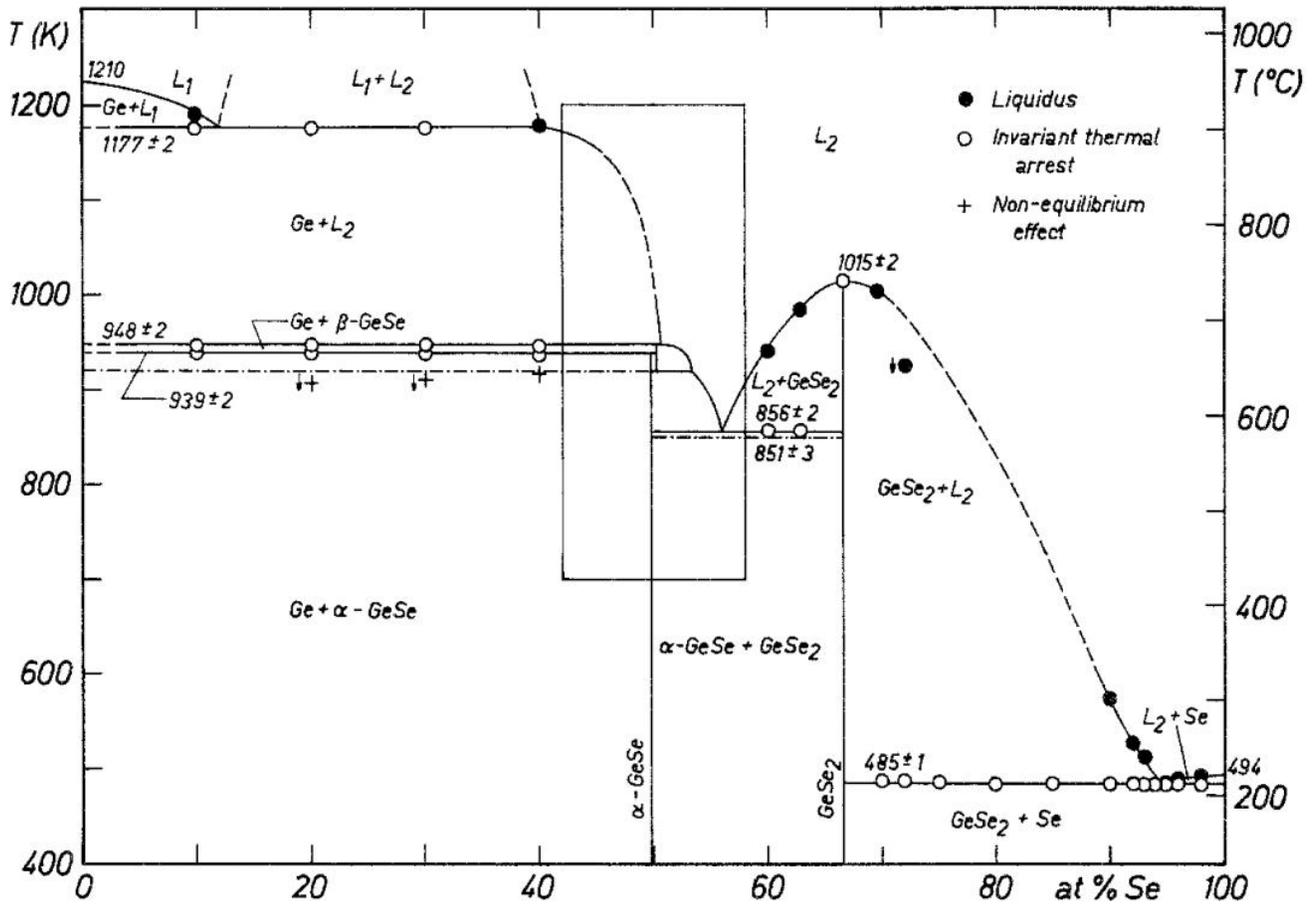

**Fig. 1** Phase Diagram of $Ge_xSe_{100-x}$ binary taken from Isper et al[38].

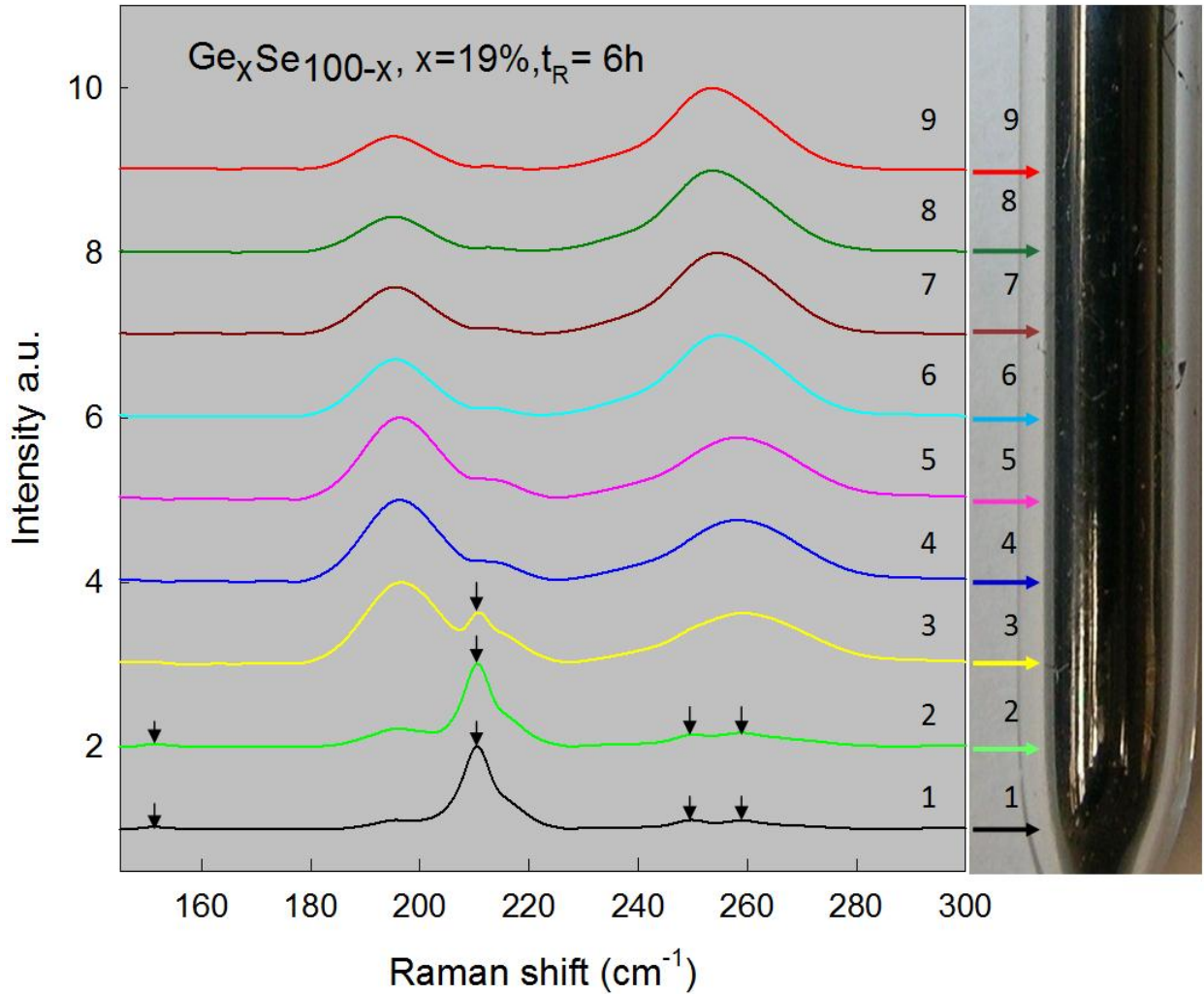

**Fig. 2** Raman scattering of a quenched $Ge_{19}Se_{81}$ melt taken along the length of a quartz tube used for reacting the melt at 950°C for 6 hours, and then lowering its temperature to $T_l$ + 50C, and water quenching. Note that spectra along the length of the tube, at indicated 9 points, show the lineshapes to systematically change. The narrow modes at arrow locations are those of α-$GeSe_2$ [32].

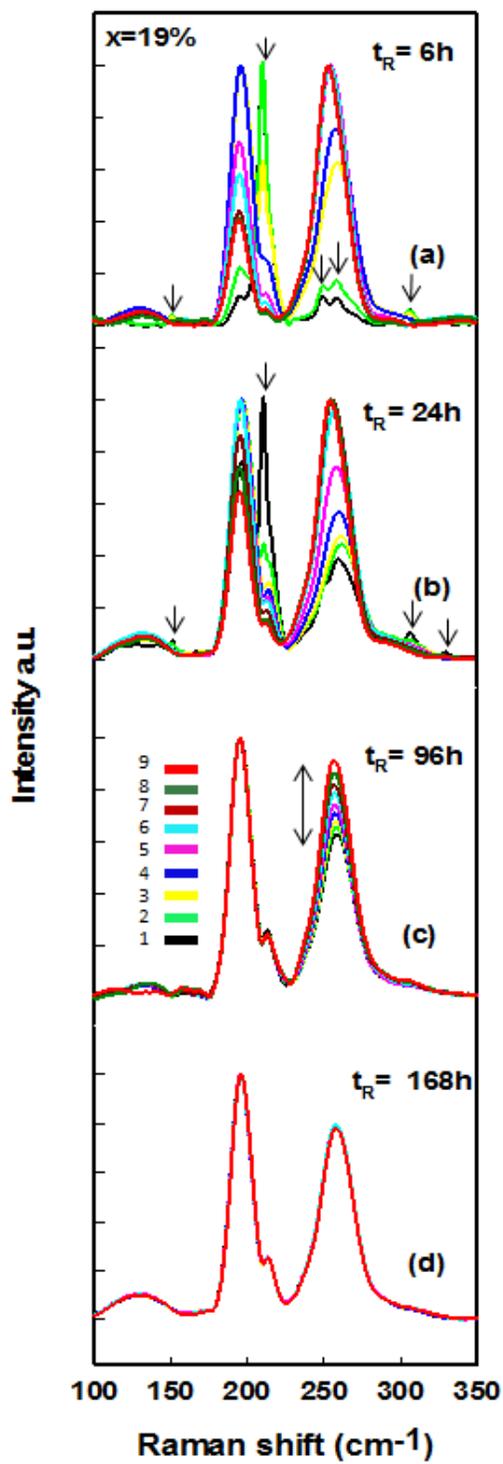

**Fig. 3** Raman profiling data providing a coalesced view of the 9 Raman spectra of Fig.2. Raman scattering of $Ge_{19}Se_{81}$ melt reacted for increasing reaction times appears in (b) $t_R = 24h$, (c) $t_R = 96h$, (d) $t_R = 168h$ (d). These data show that melts homogenize after reacting for 168h.

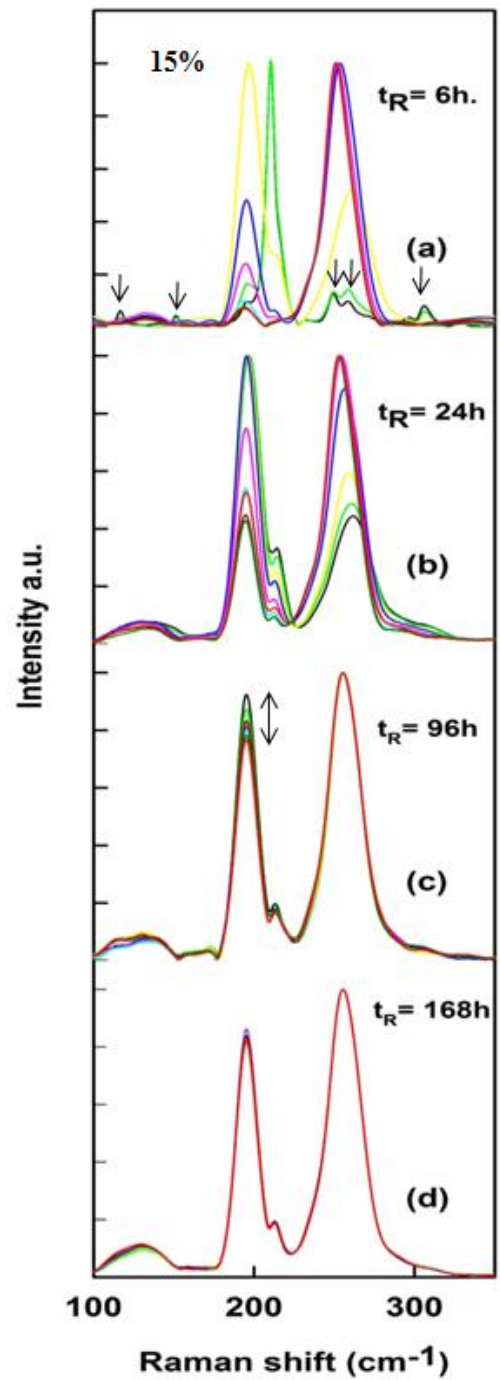

**Fig. 4** Raman profiling of $Ge_{15}Se_{85}$ melt reacted for increasing reaction times appears in (a) $t_R = 6h$, (b) $t_R = 24h$, (c) $t_R = 96h$, (d) $t_R = 168h$, demonstrating that melts homogenize after reacting for 168h.

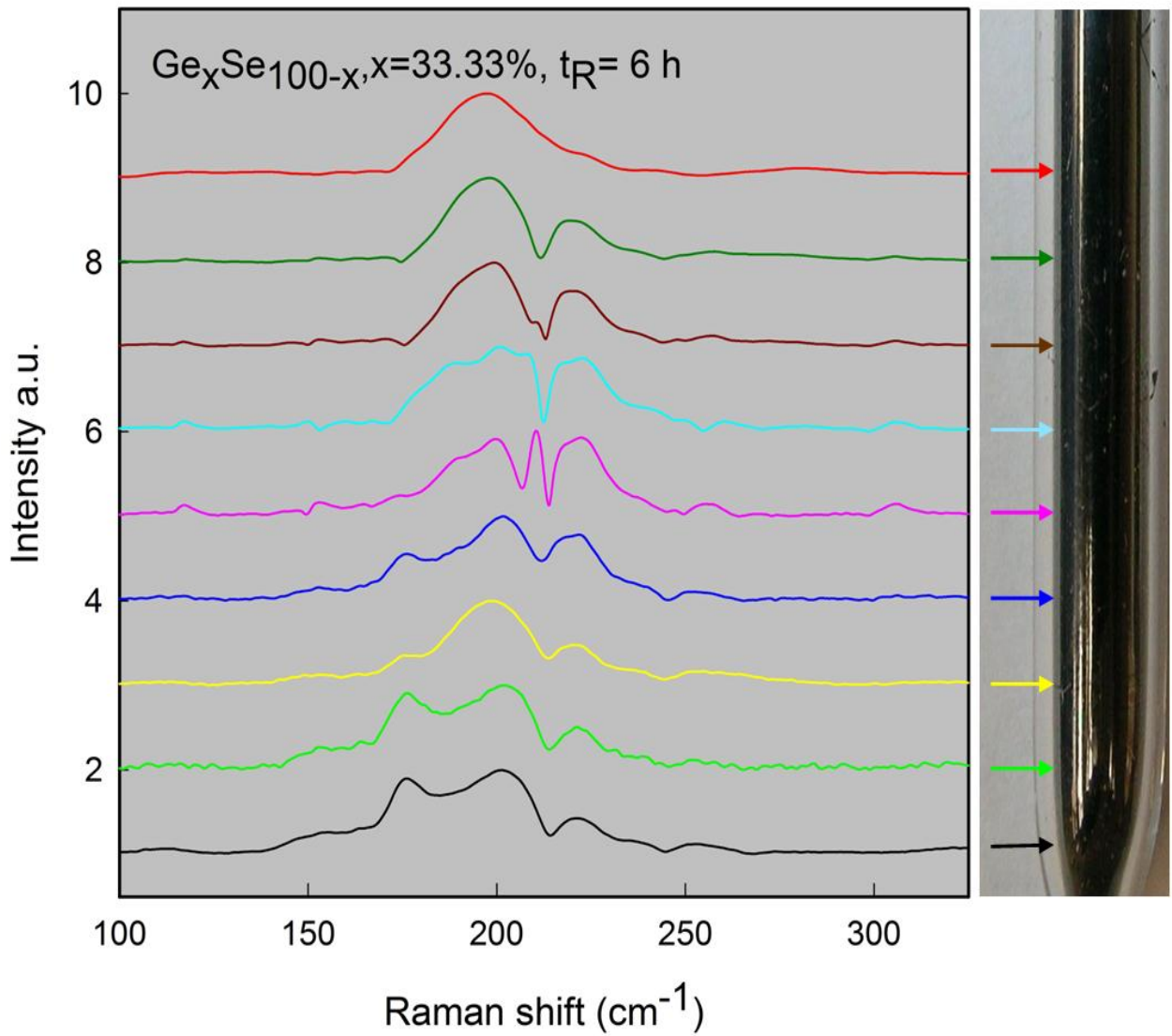

**Fig. 5** Raman profiling of GeSe$_2$ melt reacted at 950°C for $t_R$ = 6h, and taken along 9 points along the length of the quart tube containing the melt. The narrow modes at arrow locations are those of α-GeSe$_2$[32].

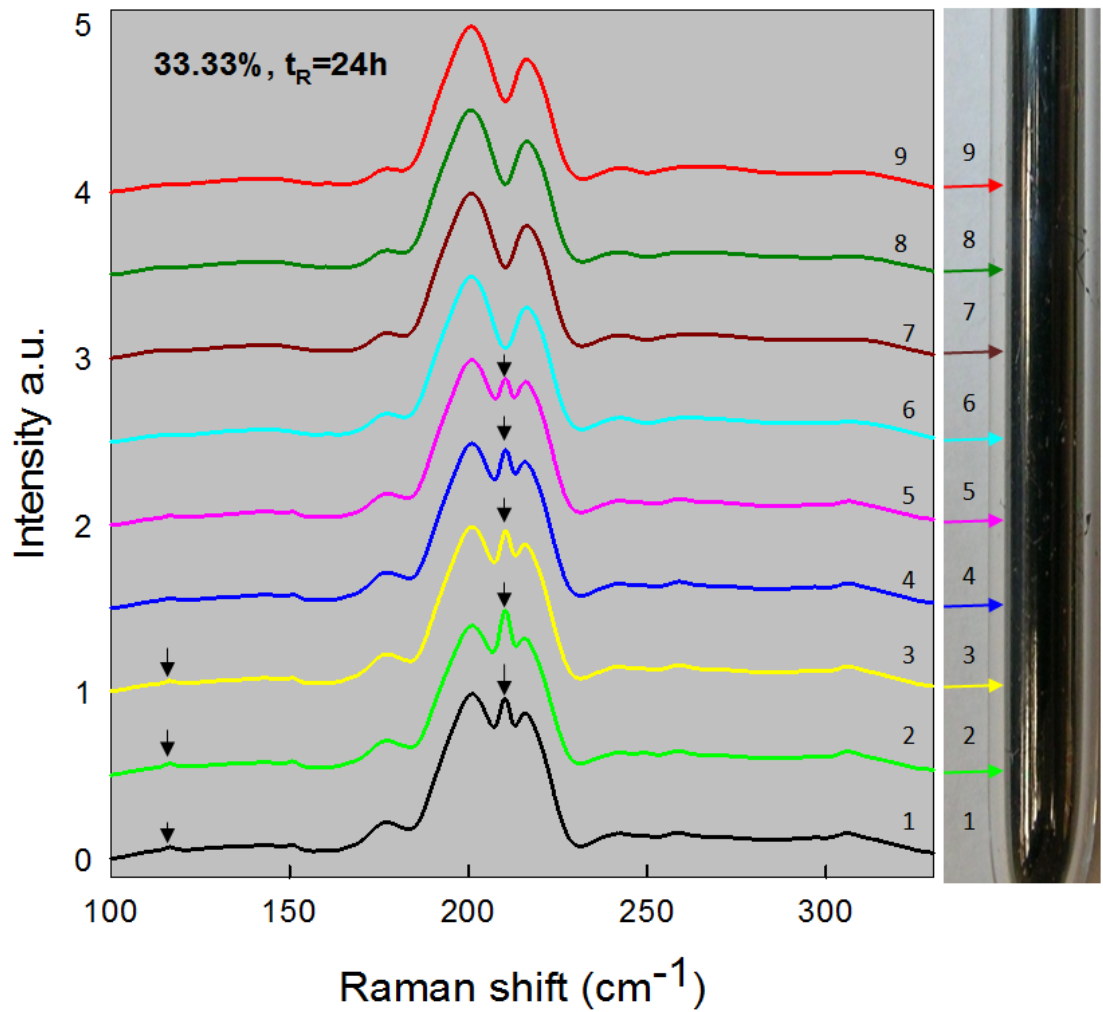

**Fig. 6** Raman profiling of GeSe$_2$ melt reacted at 950°C for $t_R$ = 24h taken along 9 points along the length of the quart tube containing the melt. The narrow modes at arrow locations are those of α-GeSe$_2$ [32].

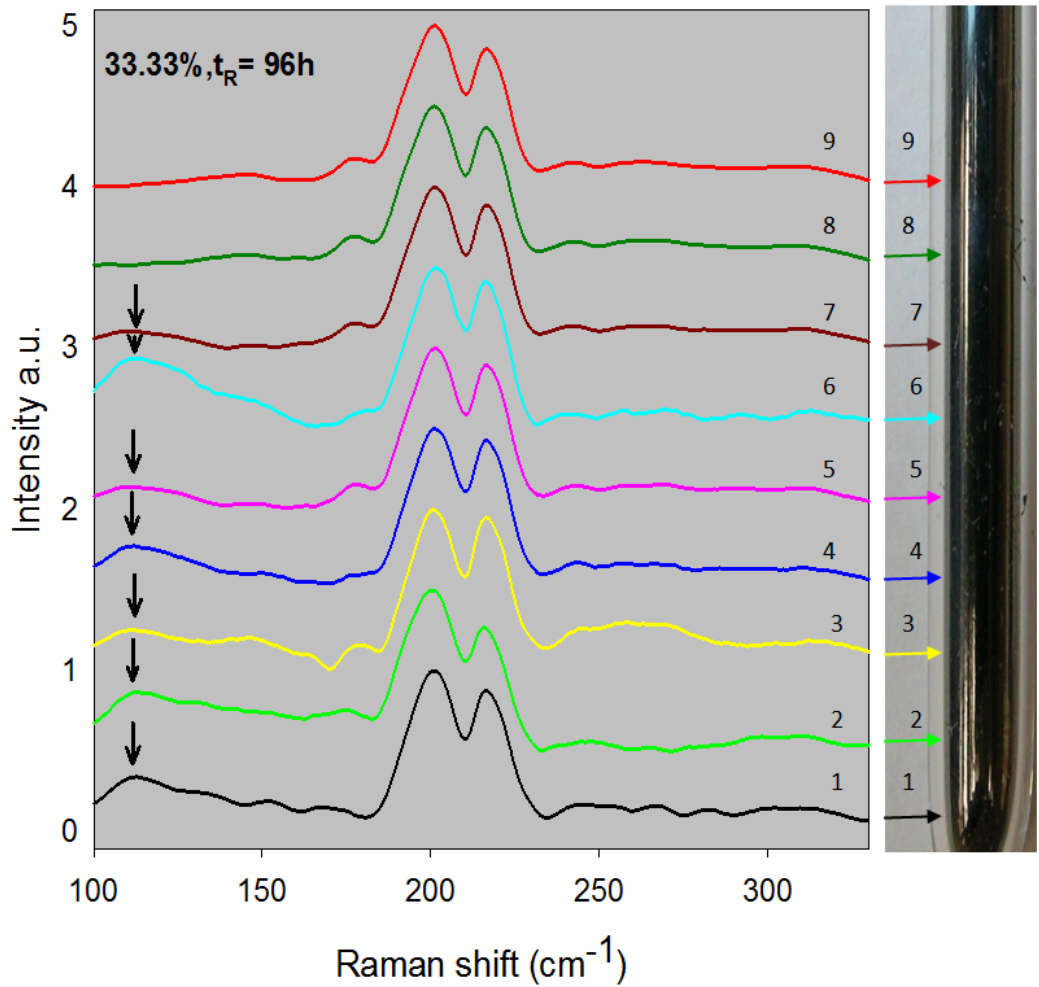

**Fig. 7** Raman profiling of GeSe$_2$ melt reacted at 950°C for t$_R$ = 96h taken along 9 points along the length of the quart tube containing the melt. The peaks at the arrow locations indicate growth in quasi elastic scattering (see text).

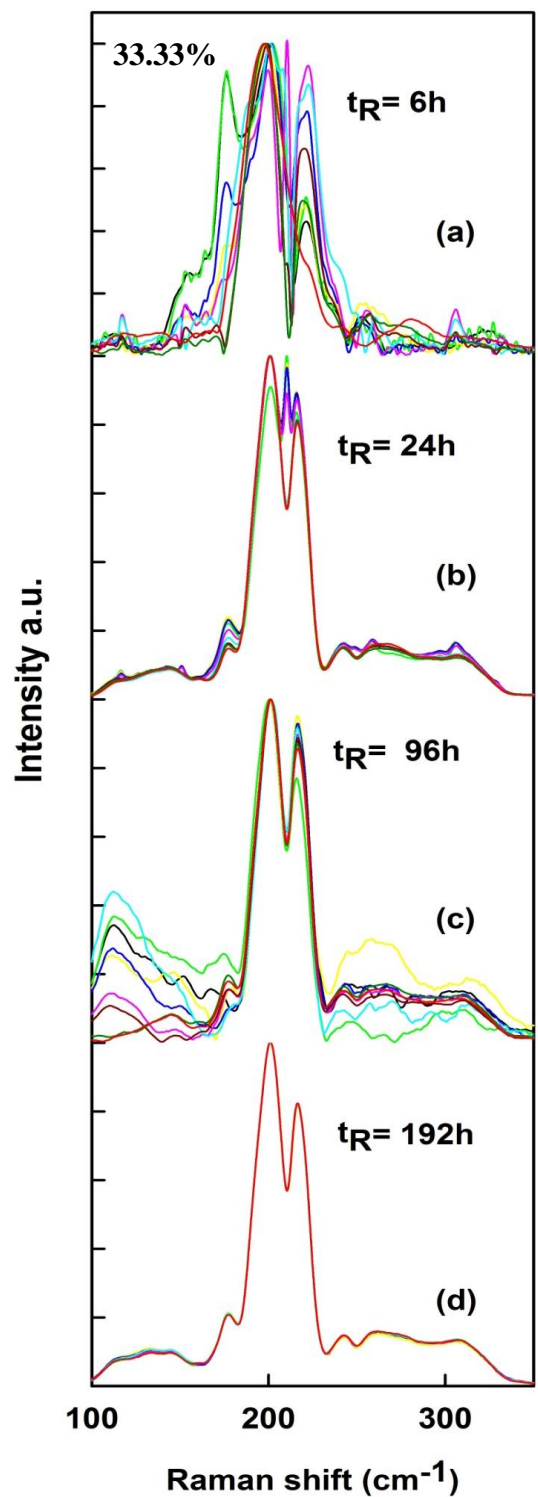

**Fig. 8** Summary of Raman profiling data on a GeSe$_2$ melt reacted for (a) 6h (b) 24h, (c) 96h, (d) 168h. These data show that after 168h of reaction melt molecular structure <u>globally</u> homogenizes across the batch composition.

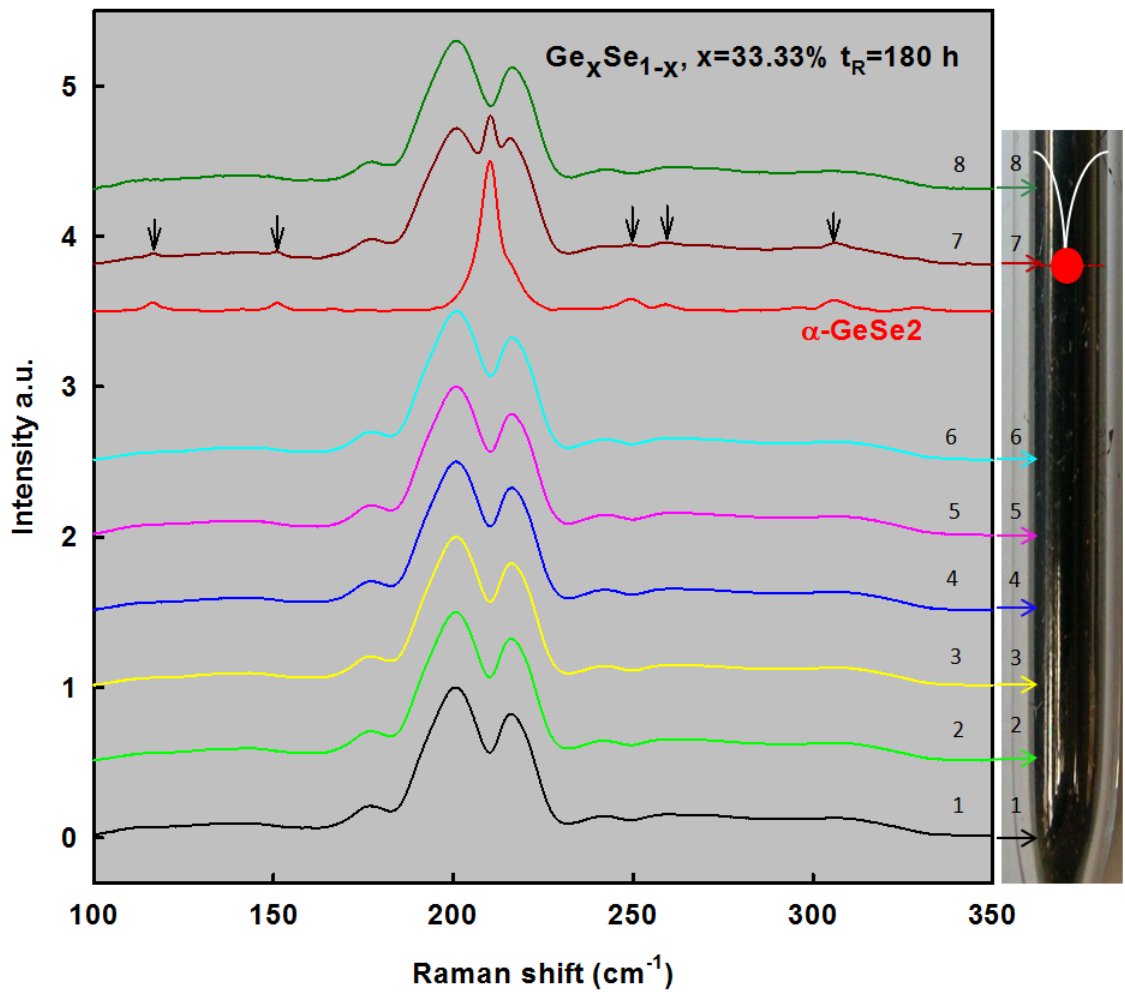

**Fig. 9** Raman scattering taken along the length of a quartz tube showing observed lineshapes at various locations of a GeSe$_2$ melt reacted for 180 h. For reference purposes, we have inserted the Raman lineshape of α-GeSe$_2$ between 6 and 7. The narrow modes at arrow locations are also observed at position 7 in the melt suggesting nucleation of α-GeSe$_2$ at the meniscus well.

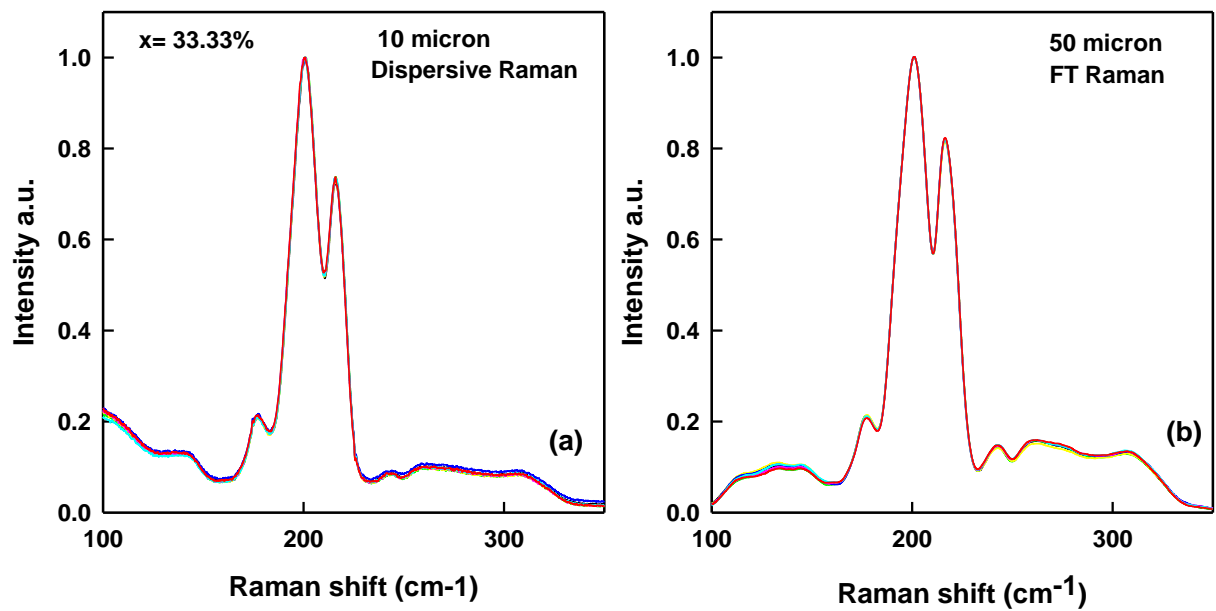

**Fig. 10** Raman profiling of a GeSe$_2$ melt taken (left) at 10 μm laser spot size in a Dispersive measurement and (right) at a 50 μm laser spot size in an FT Raman measurement. The data suggest that melts homogenized on a 50 μm scale in an FT measurement are actually homogeneous on a finer scale of 10 μm scale.

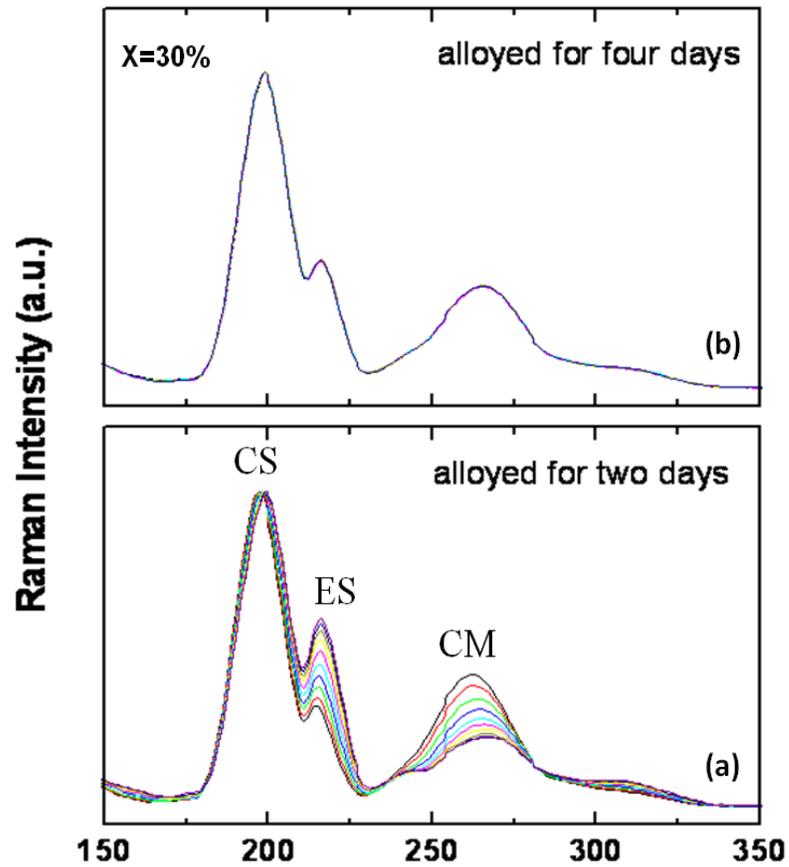

**Fig. 11** FT Raman profiling data on a $Ge_{30}Se_{70}$ melt taken with a 250 μm spot size (a) 2 days and (b) 4 days after reacting at 950°C. Note that after 4 days of reaction the melt has homogenized. Melts homogenized on a coarser spatial scale (250 μm), homogenize quicker than those homogenized on a finer scale (50 μm as seen in Fig 8), but display a reversibility window that is not as sharp as in the finely homogenized melts. See text.

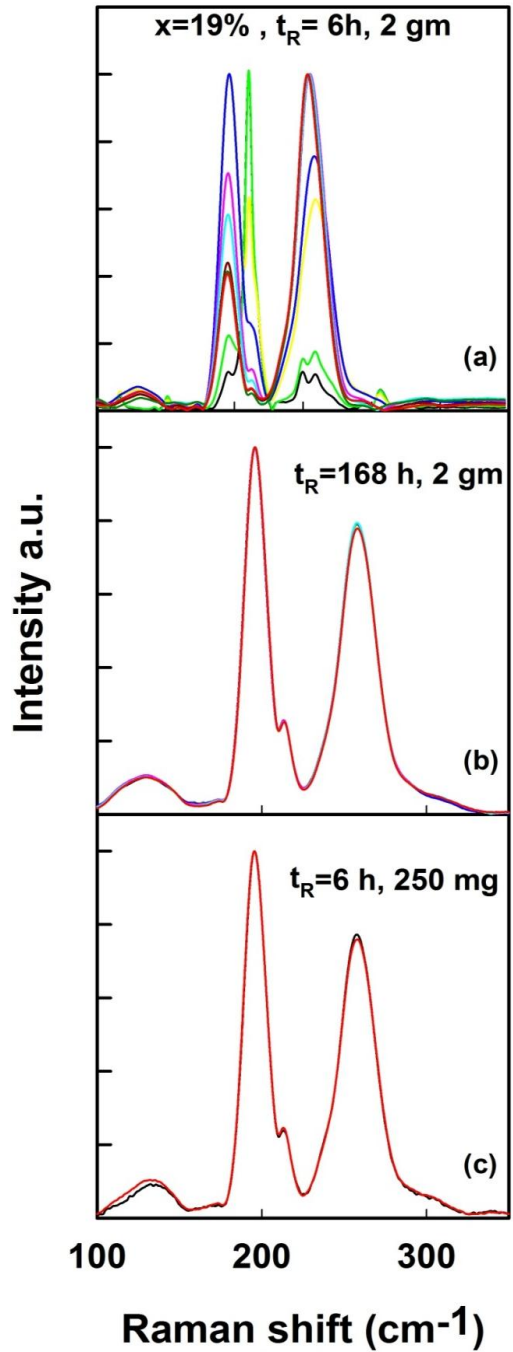

**Fig. 12** Raman profiling data on a $Ge_{19}Se_{81}$ melts (a) 2 gram melt reacted for 6h, (b) 2 gram melt reacted for 168h and (c)1/4 gm melt reacted for 6h. Smaller size melts homogenize much quicker than larger ones.

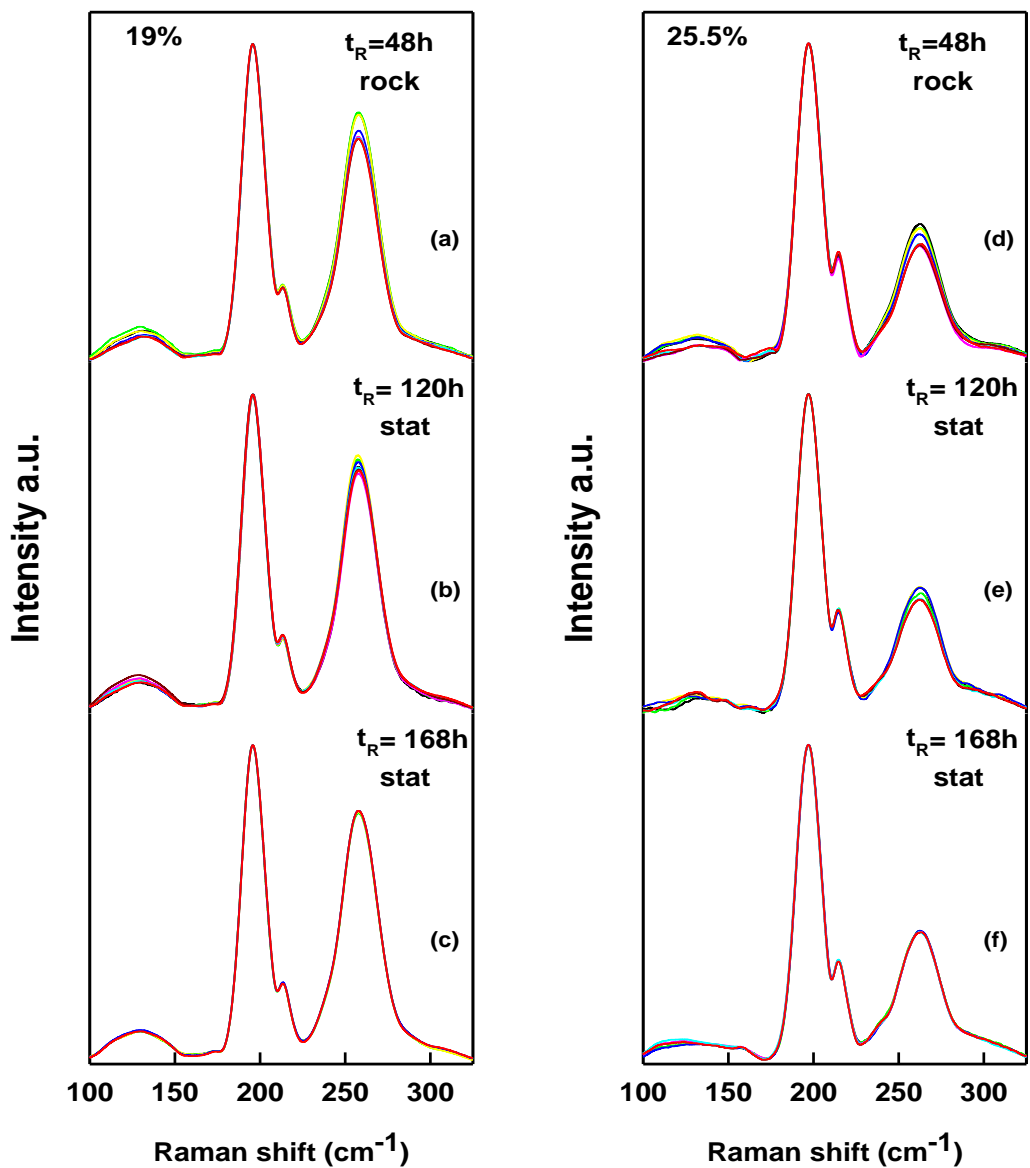

**Fig. 13** Raman profiling data on Ge$_{19}$Se$_{81}$ melts shown in the left panel after (a) being rocked for 48h (b) held stationary for 120h (c) held stationary for 168h. Parallel results are shown in the right panel for Ge$_{25.5}$Se$_{74.5}$ melts. Rocking melts assists homogenization incrementally.

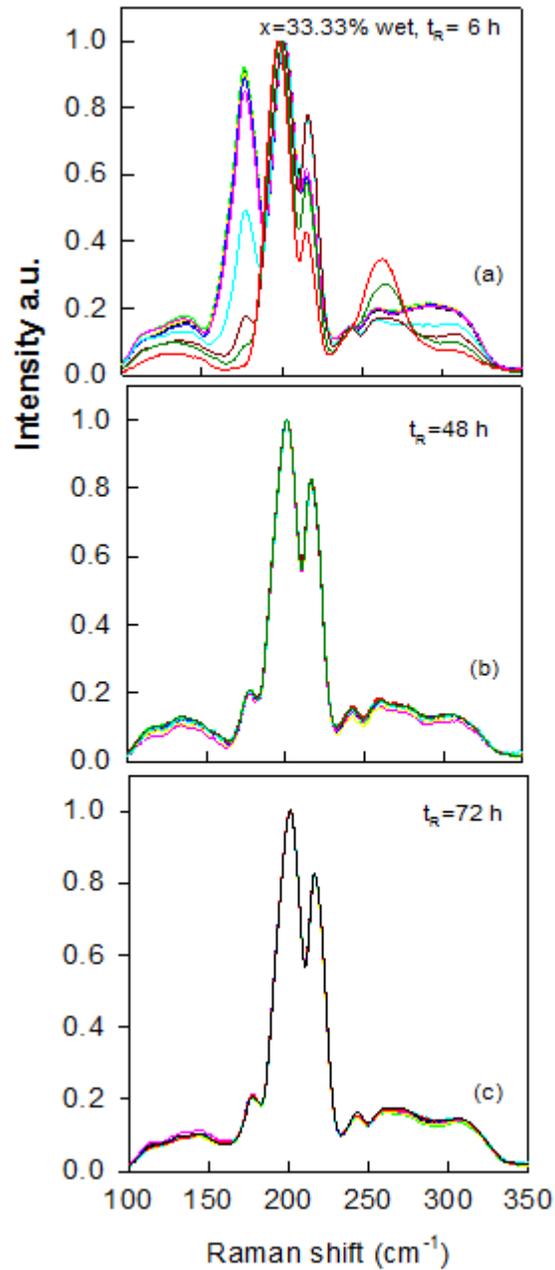

**Fig. 14** Raman profiling data on (a) a wet Ge$_{19}$Se$_{81}$ melt reacted for 42h, a wet GeSe$_2$ melt reacted for (b) 6h (c) 48 h (c) 72h, showing the more dramatic effect of homogenization assisted by water traces in melts. Results on corresponding dry melts are given in Fig 3 and 8.

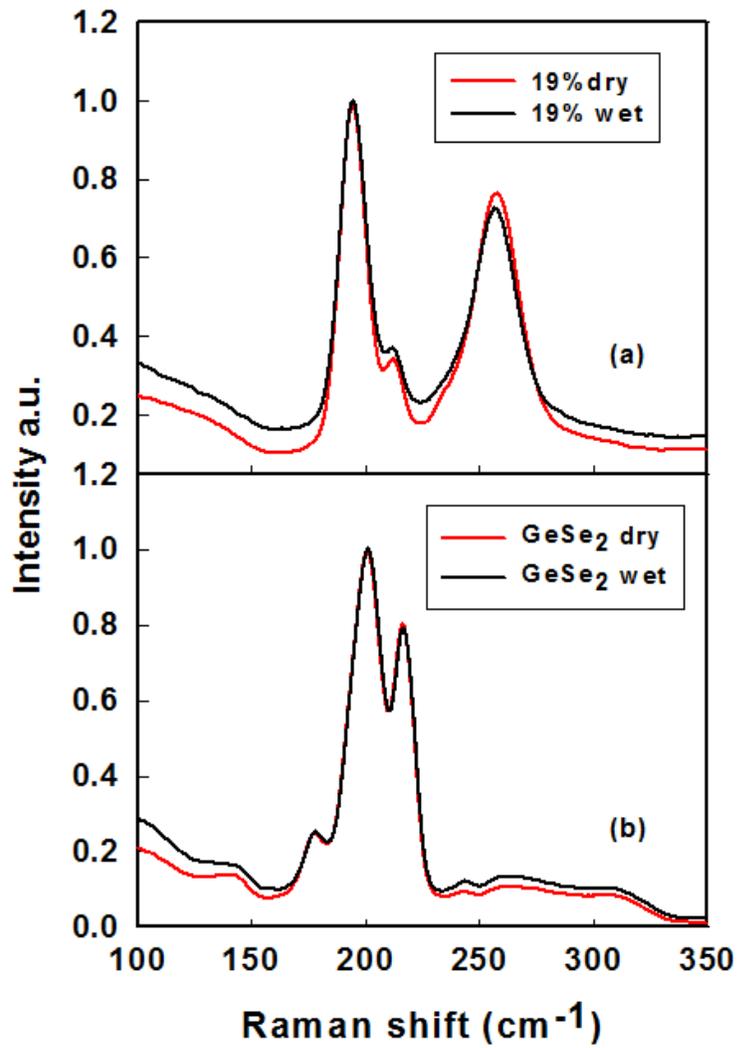

**Fig. 15** Dispersive Raman scattering of a wet melt compared to a dry one (a) at $Ge_{19}Se_{81}$ and (b) at $GeSe_2$. Note that the residual scattering in the wet melt exceeds that in the dry one.

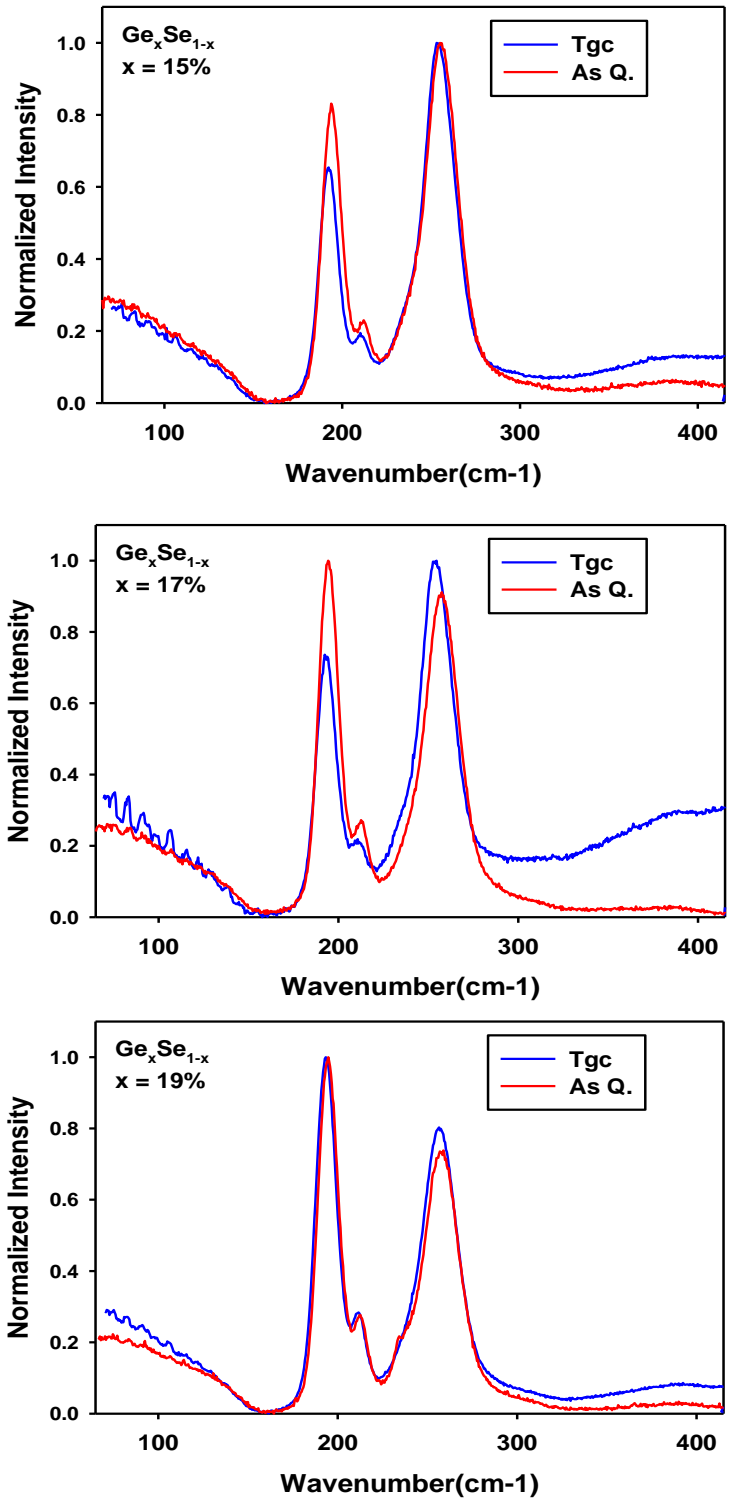

**Fig. 16** Dispersive Raman scattering lineshape in quenched melts compared to $T_g$ cycled ones (a) x = 15% (b) x=17% and (c) x= 19%. Note that changes in lineshape between the two types of samples decreases considerably at x = 19%.

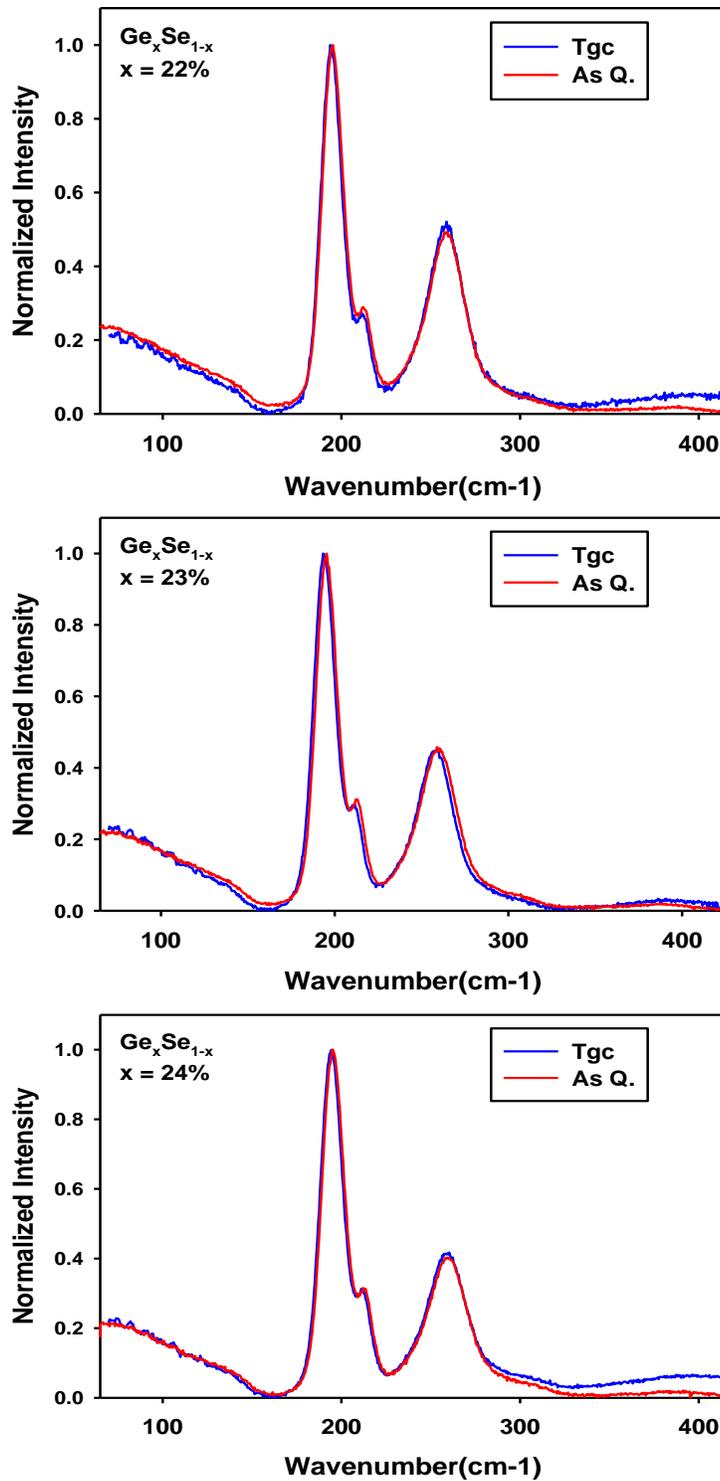

**Fig. 17** Dispersive Raman scattering lineshape in quenched melts compared to $T_g$ cycled ones (a) x = 22% (b) x=23% and (c) x= 24%. Note that changes in lineshape between the two types of samples is minuscule for these compositions in the thermally reversing window, suggesting that the melt and glass molecular structures are nearly the same in each case.

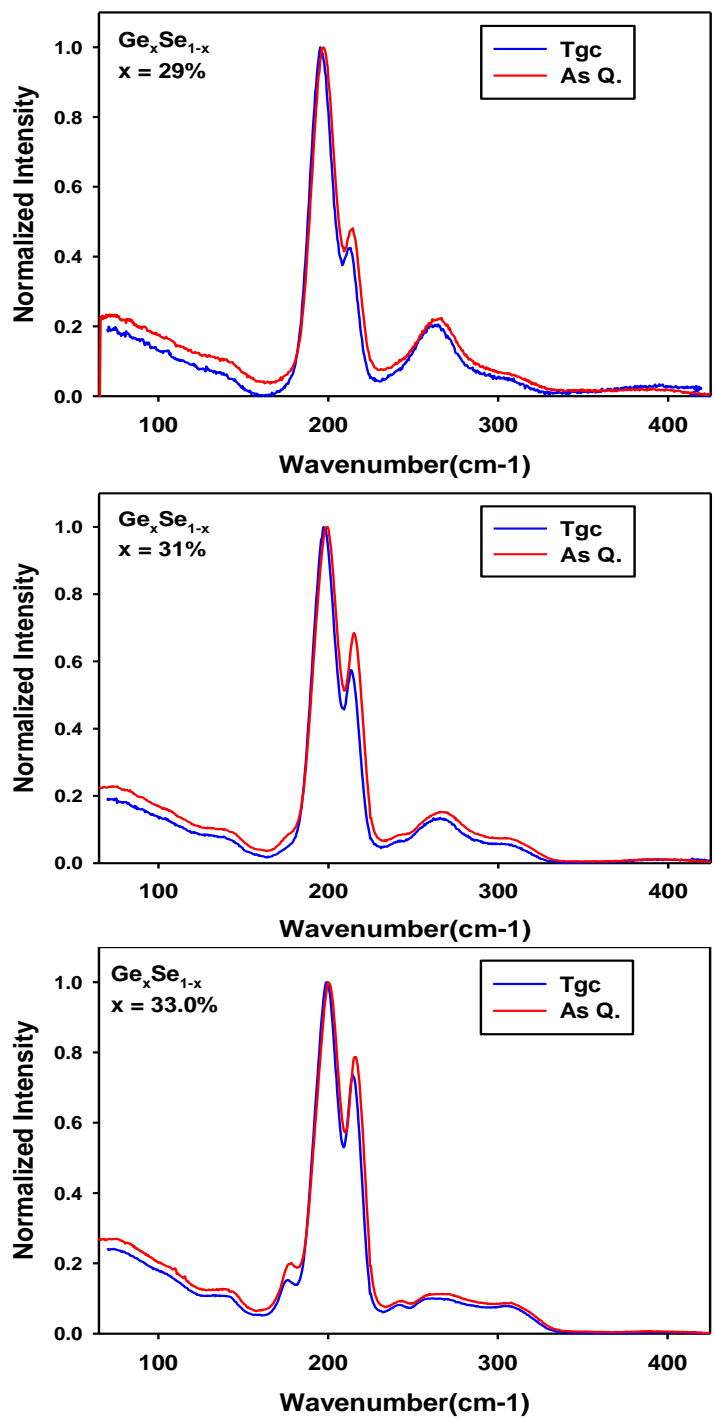

**Fig. 18** Dispersive Raman scattering lineshape in quenched melts compared to $T_g$ cycled ones (a) x = 29% (b) x=31% and (c) x= 33.33%. Note that changes in lineshape between the two types of samples increases as x > 26%. Note that the residual scattering in the as quenched melts is greater than in $T_g$ cycled ones for all these compositions in the stressed-rigid phase.

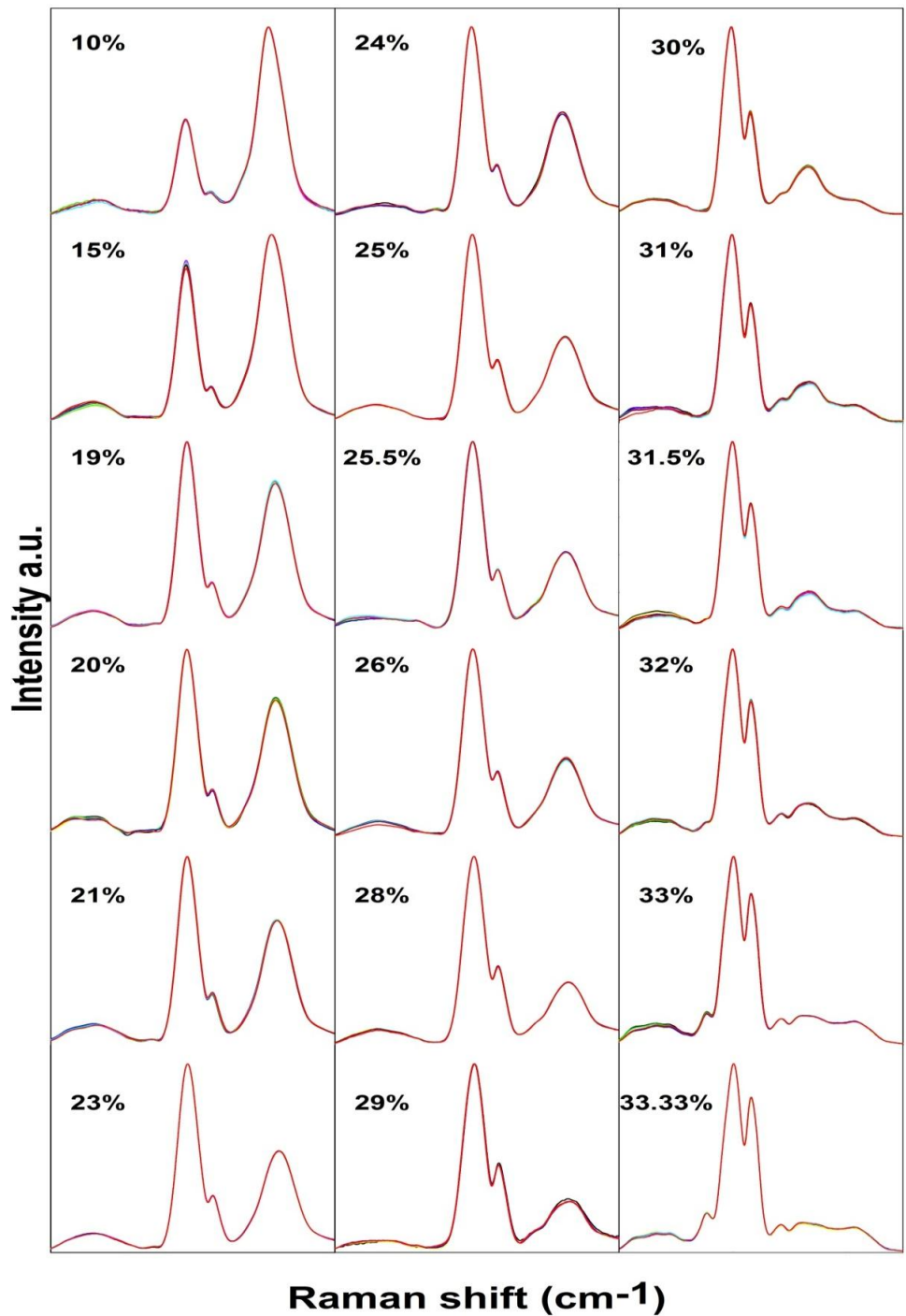

**Fig. 19** A summary of FT Raman profiling data on 18 of the 21 As-quenched $Ge_xSe_{100-x}$ melt compositions homogenized in the present study. The glass compositions are indicated as Ge content in %. We can see the growth in scattering strength of the of ES and CS tetrahedra at the expense of the Selenium chain mode as x increases from top left to bottom right.

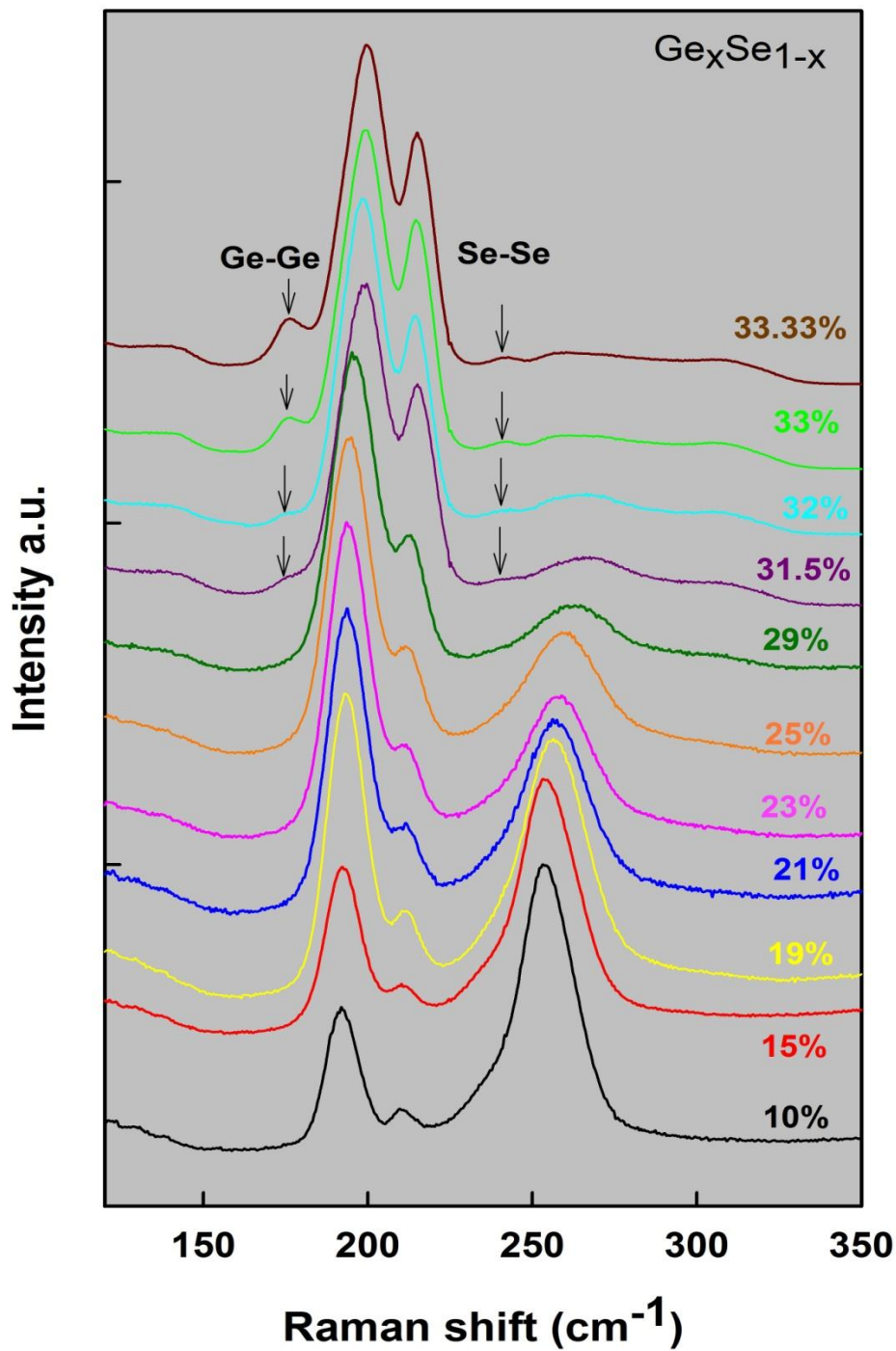

**Fig. 20** Dispersive Raman scattering in indicated Tg cycled $Ge_xSe_{100-x}$ glass compositions showing evolution of lineshapes with increasing x from the bottom to the top. Of special interest are the modes near 247 cm$^{-1}$ nd 180 cm$^{-1}$ at the arrow locations once x > 31.5%. These are associated with Se-rich and Ge-rich moiety in the glasses once they segregate.

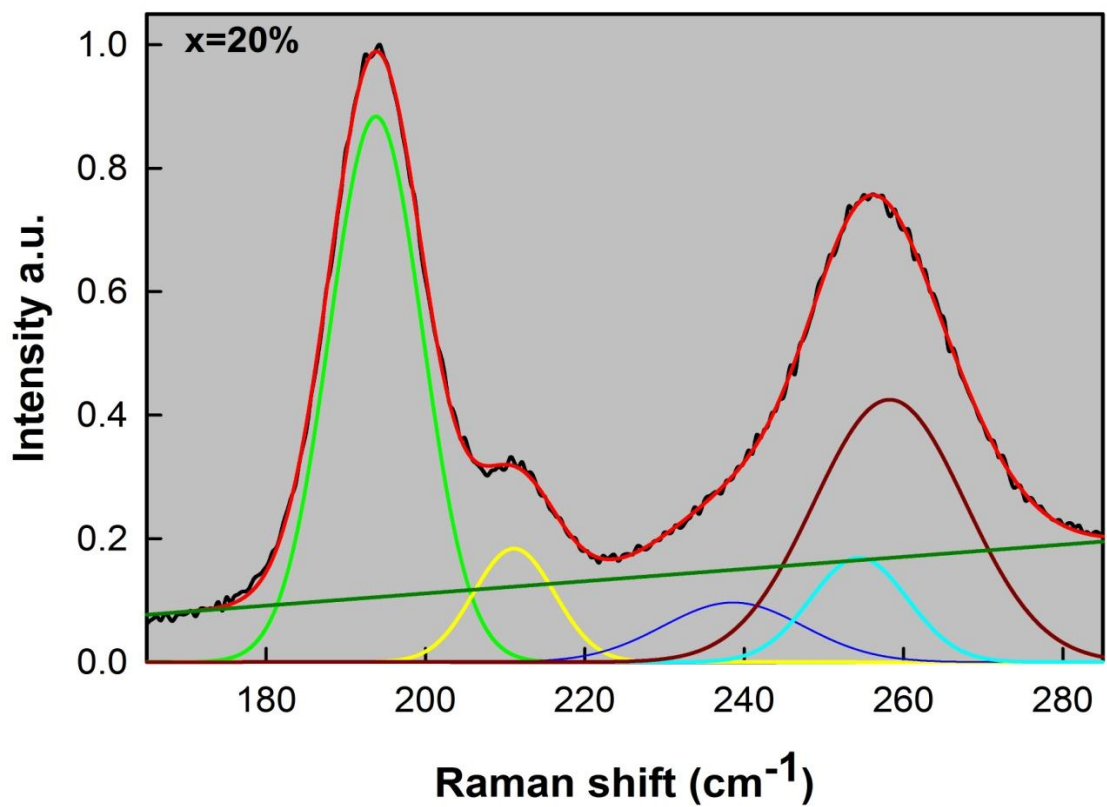

**Fig. 21** An example of Dispersive Raman lineshape deconvolution of a glass sample at x = 20% in terms of requisite number of Gaussian profiles using Peak Fit software. The corner sharing mode is shown in green, the edge sharing in yellow and the Selenium chain mode is identified in brown.

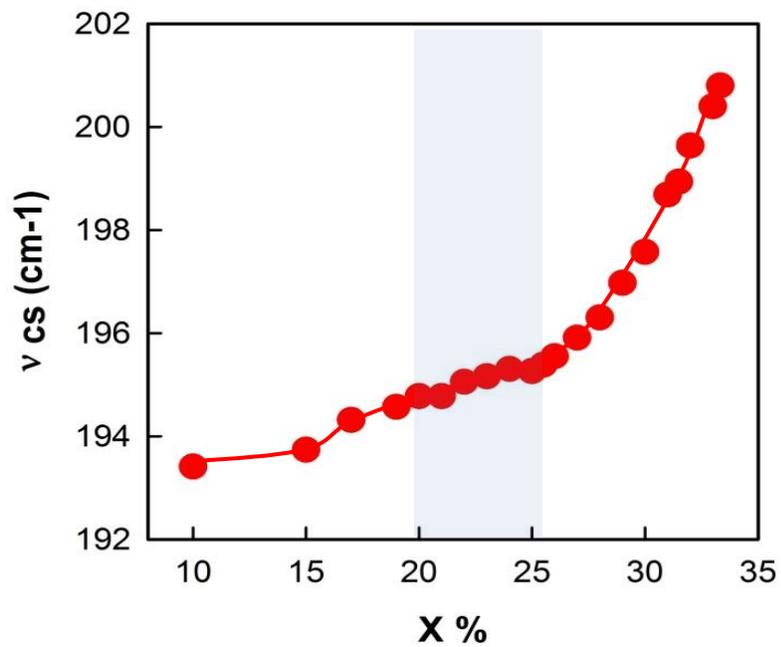

**Fig. 22** Variation in the frequency of the Corner sharing Ge tetrahedral mode as a function of x displaying three regimes, one at x < 20%, second between 20 % < x < 26%, and a third at x > 26%. See text for details.

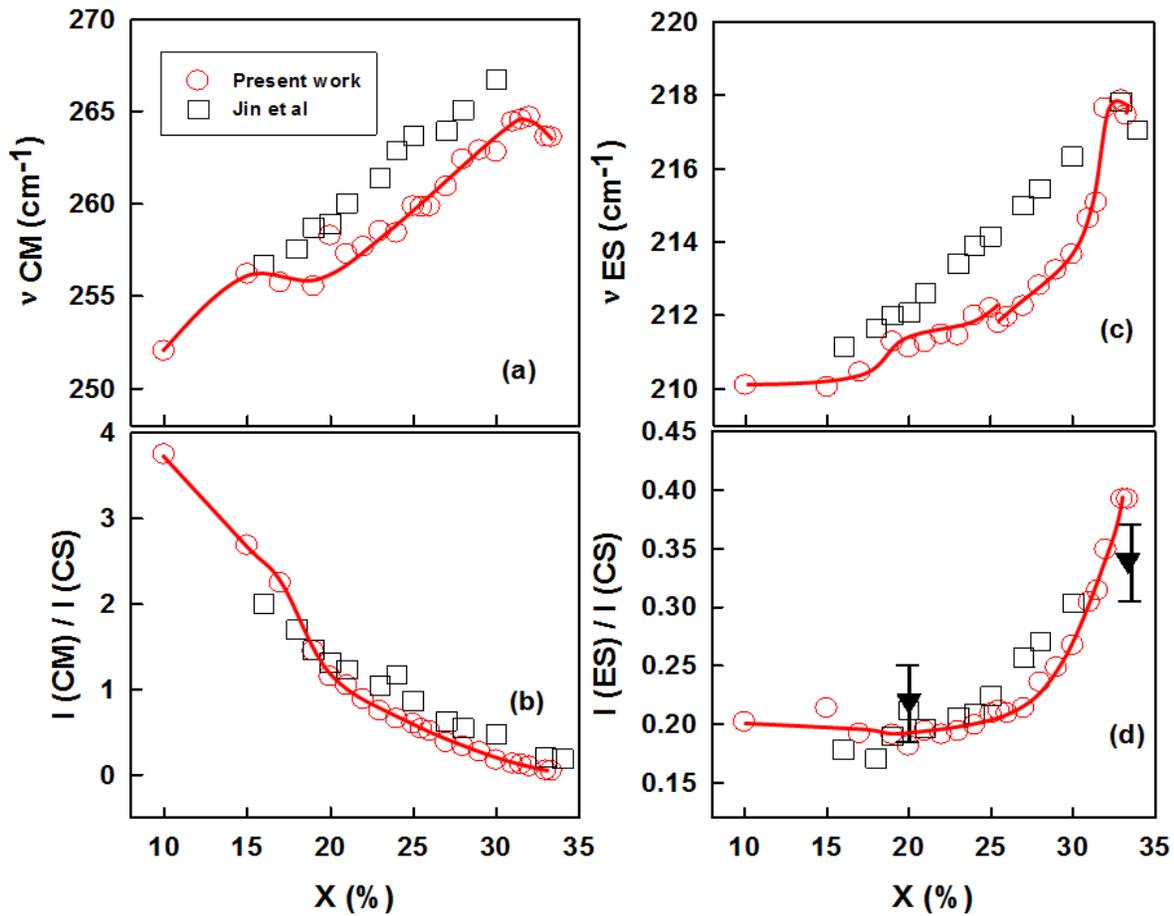

**Fig. 23** Variations in (a) Sn chain mode frequency (b) scattering strength ratio of CM/CS (c) ES mode frequency and (d) ES/CS mode scattering strength as a function of $Ge_xSe_{100-x}$ glass composition. Earlier work (□) taken form Jin et al [49].The neutron structure factor determined [51] ES/CS fraction (▼) at x = 20% and 33.33% are plotted in panel (d).

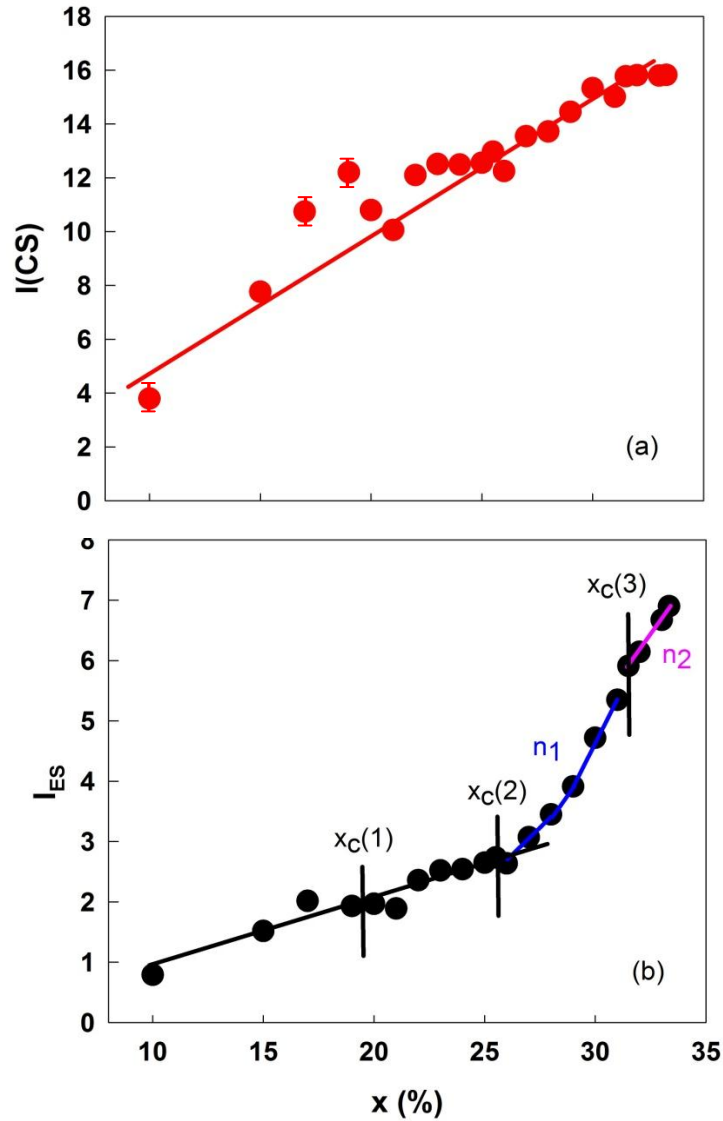

**Fig. 24** Observed variation in the observed Raman integrated intensity $I_{CS}(x)$ and $I_{ES}(x)$ of the (a) CS and (b) ES mode in $Ge_xSe_{100-x}$ glasses with x is plotted. These data were obtained using 647nm excitation, keeping the power fixed at 5 mW in a macro mode with glass sample wetting in a quartz tube. $I_{CS}(x)$ fits well to a linear variation in x across the examined range. $I_{ES}(x)$ variation is linear in the 10% < x < 26% range, power law ($\alpha\, x^{n_1}$) in the stressed-rigid range 27% < x < 31% with $n_1 = 2.30\,(5)$ and power law in the 31.5% < x < 33.33% ($\alpha\, x^{n_2}$) with $n_2 = 1.40\,(5)$ in the NSPS range.

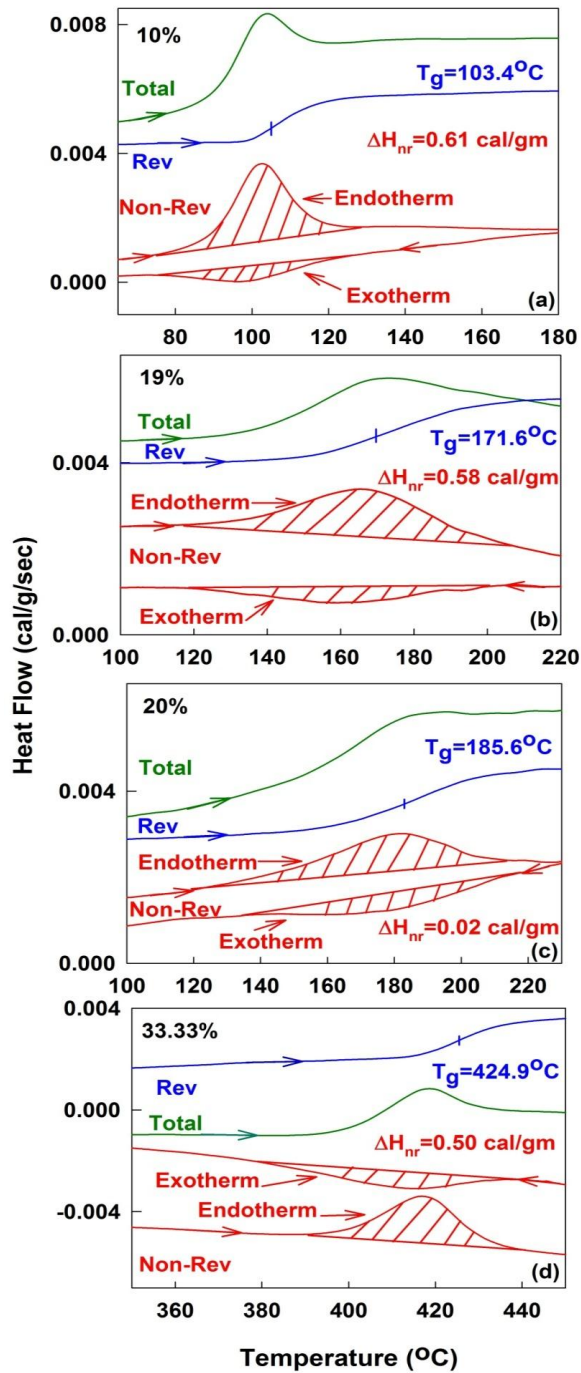

**Fig. 25** Typical mDSC scans of bulk $Ge_xSe_{100-x}$ glasses obtained at (a) x = 10%, (b) x = 19% (c) x = 20% and (d) x = 33.33% aged for 2 weeks at room temperature. Each panel shows 4 signals; the total, reversing and non-reversing heat flows in the heating cycle, and the non-reversing heat flow in the cooling cycle. Note that the enthalpy of relaxation, $\Delta H_{nr}$ term at x = 20% is minuscule (0.02 cal/gm).

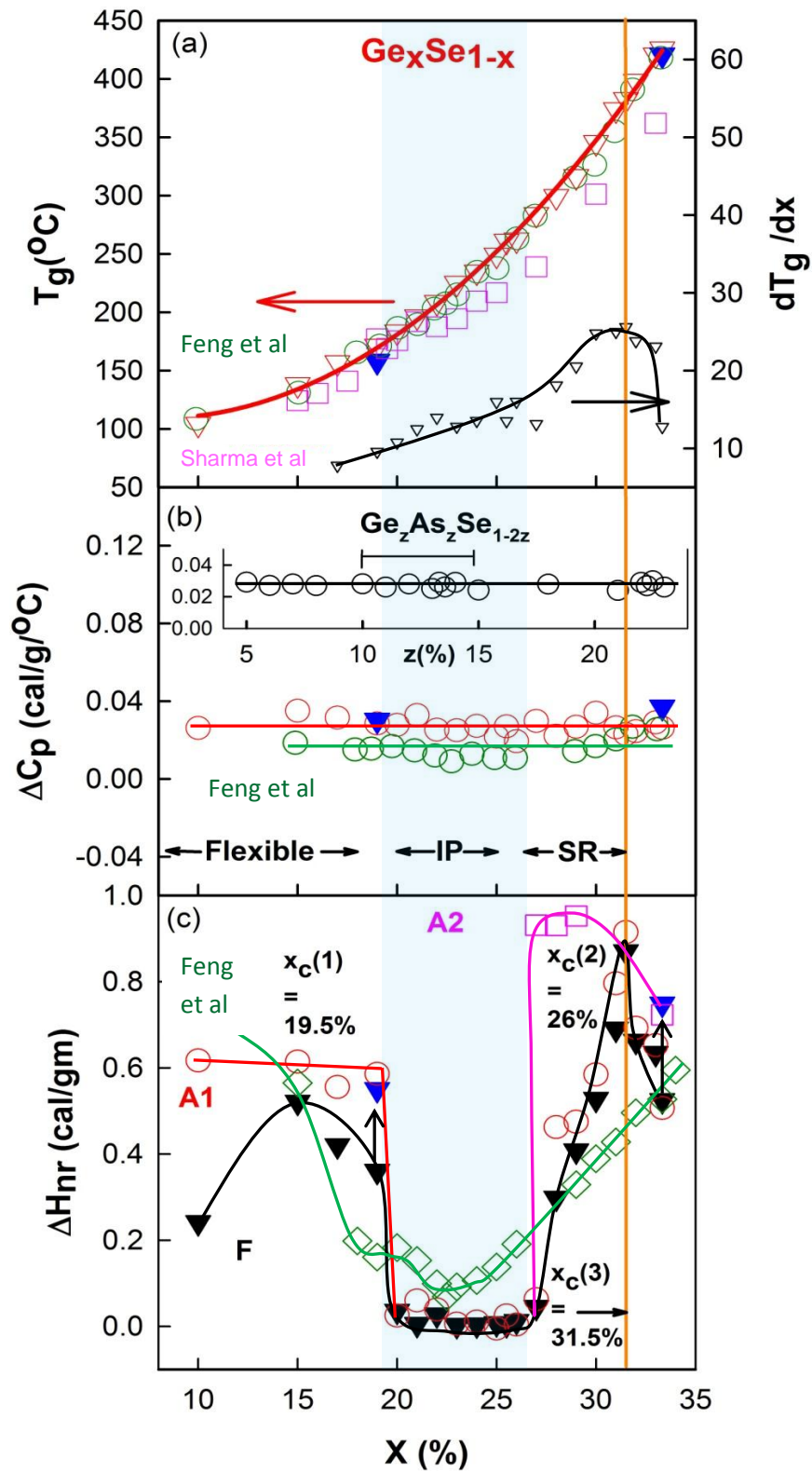

**Fig. 26** Summary of mDSC results on all samples showing variations in (a) $T_g(x)$ and $dT_g/dx$ ( $\nabla$ ). (b) $\Delta C_p(x)$ and (c) non-reversing heat flow $\Delta H_{nr}(x)$. In panel (a) $T_g(x)$ from the work of Feng

et al[7]. (○), Sharma et al[19] (□) are included for comparison. $T_g$ of wet samples (▼) at x = 19% and x = 33.33% are included. In panel (b) $\Delta C_p$ (x) trends from Feng et al (○) and Wang et al[55] on the $Ge_xAs_xSe_{100-2x}$ ternary (○) are included. In panel (c), $\Delta H_{nr}(x)$ trends in fresh (F) glasses (▼), glasses aged (A1) for 2 weeks at RT (○), glasses aged (A2) at 240°C for 2 weeks (□) are included. Trends in $\Delta H_{nr}(x)$ reported by Feng et al. (◊), displaying a near triangular variation with x is included for comparison. The increase in $\Delta H_{nr}$ term in wet (▼) glasses compared to dry ones is shown by an arrow. See text.

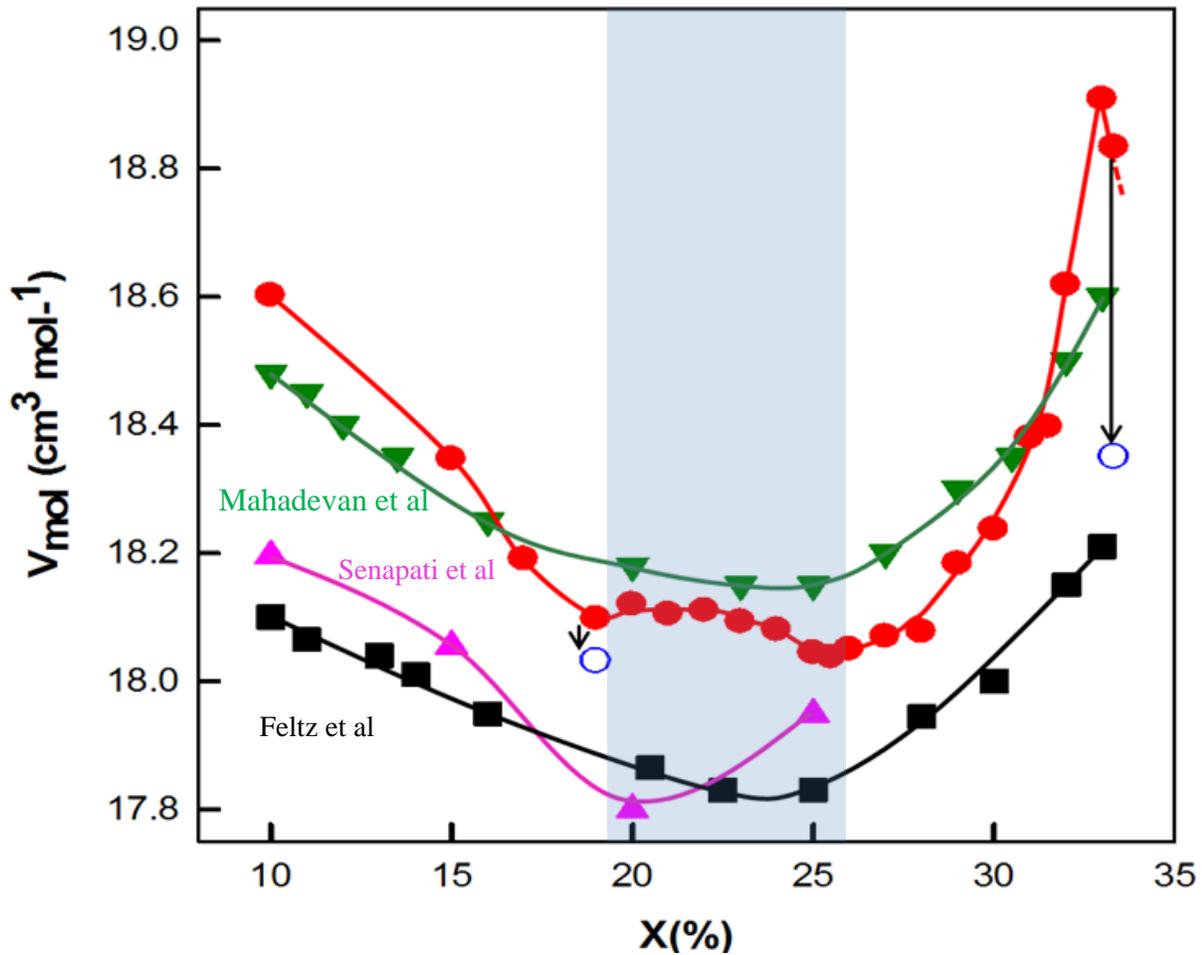

**Fig. 27** Variation in molar volumes ($V_m(x)$) of present dry (●) and wet (○) $Ge_xSe_{100-x}$ glasses is compared to earlier reports by Mahadevan et al[20]. (▼),Feltz et al[30]. (■) , and Senapati et al[31] (▲) are included. Note the larger variation IN $V_M(x)$ at x > 26% and x < 20% in the present set of samples than in earlier reports, probably related to sample homogeneity.

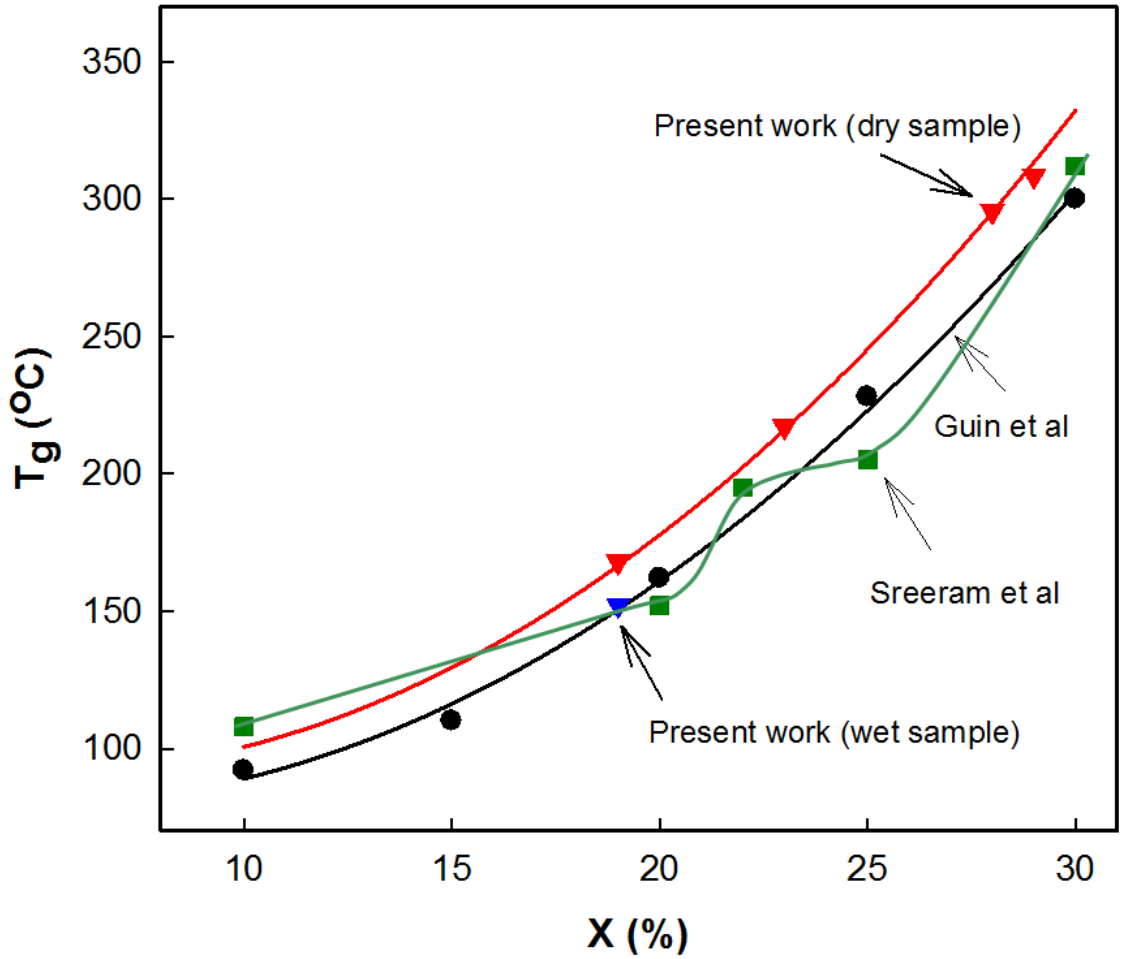

**Fig. 28** Variation in DSC measured $T_g(x)$ in present dry (▼) and wet (▼) glasses compared with those reported earlier by Guin et al[63] (●), Sreeram et al[64] (■). The DSC scan rate in all measurements was 10°C/min. The $T_g(x)$ of Guinn are 19°C less than our dry samples but coincide with our wet sample.

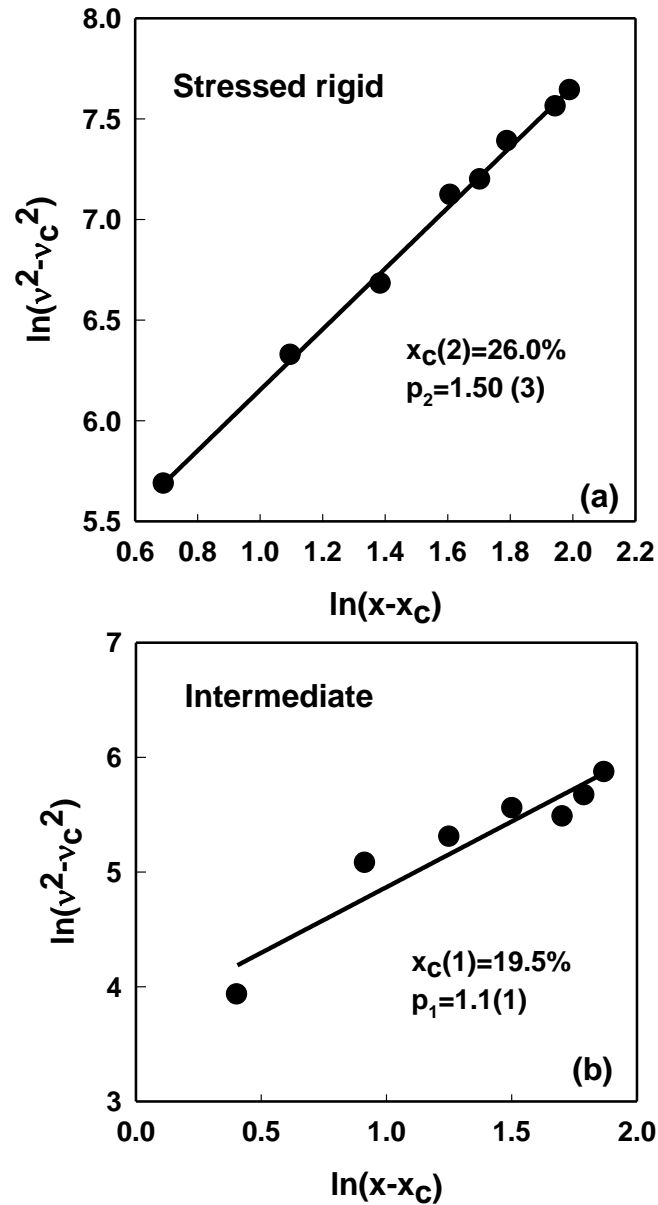

**Fig. 29** Elastic power-law (p) and thresholds ($x_c$) in the (a) stressed rigid and (b) the intermediate Phase obtained from the data of Fig 22. Here of $v_{cs}$ (x) represents the mode frequency variation of the CS mode. See text for details.

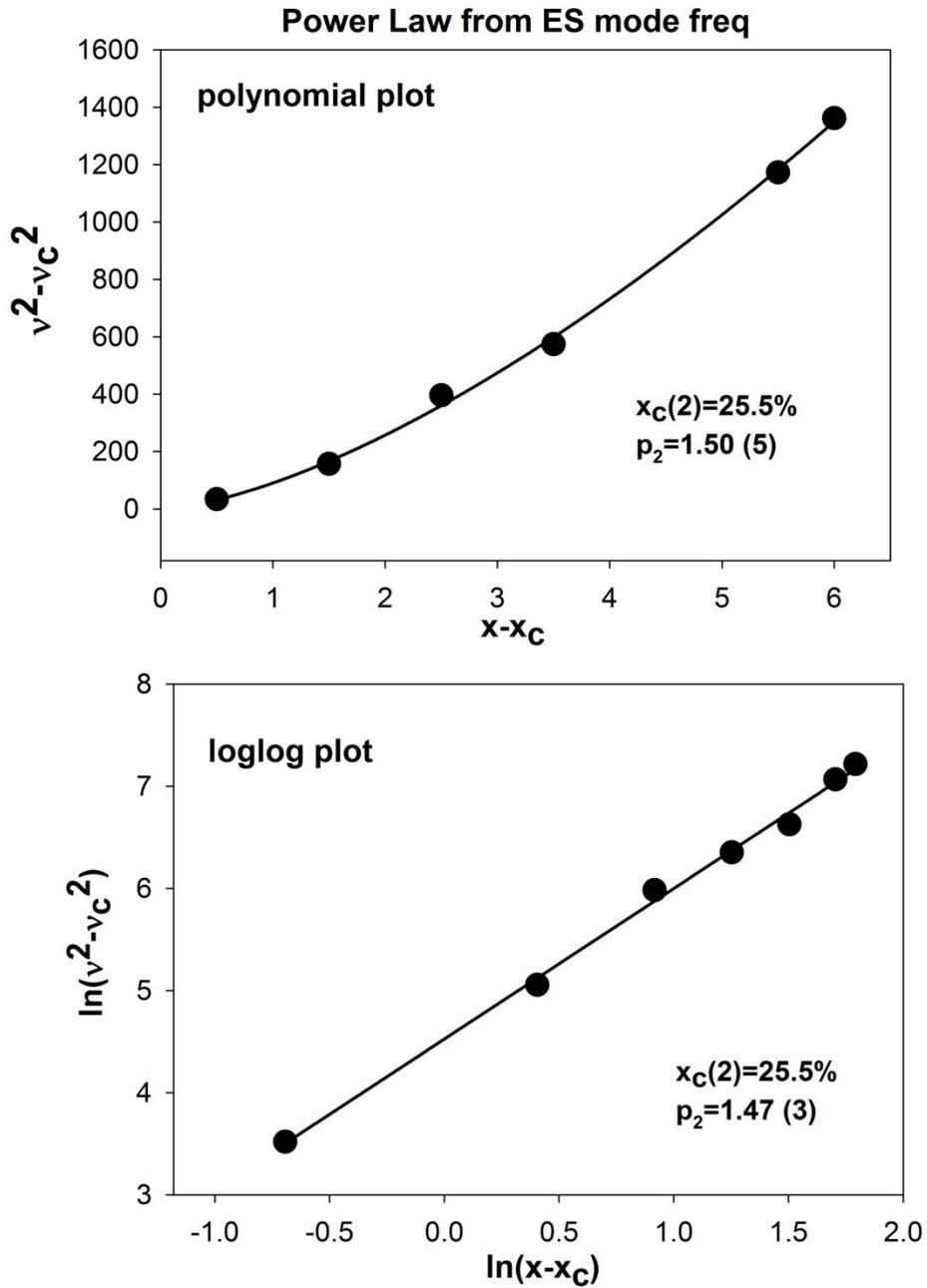

**Fig. 30** Elastic power-law (p) and threshold ($x_c$) of ES mode deduced from the data of Fig 23c., The deduced power-law was obtained using both a polynomial fit and a log-log plot of the ES mode frequency variation $\nu_{ES}(x)$.

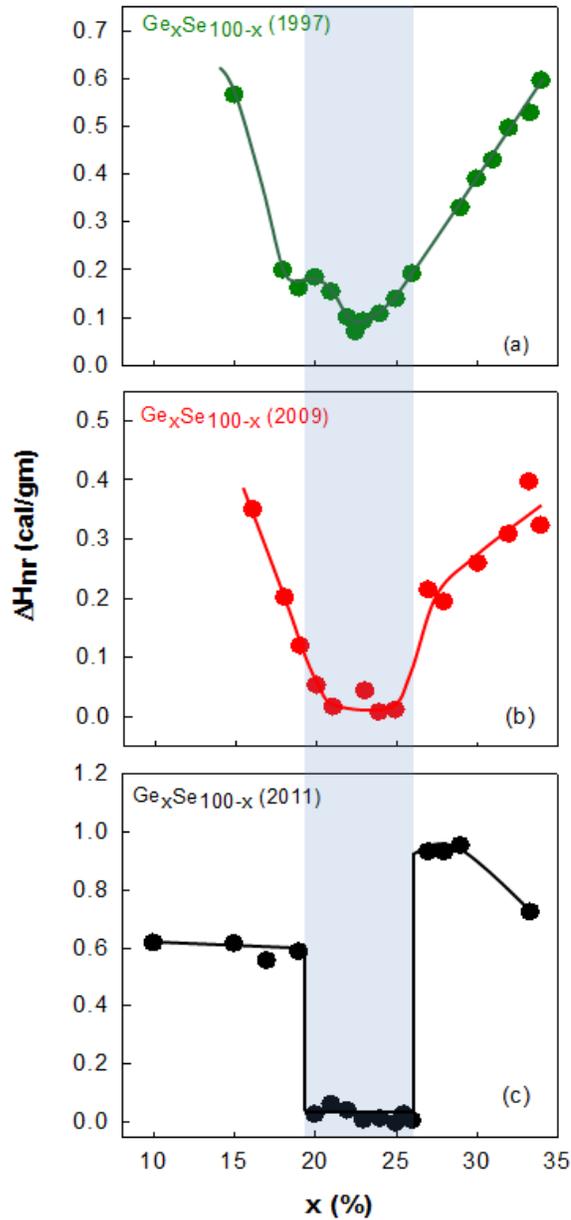

**Fig. 31** Reversibility window in GexSe100-x glasses reported (a) by Feng et al [7]. (1997), (b) Earlier report (a) (1997) and (b) Boolchand et al[80] (2009) and (c) in present work on glass samples of unprecedented homogeneity. It appears that the reversibility window in GexSe100-x glasses is intrinsically square-well like, resulting in *abrupt rigidity* and *stress* transitions. In experiments on glass samples possessing some heterogeneity of stoichiometry across a batch preparation, the window narrows and the walls broaden.

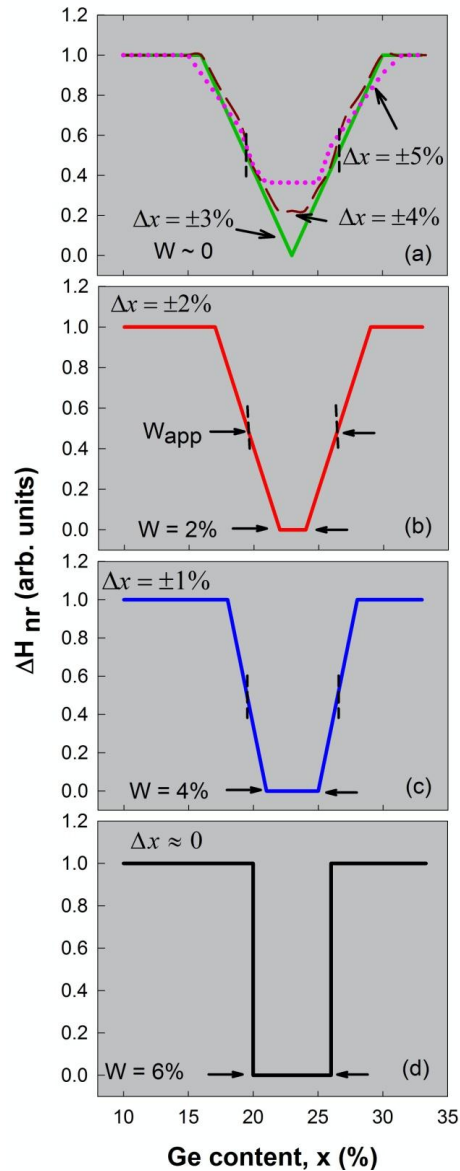

**Fig. 32** Variation of the reversibility window width W as a function of the heterogeneity (Δx) of Ge-fraction across a melt batch composition. The model assumes (d) an intrinsic square-well like reversibility window of W=6.5% in completely homogeneous melts (glasses) possessing vanishing Δx=0. With increasing heterogeneity of melts (glasses), Δx>0, the reversibility window width W and the slope of window walls steadily decreases as in (c), (b) and (a). When Δx=3%, the window becomes triangular in shape. Simulations show that for samples that are heterogeneous (Δx>0) the separation between midpoints of window walls, $W_{app}$, is a good measure of the intrinsic reversibility window width.

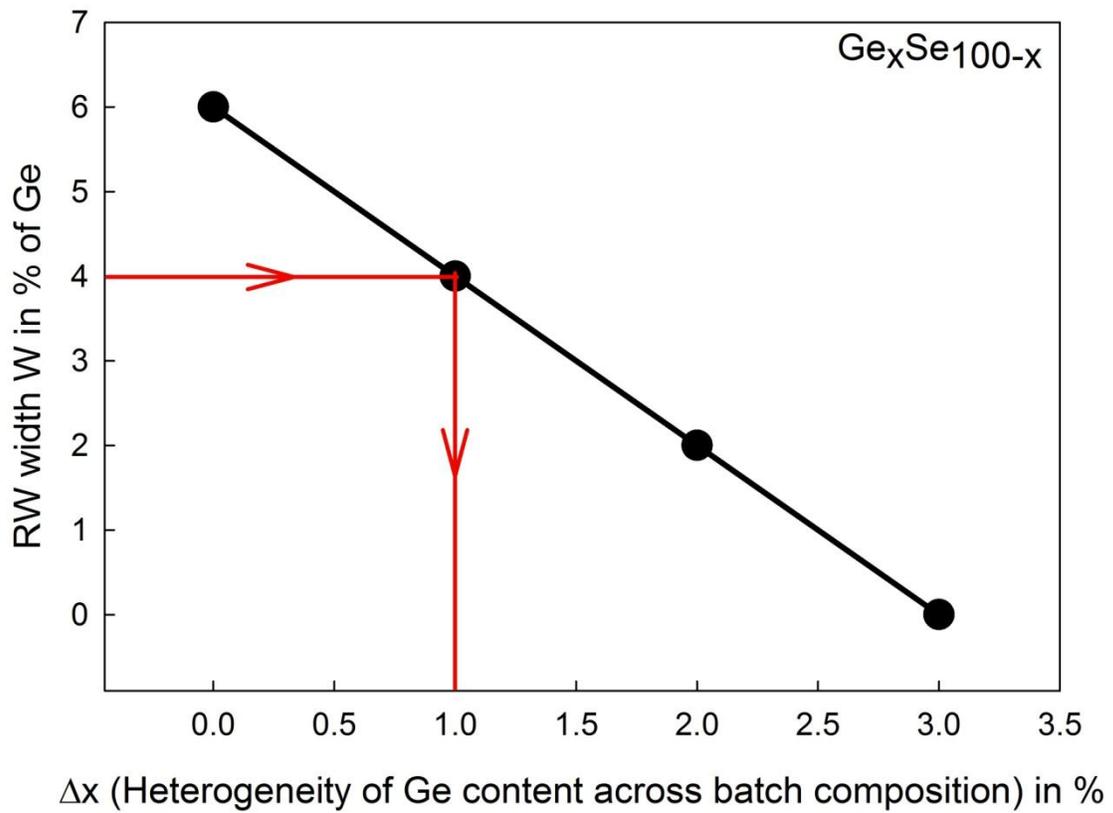

**Fig. 33.** Model prediction of reversibility window width W in $Ge_xSe_{100-x}$ glasses as a function of heterogeneity of glass samples characterized by a Ge-content spread, $\Delta x$, across a batch preparation.

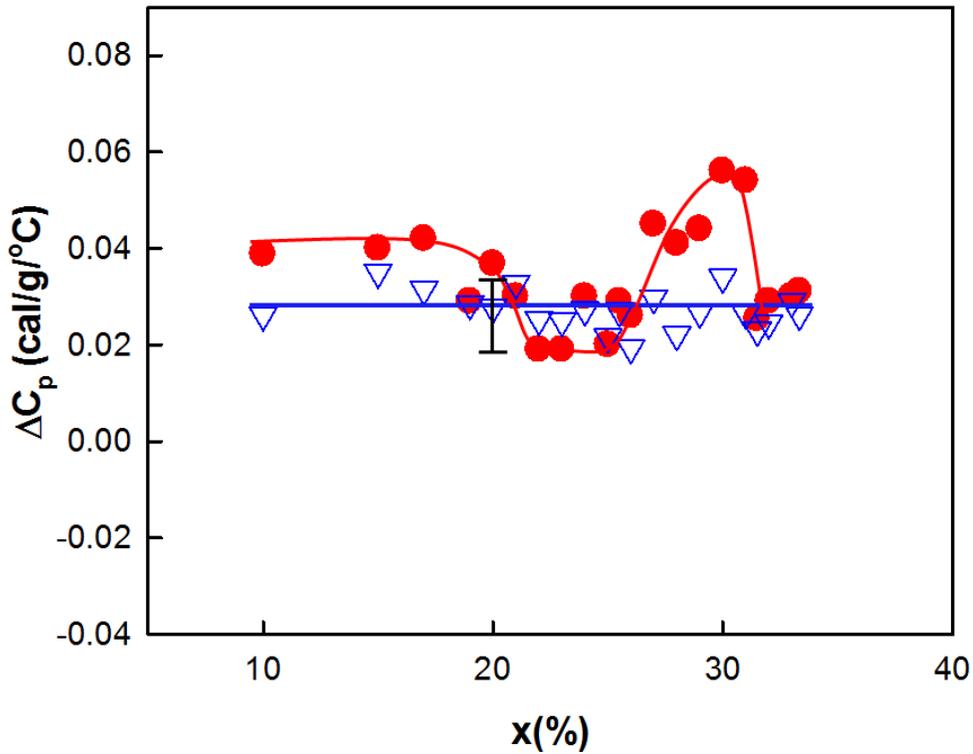

**Fig. 34** Variation in the jump $\Delta C_p$ at Tg deduced from the reversing heat flow ($\triangledown$) and the total heat flow ($\bullet$) in mDSC measurements. The former reveals a variation with x that is almost flat, the latter on the other hand shows increases at x > 26% and x < 20% probably due to the overshoot in the heat flow endotherm for indicated compositions.

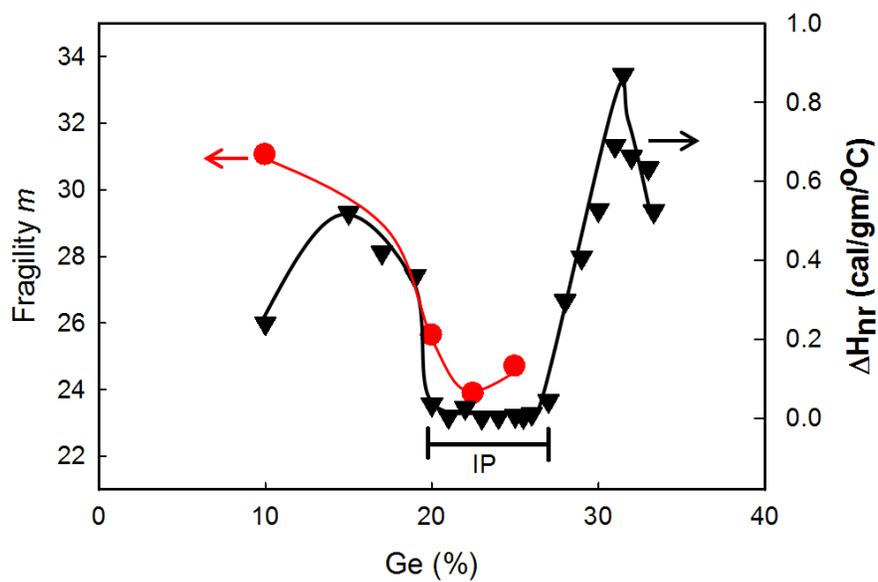

**Fig. 35** Variation in melt fragility m(x) (●) reported by Stolen et al [41] and the presently reported $\Delta H_{nr}$ term (▼) in the present glasses as a function of x. Note that both terms show a minimum in the Intermediate Phase.

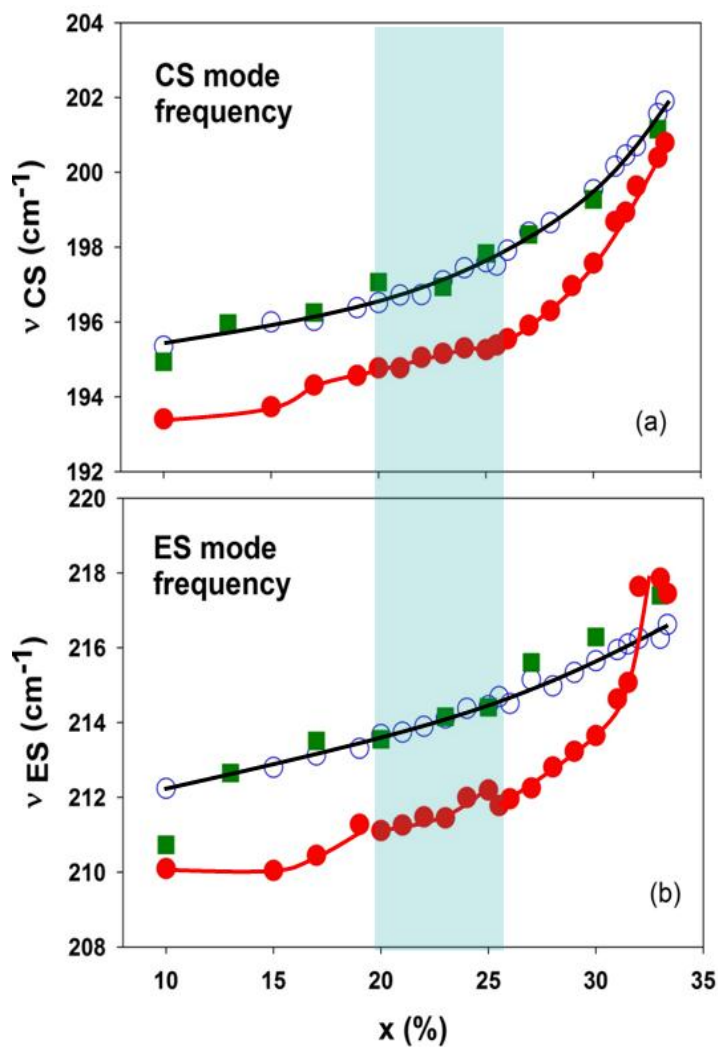

**Fig. 36** Variation in (a) the CS and (b) ES mode frequency with Ge content in bulk $Ge_xSe_{100-x}$ glasses deduced from FT-Raman (○) and Dispersive Raman (●) measurements. Note that in both the cases mode frequencies blue shift with x, however there is an offset between the two mode frequencies that steadily vanishes as x increases to x = 33.33%. In general, the dispersive measurements yield a lower frequency than the FT-measurement. Results from an FT-Raman study on bulk $Ge_xSe_{100-x}$ by Sen et al[88] are also projected in the fig as (■) data points for comparison to our data. We find that our FT-Raman results are in reasonable agreement with those of Sen et al.

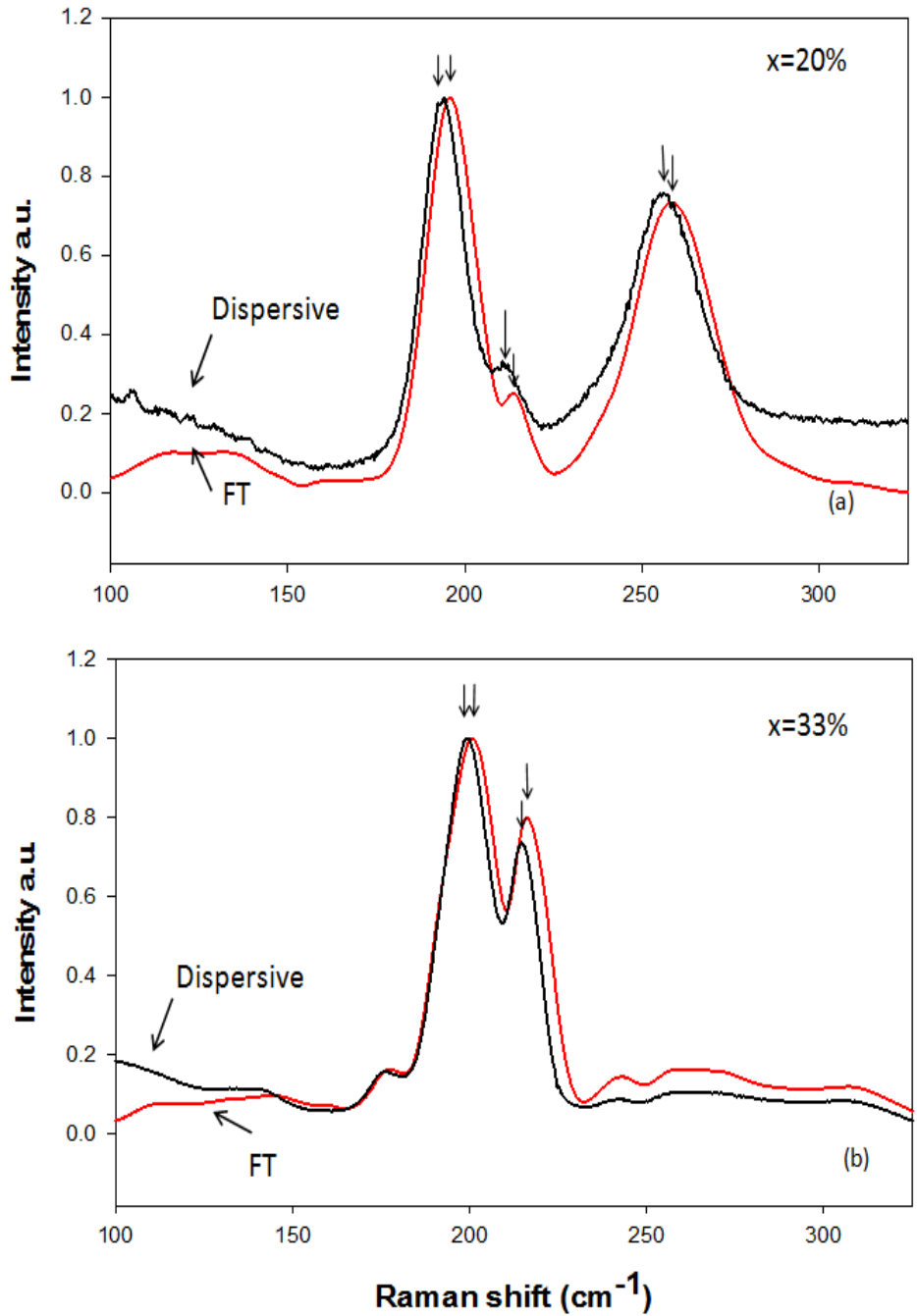

**Fig. 37** Comparison of the FT-Raman and Dispersive-Raman lineshapes of bulk $Ge_xSe_{100-x}$ glasses at (a) x = 20% and (b) x = 33% glass. Note that at x = 20% glass the peaks in FT-Raman spectrum are shifted to higher frequency than in dispersive Raman measurements as shown by the arrow locations. On the other hand at x = 33%, two lineshapes, FT-Raman and dispersive Raman, almost coincide near 180 $cm^{-1}$ but steadily deviate at larger frequency shifts.

**Figure Captions**

**Fig. 1** Phase Diagram of $Ge_xSe_{100-x}$ binary taken from Isper et al[38].

**Fig. 2** Raman scattering of a quenched $Ge_{19}Se_{81}$ melt taken along the length of a quartz tube used for reacting the melt at 950°C for 6 hours, and then lowering its temperature to $T_1 + 50C$, and water quenching. Note that spectra along the length of the tube, at indicated 9 points, show the lineshapes to systematically change. The narrow modes at arrow locations are those of α-$GeSe_2$ [32].

**Fig. 3** Raman profiling data providing a coalesced view of the 9 Raman spectra of Fig.2. Raman scattering of $Ge_{19}Se_{81}$ melt reacted for increasing reaction times appears in (b) $t_R = 24h$, (c) $t_R = 96h$, (d) $t_R = 168h$ (d). These data show that melts homogenize after reacting for 168h.

**Fig. 4** Raman profiling of $Ge_{15}Se_{85}$ melt reacted for increasing reaction times appears in (a) $t_R = 6h$, (b) $t_R = 24h$, (c) $t_R = 96h$, (d) $t_R = 168h$, demonstrating that melts homogenize after reacting for 168h.

**Fig. 5** Raman profiling of $GeSe_2$ melt reacted at 950°C for $t_R = 6h$, and taken along 9 points along the length of the quart tube containing the melt. The narrow modes at arrow locations are those of α-$GeSe_2$[32].

**Fig. 6** Raman profiling of $GeSe_2$ melt reacted at 950°C for $t_R = 24h$ taken along 9 points along the length of the quart tube containing the melt. The narrow modes at arrow locations are those of α-$GeSe_2$ [32].

**Fig. 7** Raman profiling of GeSe$_2$ melt reacted at 950°C for $t_R$ = 96h taken along 9 points along the length of the quart tube containing the melt. The peaks at the arrow locations indicate growth in quasi elastic scattering (see text).

**Fig. 8** Summary of Raman profiling data on a GeSe$_2$ melt reacted for (a) 6h (b) 24h, (c) 96h, (d) 168h. These data show that after 168h of reaction melt molecular structure <u>globally</u> homogenizes across the batch composition.

**Fig. 9** Raman scattering taken along the length of a quartz tube showing observed lineshapes at various locations of a GeSe$_2$ melt reacted for 180 h. For reference purposes, we have inserted the Raman lineshape of α-GeSe$_2$ between 6 and 7. The narrow modes at arrow locations are also observed at position 7 in the melt suggesting nucleation of α-GeSe$_2$ at the meniscus well.

**Fig. 10** Raman profiling of a GeSe$_2$ melt taken (left) at 10 μm laser spot size in a Dispersive measurement and (right) at a 50 μm laser spot size in an FT Raman measurement. The data suggest that melts homogenized on a 50 μm scale in an FT measurement are actually homogeneous on a finer scale of 10 μm scale.

**Fig. 11** FT Raman profiling data on a Ge$_{30}$Se$_{70}$ melt taken with a 250 μm spot size (a) 2 days and (b) 4 days after reacting at 950°C. Note that after 4 days of reaction the melt has homogenized. Melts homogenized on a coarser spatial scale (250 μm), homogenize quicker than those homogenized on a finer scale (50 μm as seen in Fig 8), but display a reversibility window that is not as sharp as in the finely homogenized melts. See text.

**Fig. 12** Raman profiling data on a $Ge_{19}Se_{81}$ melts (a) 2 gram melt reacted for 6h, (b) 2 gram melt reacted for 168h and (c)1/4 gm melt reacted for 6h. Smaller size melts homogenize much quicker than larger ones.

**Fig. 13** Raman profiling data on $Ge_{19}Se_{81}$ melts shown in the left panel after (a) being rocked for 48h (b) held stationary for 120h (c) held stationary for 168h. Parallel results are shown in the right panel for $Ge_{25.5}Se_{74.5}$ melts. Rocking melts assists homogenization incrementally.

**Fig. 14** Raman profiling data on (a) a wet $Ge_{19}Se_{81}$ melt reacted for 42h, a wet $GeSe_2$ melt reacted for (b) 6h (c) 48 h (c) 72h, showing the more dramatic effect of homogenization assisted by water traces in melts. Results on corresponding dry melts are given in Fig 3 and 8.

**Fig. 15** Dispersive Raman scattering of a wet melt compared to a dry one (a) at $Ge_{19}Se_{81}$ and (b) at $GeSe_2$. Note that the residual scattering in the wet melt exceeds that in the dry one.

**Fig. 16** Dispersive Raman scattering lineshape in quenched melts compared to $T_g$ cycled ones (a) x = 15% (b) x=17% and (c) x= 19%. Note that changes in lineshape between the two types of samples decreases considerably at x = 19%.

**Fig. 17** Dispersive Raman scattering lineshape in quenched melts compared to $T_g$ cycled ones (a) x = 22% (b) x=23% and (c) x= 24%. Note that changes in lineshape between the two types of samples is minuscule for these compositions in the thermally reversing window, suggesting that the melt and glass molecular structures are nearly the same in each case.

**Fig. 18** Dispersive Raman scattering lineshape in quenched melts compared to $T_g$ cycled ones (a) x = 29% (b) x=31% and (c) x= 33.33%. Note that changes in lineshape between the two types of samples increases as x > 26%. Note that the residual scattering in the as quenched melts is greater than in $T_g$ cycled ones for all these compositions in the stressed-rigid phase.

**Fig. 19** A summary of FT Raman profiling data on 18 of the 21 As-quenched $Ge_xSe_{100-x}$ melt compositions homogenized in the present study. The glass compositions are indicated as Ge content in %. We can see the growth in scattering strength of the of ES and CS tetrahedra at the expense of the Selenium chain mode as x increases from top left to bottom right.

**Fig. 20** Dispersive Raman scattering in indicated Tg cycled $Ge_xSe_{100-x}$ glass compositions showing evolution of lineshapes with increasing x from the bottom to the top. Of special interest are the modes near 247 $cm^{-1}$ nd 180 $cm^{-1}$ at the arrow locations once x > 31.5%. These are associated with Se-rich and Ge-rich moiety in the glasses once they segregate.

**Fig. 21** An example of Dispersive Raman lineshape deconvolution of a glass sample at x = 20% in terms of requisite number of Gaussian profiles using Peak Fit software. The corner sharing mode is shown in green, the edge sharing in yellow and the Selenium chain mode is identified in brown.

**Fig. 22** Variation in the frequency of the Corner sharing Ge tetrahedral mode as a function of x displaying three regimes, one at x < 20%, second between 20 % < x < 26%, and a third at x > 26%.See text for details.

**Fig. 23** Variations in (a) Sn chain mode frequency (b) scattering strength ratio of CM/CS (c) ES mode frequency and (d) ES/CS mode scattering strength as a function of $Ge_xSe_{100-x}$ glass composition. Earlier work (□) taken form Jin et al [49].The neutron structure factor determined [51] ES/CS fraction (▼) at x = 20% and 33.33% are plotted in panel (d).

**Fig. 24** Observed variation in the observed Raman integrated intensity $I_{CS}(x)$ and $I_{ES}(x)$ of the (a) CS and (b) ES mode in $Ge_xSe_{100-x}$ glasses with x is plotted. These data were obtained using 647nm excitation, keeping the power fixed at 5 mW in a macro mode with glass sample wetting in a quartz tube. $I_{CS}(x)$ fits well to a linear variation in x across the examined range. $I_{ES}(x)$ variation is linear in the 10% < x < 26% range, power law ($\alpha\ x^{n_1}$) in the stressed-rigid range 27% < x < 31% with $n_1$ = 2.30 (5) and power law in the 31.5% < x < 33.33% ($\alpha\ x^{n_2}$) with $n_2$ = 1.40 (5) in the NSPS range.

**Fig. 25** Typical mDSC scans of bulk $Ge_xSe_{100-x}$ glasses obtained at (a) x = 10%, (b) x = 19% (c) x = 20% and (d) x = 33.33% aged for 2 weeks at room temperature. Each panel shows 4 signals; the total, reversing and non-reversing heat flows in the heating cycle, and the non-reversing heat flow in the cooling cycle. Note that the enthalpy of relaxation, $\Delta H_{nr}$ term at x = 20% is minuscule (0.02 cal/gm).

**Fig. 26** Summary of mDSC results on all samples showing variations in (a) $T_g(x)$ and $dT_g/dx$ (▽). (b) $\Delta C_p(x)$ and (c) non-reversing heat flow $\Delta H_{nr}(x)$. In panel (a) $T_g(x)$ from the work of Feng et al[7]. (○), Sharma et al [19] (□) are included for comparison. $T_g$ of wet samples (▼) at x = 19% and x = 33.33% are included. In panel (b) $\Delta C_p(x)$ trends from Feng et al (○) and Wang et al[55] on the $Ge_xAs_xSe_{100-2x}$ ternary (○) are included. In panel (c), $\Delta H_{nr}(x)$ trends in fresh (F) glasses (▼), glasses aged (A1) for 2 weeks at RT (○), glasses aged (A2) at 240°C for 2 weeks (□) are included. Trends in $\Delta H_{nr}(x)$ reported by Feng et al. (◊), displaying a near triangular variation with x is included for comparison. The increase in $\Delta H_{nr}$ term in wet (▼) glasses compared to dry ones is shown by an arrow. See text.

**Fig. 27** Variation in molar volumes ($V_m(x)$) of present dry (●) and wet (○) $Ge_xSe_{100-x}$ glasses is compared to earlier reports by Mahadevan et al[20]. (▼),Feltz et al[30]. (■), and Senapati et al[31] (▲) are included. Note the larger variation IN $V_M(x)$ at x > 26% and x < 20% in the present set of samples than in earlier reports, probably related to sample homogeneity.

**Fig. 28** Variation in DSC measured $T_g(x)$ in present dry (▼) and wet (▼) glasses compared with those reported earlier by Guin et al[63] (●), Sreeram et al[64] (■). The DSC scan rate in all measurements was 10°C/min. The $T_g(x)$ of Guinn are 19°C less than our dry samples but coincide with our wet sample.

**Fig. 29** Elastic power-law (p) and thresholds ($x_c$) in the (a) stressed rigid and (b) the intermediate Phase obtained from the data of Fig 22. Here of $v_{cs}$ (x) represents the mode frequency variation of the CS mode. See text for details.

**Fig. 30** Elastic power-law (p) and threshold ($x_c$) of ES mode deduced from the data of Fig 23c. , The deduced power-law was obtained using both a polynomial fit and a log-log plot of the ES mode frequency variation $v_{ES}(x)$.

**Fig. 31** Reversibility window in GexSe100-x glasses reported (a) by Feng et al [7]. (1997), (b) Earlier report (a) (1997) and (b) Boolchand et al[80] (2009) and (c) in present work on glass samples of unprecedented homogeneity. It appears that the reversibility window in GexSe100-x glasses is intrinsically square-well like, resulting in *abrupt rigidity* and *stress* transitions. In experiments on glass samples possessing some heterogeneity of stoichiometry across a batch preparation, the window narrows and the walls broaden.

**Fig. 32** Variation of the reversibility window width W as a function of the heterogeneity ($\Delta x$) of Ge-fraction across a melt batch composition. The model assumes (d) an intrinsic square-well like reversibility window of W=6.5% in completely homogeneous melts (glasses) possessing vanishing $\Delta x$=0. With increasing heterogeneity of melts (glasses), $\Delta x$>0, the reversibility window width W and the slope of window walls steadily decreases as in (c), (b) and (a). When $\Delta x$=3%, the window becomes triangular in shape. Simulations show that for samples that are heterogeneous ($\Delta x$>0) the separation between midpoints of window walls, $W_{app}$, is a good measure of the intrinsic reversibility window width.

**Fig. 33.** Model prediction of reversibility window width W in $Ge_xSe_{100-x}$ glasses as a function of heterogeneity of glass samples characterized by a Ge-content spread, $\Delta x$, across a batch preparation.

**Fig. 34** Variation in the jump $\Delta C_p$ at Tg deduced from the reversing heat flow ($\triangledown$) and the total heat flow ($\bullet$) in mDSC measurements. The former reveals a variation with x that is almost flat, the latter on the other hand shows increases at x > 26% and x < 20% probably due to the overshoot in the heat flow endotherm for indicated compositions.

**Fig. 35** Variation in melt fragility m(x) ($\bullet$) reported by Stolen et al [41] and the presently reported $\Delta H_{nr}$ term ($\blacktriangledown$) in the present glasses as a function of x. Note that both terms show a minimum in the Intermediate Phase.

**Fig. 36** Variation in (a) the CS and (b) ES mode frequency with Ge content in bulk $Ge_xSe_{100-x}$ glasses deduced from FT-Raman (○) and Dispersive Raman (●) measurements. Note that in both the cases mode frequencies blue shift with x, however there is an offset between the two mode frequencies that steadily vanishes as x increases to x = 33.33%. In general, the dispersive measurements yield a lower frequency than the FT-measurement. Results from an FT-Raman study on bulk $Ge_xSe_{100-x}$ by Sen et al [88] are also projected in the fig as (■) data points for comparison to our data. We find that our FT-Raman results are in reasonable agreement with those of Sen et al.

**Fig. 37** Comparison of the FT-Raman and Dispersive-Raman lineshapes of bulk $Ge_xSe_{100-x}$ glasses at (a) x = 20% and (b) x = 33% glass. Note that at x = 20% glass the peaks in FT-Raman spectrum are shifted to higher frequency than in dispersive Raman measurements as shown by the arrow locations. On the other hand at x = 33%, two lineshapes, FT-Raman and dispersive Raman, almost coincide near 180 $cm^{-1}$ but steadily deviate at larger frequency shifts.

**Table 1**

|  | $T_g$ (°C) | $\Delta H_{nr}$ (cal/g) | $V_{mol}$ (cm$^3$mol$^{-1}$) | Comments |
|---|---|---|---|---|
| **19% Dry** | 171.6 | 0.360 | 18.34 | $t_R$= 168 h |
| **19% Wet** | 158.0 | 0.55 | 18.03 | $t_R$= 42 h |
| **33.33% Dry** | 425.7 | 0.52 | 18.87 | $t_R$= 192 h |
| **33.33% Wet** | 420.6 | 0.74 | 18.14 | $t_R$=72 h |